\documentclass{article}

\usepackage{arxiv}

\usepackage[utf8]{inputenc} 
\usepackage[T1]{fontenc}    
\usepackage{hyperref}       
\usepackage{url}            
\usepackage{booktabs}       
\usepackage{amsfonts}       
\usepackage{nicefrac}       
\usepackage{microtype}      
\usepackage{graphicx}
\usepackage{natbib}
\usepackage{doi}
\RequirePackage{amsthm,amsmath,amsfonts,amssymb}
\usepackage{float}
\usepackage{subfigure}
\theoremstyle{plain}
\newtheorem{p1}{PROPOSITION}[section]
\newtheorem{p2}[p1]{PROPOSITION}
\newtheorem{p3}[p1]{PROPOSITION}
\newtheorem{p9}[p1]{PROPOSITION}
\newtheorem{p4}[p1]{PROPOSITION}
\newtheorem{p5}[p1]{PROPOSITION}
\newtheorem{p10}[p1]{PROPOSITION}
\newtheorem{p8}[p1]{PROPOSITION}
\newtheorem{p11}[p1]{PROPOSITION}
\newtheorem{c1}[p1]{Corollary}
\newtheorem{c2}[p1]{Corollary}
\newtheorem{l7}{LEMMA}
\newtheorem{l8}[l7]{LEMMA}
\newtheorem{CLT}[l7]{LEMMA}
\newtheorem{l1}[l7]{LEMMA}
\newtheorem{l2}[l7]{LEMMA}
\newtheorem{l9}[l7]{LEMMA}
\newtheorem{l3}[l7]{LEMMA}
\newtheorem{l4}[l7]{LEMMA}
\newtheorem{l5}[l7]{LEMMA}
\newtheorem{l6}[l7]{LEMMA}
\newtheorem{l10}[l7]{LEMMA}

\theoremstyle{remark}
\newtheorem{assumption1}{Assumption}
\newtheorem{assumption2}[assumption1]{Assumption}
\newtheorem{assumption3}[assumption1]{Assumption}

\newtheorem*{remark}{Remark}
\newtheorem*{remark2}{Remark}
\newtheorem*{remark4}{Remark}
\newtheorem*{remark5}{Remark}
\newtheorem*{remark6}{Remark}

\newtheorem*{remark8}{Remark}
\newtheorem{definition1}{Definition}
\newtheorem{definition2}[definition1]{Definition}
\newtheorem*{notation}{Notation}

\title{A General Modeling Framework for \\ Network Autoregressive Processes}

\date{} 					

\author{ {Hang Yin} \\
	Department of Statistics\\
	University of Florida\\
	Gainesville, FL\\
	\texttt{hyin@ufl.edu} \\
	\And
	{Abolfazl Safikhani} \\
	Department of Statistics \& Informatics Institute\\
	University of Florida\\
	Gainesville, FL\\
	\texttt{a.safikhani@ufl.edu} \\
	\And
	{George Michailidis} \\
	Department of Statistics \& Informatics Institute\\
	University of Florida\\
	Gainesville, FL\\
	\texttt{gmichail@ufl.edu} \\

}

\renewcommand{\shorttitle}{A General Modeling Framework for Network Autoregressive Processes}

\hypersetup{
pdftitle={A General Modeling Framework for Network Autoregressive Processes},
pdfsubject={stat.ME, stat.TH},
pdfauthor={Hang Yin, Abolfazl Safikhani, George Michailidis},
pdfkeywords={Network autoregressive process, Stability, Inference, Misspecified network connectivity, Ridge penalty},
}

\begin{document}
\maketitle

\begin{abstract}
	The paper develops a general flexible framework for Network Autoregressive Processes (NAR), wherein the response of each node linearly depends on its past values, a prespecified linear combination of neighboring nodes and a set of node-specific covariates. The corresponding coefficients are node-specific, while the framework can accommodate heavier than Gaussian errors with both spatial-autorgressive and factor based covariance structures. We provide a sufficient condition that ensures the stability (stationarity) of the underlying NAR that is significantly weaker than its counterparts in previous work in the literature. Further, we develop ordinary and generalized least squares estimators for both a fixed, as well as a diverging number of network nodes, and also provide their ridge regularized counterparts that exhibit better performance in large network settings, together with their asymptotic distributions. We also address the issue of misspecifying the network connectivity and its impact on the aforementioned asymptotic distributions of the various  NAR parameter estimators. The framework is illustrated on both synthetic and real air pollution data.
\end{abstract}


\section{Introduction}
\label{sec:introduction}

Consider a network comprising of $N$ nodes, for which we collect measurements over $T$ time periods for a variable $X$; i.e. $X_{it}, i=1,\cdots,N, t=1,\cdots,T$. Depending on the application of interest, these nodes may correspond to agents/actors in a social network, companies in an economic network, sensors in an environmental network and even physical sites or devices in an engineering network. Further, for each node $i$ we also observe $p$ covariates $Y_{i,t}\in \mathbb{R}^p$ that are also time-varying. 
The model posited next assumes that the measurements $X_{it}$ for node $i$ are influenced by their past values (self-lags), plus past values of "related" nodes (network lags), after adjusting for the effect of covariates. Henceforth, we refer to this model as the Network Autoregressive (NAR) model. The corresponding NAR($q_1$,$q_2$) process with $q_1$ self-lags and $q_2$ network lags takes the form:
\begin{equation} 
    \label{eq:model-node}
    X_{i,t}=\sum\limits_{j=1}^{q_1}a_i^{(j)} X_{i(t-j)} + \sum\limits_{j=1}^{q_2}b_i^{(j)} \sum_{k=1}^N w_{i,k} X_{k(t-j)} +\gamma_i^TY_{i,(t-1)}+\epsilon_{i,t}, \ \ i=1,\cdots,N,
\end{equation}
where $a_i^{(j)} \in \mathbb{R}, b_i^{(j)}\in\mathbb{R}, \gamma_i\in\mathbb{R}^p$ are regression coefficients for the self-lags, the network lags and the covariates, respectively; further, $w_{i,k}\in [0,1]$ are weights capturing the degree of dependence among node $i$ and other nodes $k\neq i$. We impose further constraints on these weights in the sequel (see Assumption \ref{A3}). Finally, $\epsilon_{i,t}$ is an error term with $E(\epsilon_{i,t})=0$ and $E(\epsilon_{i,t})^{4}<\infty$, which is assumed to be independent of the covariates $Y_{i,t}$.

The posited model encompasses as special cases a number of models that appeared in recent literature, and also extends other related models, as discussed next. Specifically, 
\cite{zhu2017network} consider an NAR model with
$a_i=a, b_i=b$ for all $=1,\cdots,N$, while \cite{zhu2018grouped} assume that the nodes belong to $K$
groups $G_k, k=1,\cdots,K$ and thus all nodes in group $G_k$ share the same coefficients; i.e., $a_i=a_k, b_i=b_k$, for all $i\in G_k$. The assignment of nodes into groups is obtained from the data, by assuming a mixture model. Further, in both cases the error term is homoskedastic, i.e., $\epsilon_k \sim N(0,\sigma_kI)$, $k=1,\cdots,K$.

A variation of the model in \cite{zhu2018grouped} is presented in \cite{2007.05521}, wherein the adjacency matrix of the network $W$ is assumed to be generated by a Stochastic Block model with $K$ communities, which allows interactions between nodes belonging to the same community, as well as belonging to different communities. Further, the covariance matrix of the error term can exhibit factor structure, while the community structure is estimated from the data through spectral clustering. \cite{knight2019generalised} allow for different coefficients for the nodes, but do not consider exogenous covariates. Further, a variant of the popular in the econometrics literature Seemingly Unrelated Regressions model is also encompassed by the NAR one; specifically, by letting 
\[y_{it}=\beta_i^Tx_{it}+\epsilon_{it}, \ \,\,\,\, \text{where}\ \epsilon_{it}\sim F(0,\Sigma),\]
and $x_{1t} := y_{i,(t-1)}$, $x_{2t}:=w_i^T y_{t-1}$ while the remaining terms are $x_{it}:=Y_{ij,(t-1)}$ where $Y_{ij,(t-1)}$ are defined to be exogenous covariates.

Specific variants of the NAR model have been employed in diverse application areas, including social media analysis \citep{zhu2017network}, pollution \cite{zhu2018grouped} and environmental monitoring \cite{knight2019generalised},  economic growth studies \cite{knight2019generalised} and predicting stock market returns \cite{2007.05521}.

In the study of the $\{X_{it}\}$ NAR processes, the following two issues need to be addressed at the technical level: (i) their stability/stationarity, and (ii) estimation of their model parameters and inference. For the first issue, the work in the literature has adopted a rather stringent sufficient condition, that this paper substantially relaxes (see Proposition \ref{stationary} and ensuing discussion in Remark \ref{remark:stationarity}). For the second issue, the nature of the posited model dictates the estimation procedure and associated inference results. Specifically, \cite{zhu2017network} use ordinary least squares and establish asymptotic normality for the fixed number of the underlying estimated parameters.\cite{zhu2018grouped} use the EM algorithm to identify the underlying group structure, and then apply the NAR model defined in \cite{zhu2017network} to each group. \cite{knight2019generalised} use a least squares criterion to fit the model and establish the asymptotic normality of the model parameters assuming that the network size $N$ is fixed. \cite{2007.05521} use a multi-step estimation procedure to first identify the community structure, then the factor structure of the error term and finally through generalized least squares obtain estimates of the model parameters. Further, asymptotic distributions for the parameters are also derived.

The posited model in \ref{eq:model-node} has a growing number of parameters as a function of the network size, which adds flexibility to capture \textit{heterogeneity} across nodes, but also imposes technical challenges. In addition, a general structure of the covariance matrix is assumed for the error term, which is also flexible, but adds to the technical challenges. Hence, the key contributions of this work are:
(i) building a general flexible modeling framework for network autoregressive data, (ii) development of a relaxed sufficient condition for stability/stationarity of the underlying NAR process, (iii) establishing inference procedures for the growing number of model parameters, including regularized variants of the ordinary and generalized least squares estimates.

The remainder of the paper is organized as follows. Section \ref{sec:stability} addresses the key issue of stability (and hence stationarity) of the NAR process, while Section \ref{sec:theory} presents various estimators for the model parameters, together with their asymptotic distributions.
Section \ref{sec:performance} presents numerical studies that evaluate the performance of the proposed estimators both in terms of their estimation accuracy, as well as the coverage of their asymptotic distributions. Section \ref{sec:application} employs the NAR model to analyze air quality data from a number of monitoring stations in China.
Finally, Section \ref{sec:discussion} draws some concluding remarks.

 \begin{notation}
 Throughout the paper, we use $||A||_{\infty}$ to denote the matrix induced infinity norm of matrix $A\in \mathbb{R}^{m\times n}$, that is, $||A||_{\infty}=\mathop{max}\limits_{1\leq i\leq m}\sum\limits_{j=1}^n|a_{ij}|$. Further, we use $||A||_{max}$, $||A||$ and $||A||_F$ to denote the (element-wise) max norm, the oprator norm and Frobenius norm of $A$, respectively. We use $e_i$ to denote the $i$-th unit vector in $\mathbb{R}^p$. For a symmetric or Hermitian matrix $A$, we denote its spectral radius by $\rho(A)$.
 \end{notation}
\section{Stability of the NAR process}
\label{sec:stability}

The first issue addressed is to derive conditions that ensure the stability/stationarity of the NAR($q_1,q_2$) process for the model posited in \eqref{eq:model-node}. 

To proceed, some additional notation is required.
Let $A_i:= \text{diag}\{a_1^{(i)},a_2^{(i)},...,a_N^{(i)}\}\in \mathbb{R}^{N\times N}$ for $i=1,2,\cdots,q_1$, $B_j:=\text{diag}\{b_1^{(j)},b_2^{(j)},...,b_N^{(j)}\}\in \mathbb{R}^{N\times N}$ for $j=1,2,\cdots,q_2$, $C_k:=\text{diag}\{c_{1k},c_{2k},...,c_{Nk}\}\in \mathbb{R}^{N\times N}$ for $k=1,2,\cdots,p$, and $G_\ell:=A_\ell+B_\ell W$, wherein $\ell=1,2,\cdots,\max\{q_1,q_2\}$, with the convention that zero matrices are included/padded for the relationship to hold; namely, if $q_1>q_2$, $B_j=0$ for $j>q_2$, whereas if $q_1<q_2$, $A_j=0$ for $j>q_1$. Let $q=\max\{q_1,q_2\}$, then we can rewrite the NAR($q_1,q_2$) model posited in matrix form as follows:
\begin{equation}
    \label{eq:matrix}
 \mathbb{X}_{t}=\sum\limits_{i=1}^{q_1} A_i \mathbb{X}_{t-i} + \sum\limits_{j=1}^{q_2} B_j W \mathbb{X}_{t-j} +\sum\limits_{k=1}^{p}C_k\mathbb{Y}_{k,(t-1)}+\epsilon_{t}=\sum\limits_{\ell=1}^{q}G_\ell\mathbb{X}_{t-\ell}+
\sum\limits_{k=1}^{p}C_k\mathbb{Y}_{k,(t-1)}+\epsilon_{t},
\end{equation}
wherein $\mathbb{X}_{t}$ and $\mathbb{X}_{t-\ell}\in \mathbb{R}^{N}$ and $\mathbb{Y}_{k,(t-1)}=\left[Y_{1k,(t-1)}\ \cdots \ Y_{Nk,(t-1)}\right]^T\in \mathbb{R}^{N}$. We impose further constraints on these weights in the sequel (see Assumption \ref{A3}). Finally, $\epsilon_{i,t}$ is an error process with $E(\epsilon_{i,t})=0$ and $E(\epsilon_{i,t})^{4}<\infty$, which is assumed to be independent of the covariates $Y_{i,t}$. Additional conditions on the error processes are discussed in the sequel.

For future technical developments, it is convenient to also express \eqref{eq:matrix} in the following form:
\setcounter{MaxMatrixCols}{20}
\begin{equation} \label{compact}
\begin{aligned}
\mathbb{X}_{t}& = \mathbb{Z}_{t-1}\beta+ \epsilon_{t},
\end{aligned}
\end{equation} 
where \[\mathbb{Z}_{t-1}:=\begin{bmatrix}
Z_{t-1} & Z_{t-2} & \cdots & Z_{t-q} & \text{diag}\{\mathbb{Y}_{1,(t-1)}\}&\cdots& \text{diag}\{\mathbb{Y}_{p,(t-1)}\}
\end{bmatrix},\] \[Z_{t-\ell}:=\begin{bmatrix}\text{diag}\{\mathbb{X}_{t-l}\}&\text{diag}\{W\mathbb{X}_{t-l}\}\end{bmatrix},l=1,\cdots,q,\] \[\beta:=\begin{bmatrix}
\beta_1^T&\beta_2^T&\cdots&\beta_q^T&\gamma_1^T&\gamma_2^T&\cdots&\gamma_p^T\end{bmatrix}^T,\] \[\beta_\ell:=\begin{bmatrix}
a_1^{(\ell)}&
a_2^{(\ell)}&
\cdots&
a_N^{(\ell)}&
b_1^{(\ell)}&
b_2^{(\ell)}&
\cdots&
b_N^{(\ell)}
\end{bmatrix}^T,l=1,\cdots,q,\]
and \[\gamma_k:=\begin{bmatrix}
c_{1k}&
c_{2k}&
\cdots&
c_{Nk}
\end{bmatrix}^T,k=1,\cdots,p.\]

Let $\Tilde{\epsilon}_{t}=\epsilon_{t}+
\sum\limits_{k=1}^{p}C_k\mathbb{Y}_{k,(t-1)}$ and rewrite \eqref{eq:matrix} as \begin{equation}
\label{varq}
\mathbb{X}_t=\sum\limits_{\ell=1}^{q}G_\ell\mathbb{X}_{t-\ell}+\Tilde{\epsilon}_{t},
\end{equation}
which can be considered as a vector autoregressive model (VAR) with transition matrix $G$ and error term $\Tilde{\epsilon}_t$. The latter model can also be expressed as a VAR(1) one model (see \citep{lutkepohl2005new}):
\begin{equation}
\label{var1}
\boldsymbol{X}_t=\boldsymbol{G}\boldsymbol{X}_{t-1}+\mathcal{E}_t
\end{equation}
with
\begin{equation}
\label{XEG}
\begin{split}
&\boldsymbol{X}_t:=\begin{bmatrix}\mathbb{X}_t^T&\mathbb{X}_{t-1}^T&\cdots&\mathbb{X}_{t-q+1}^T\end{bmatrix}^T,\ \mathcal{E}_t:=\begin{bmatrix}\Tilde{\epsilon}_t^T&0^T&\cdots&0^T \end{bmatrix}^T, \text{and} \\
&\boldsymbol{G}:=\begin{bmatrix}G_1&G_2&\cdots&G_{q-1}&G_q\\I_N&0&\cdots&0&0\\0&I_N&\cdots&0&0\\\vdots&\vdots&\ddots&\vdots&\vdots\\0&0&\cdots&I_N&0 \end{bmatrix}.
\end{split}
\end{equation}

Before stating the main result, we introduce the following assumptions:
\begin{assumption1}
\label{A1}
 Moment conditions on $\epsilon_t$ and $\mathbb{Y}_{t}$:
    \begin{itemize}
\item[(i)]$\{\epsilon_t,t\in \mathbb{N}\}$ is a sequence of random vectors satisfying $E(\epsilon_t)=0$, $\Sigma_\epsilon=E(\epsilon_t\epsilon_t^T)$ is nonsingular, $\epsilon_t$ and $\epsilon_s$ are independent for $s\not=t$, and for some finite constant $c_1$, the following relationship holds
\[E|\epsilon_{it}\epsilon_{jt}\epsilon_{kt}\epsilon_{mt}|\leq c_1\ \text{for} \ i,j,k,m=1,\cdots,N, \text{and all} \  t.\]
    \item[(ii)] 
$\{\mathbb{Y}_{t},t\in \mathbb{N}\}$ where $\mathbb{Y}_{t}:=\begin{bmatrix}\mathbb{Y}_{1,t}^T&\mathbb{Y}_{2,t}^T&\cdots&\mathbb{Y}_{p,t}^T\end{bmatrix}^T$ is a sequence of i.i.d. random vectors with $E(\mathbb{Y}_{t})=0$ and $E(\mathbb{Y}_t\mathbb{Y}_t^T)=\Sigma_Y$, and for some finite constant $c_2$, the following relationship holds:
\begin{multline*}
E|Y_{i_1j_1,t}Y_{i_2j_2,t}Y_{i_3j_3,t}Y_{i_4j_4,t}|\leq c_2, \\ \text{for}\ i_1,i_2,i_3,i_4=1,\cdots,N,\ j_1,j_2,j_3,j_4=1,\cdots,p \text{ and all}\ t.
\end{multline*}
   \item[(iii)] $\{\epsilon_t,t\in \mathbb{N}\}$ is independent of $\{\mathbb{Y}_{t},t\in \mathbb{N}\}$. 
    \end{itemize}
\end{assumption1}
\begin{assumption2}
    \label{A3} $W\in \mathbb{R}^{N\times N}$ is a row-normalized matrix; i.e., $\sum_{j=1}^N w_{ij}=1$ with $w_{ij}\geq0$.
\end{assumption2}
\begin{assumption3}
  For diverging network size $N$ as a function of time $T$, $\{\epsilon_t,t\in \mathbb{N}\}$ and $\{\mathbb{Y}_{t-1},t\in \mathbb{N}\}$ are assumed to be sub-Weibull (sub-Weibull random vectors are defined next in Definition \ref{d:swrv} following along the lines in \cite{wong2020lasso}). \label{A2}  
\end{assumption3}

Assumptions~\ref{A1} (i)-(ii) requires existence of fourth moments for the error process, as well as the covariate processes. While finiteness of second moments is sufficient to ensure the existence of a unique stationary solution to the recursive equations \eqref{eq:matrix}, finiteness of fourth moments are needed to establish the asymptotic normality of the various estimators presented in Section \ref{sec:theory}. Further, Assumption \ref{A1} (iii) requires independence between the error and the covariate process, which makes the latter process \textit{exogenous}. Assumption \ref{A3} is required for the identifiability of the network coefficients and is needed both for establishing the stability/stationarity of the NAR process and for the asymptotic properties of the estimators of the model parameters. Finally, Assumption \ref{A2} imposes a mild condition on the tail behaviour of the distribution of the error and the covariate processes that encompasses a wide range of possibilities, including sub-Gaussian and sub-exponential random variables. 

\begin{remark}
Note that all prior work in the literature \citep{zhu2017network,zhu2018grouped,2007.05521,knight2019generalised} 
assumes that both the exogenous variables $Y_t$, as well as the error terms $\epsilon_t$ are normally distributed. 
Assumption \ref{A2} relaxes significantly this requirement.
\end{remark}

\begin{l10}[Sub-Weibull properties]
\label{l:sw}
Let $X$ be a random variable. Then, the following statements are equivalent for every $\gamma>0$. The constants $K_1$, $K_2$, $K_3$ differ from each other at most by a constant depending only on $\gamma$.
\begin{itemize}
    \item 
    The tails of $X$ satisfies 
    \[P(|X|>t)\leq 2\mathop{exp}\{-(t/K_1)^\gamma\},\forall t\geq 0.\]
    \item
    The moments of $X$ satisfy,
    \[||X||_p:=(E|X|^p)^{1/p}\leq K_2p^{1/\gamma},\forall p\geq 1\wedge \gamma.\]
    \item 
    The moment generating function of $|X|^\gamma$ is finite at some point; namely
    \[E(\mathop{exp}(|X|/K_3)^\gamma\leq 2).\]
\end{itemize}
\end{l10}

\begin{definition1}[Subweibull($\gamma$) Random Variable and Norm]
A random variable $X$ that satisfies any property in Lemma \ref{l:sw} is called a sub-Weibull($\gamma$) random variable. The sub-Weibull($\gamma$) norm associated with X, denoted $||X||_{\psi_\gamma}$, is defined to be the smallest constant such that the moment condition in Lemma \ref{l:sw} holds. In other words, for every $\gamma>0$,
\[||X||_{\psi_{\gamma}}:=\mathop{sup}_{p\geq 1}(E|X|^p)^{1/p}p^{-1/\gamma}.\]
\end{definition1}

\begin{definition2}
\label{d:swrv}
Let $\gamma \in (0,\infty).$ A random vector $X\in \mathbb{R}^p$ is said to be a sub-Weibull($\gamma$) random vector if all of its one dimensional projections are sub-Weibull($\gamma$) random variables. We define the sub-Weibull($\gamma$) norm of a random vector as 
\[||X||_{\psi_{\gamma}}:=\mathop{sup}_{v\in S^{p-1}}||v^TX||_{\psi_{\gamma}},\]
where $S^{p-1}$ is the unit sphere in $\mathbb{R}^p.$
\end{definition2}

\begin{p1}
\label{stationary}
Consider the NAR($q_1,q_2$) process defined recursively by $\mathbb{X}_t=\sum\limits_{\ell=1}^{q}G_\ell\mathbb{X}_{t-\ell}+\Tilde{\epsilon}_{t}$ where $G_\ell=A_\ell+B_\ell W$ and $\Tilde{\epsilon}_{t}=\epsilon_{t}+\sum\limits_{k=1}^{p}C_k\mathbb{Y}_{k,(t-1)}$.
Assume Assumptions \ref{A1}-\ref{A3} hold. Then, $\mathbb{X}_t$ is a stationary process with a finite first order moment that can be expressed as:
\[\mathbb{X}_t=\begin{bmatrix}
I_N& \boldsymbol{0}_{N\times N(q-1)}
\end{bmatrix}\sum_{j=0}^{\infty}\boldsymbol{G}^j\mathcal{E}_{t-j},\]  
if $\rho(\boldsymbol{G})<1$. Equivalently, $\mathbb{X}_t$ is a stationary process if
\[det(I_{Nq}-\boldsymbol{Gz})=det(I_{N}-G_1z-\cdots-G_q z^q)\not=0\ \text{for} \ |z|\leq 1.\] 
\end{p1}

\begin{proof}
The proof of Proposition \ref{stationary} is provided in Appendix \ref{pos}.
\end{proof}

\begin{remark4}\label{remark:stationarity}
For row-normalized $W$, $\mathop{max}\limits_{1\leq i\leq N}\{\sum\limits_{l=1}^q(|a_i^{(l)}|+|b_i^{(l)}|)\}<1$ is only a sufficient condition for an NAR$(q_1,q_2)$ model to be stationary.
\end{remark4}

\begin{remark5}\label{remark:comparisons-stationarity}
The stability/stationarity condition in Proposition \ref{stationary} is significantly weaker than those in the literature for even special cases of the posited model as illustrated next.
\begin{itemize}
    
\item[1.]
\noindent
\textit{Homogeneous NAR Process:}

\cite{zhu2017network} require that $|a|+|b|<1$ for an NAR(1,1) process with $A_1=aI$ and $B_1=bI$, which is both necessary and sufficient. They also require $\sum\limits_{l=1}^q(|a^{(l)}|+|b^{(l)}|)<1$ for an NAR(q,q) process with $A_l=a^{(l)}I$, $B_l=b^{(l)}I$; however, this condition guarantees stationarity of the process, but there exist processes that are stationary and violate the condition as the next simple numerical example shows.

Let $N=2$, thus $X_t=\begin{bmatrix}X_{1t}\\X_{2t}\end{bmatrix}$.

\begin{itemize}
\item 
\textit{Sufficiency.} Consider an NAR$(1,1)$ process with $W=\begin{bmatrix}0&1\\1&0\end{bmatrix}$. Further, let $A_1=aI_2$ and $B_1=bI_2$, so that  $\boldsymbol{G}=\begin{bmatrix}a&b\\b&a\end{bmatrix}.$ 
    It is then easy to see that $\rho(\boldsymbol{G})<1$, if and only if $|a|+|b|<1$.
    \item  
\textit{Non-necessity:} For an NAR$(2,2)$ process with $W=\begin{bmatrix}0&1\\1&0\end{bmatrix}$, let $A_1=1.5I_2$, $A_2=-0.8I_2$, $B_1=0.1I_2$ and $A_2=0.1I_2$, so that $G=\begin{bmatrix}1.5&0.1&-0.8&0.1\\0.1&1.5&0.1&-0.8\\1&0&0&0\\0&1&0&0\end{bmatrix}$. In this case, $\sum_{i=1}^2(|a^{(i)}|+|b^{(i)}|)>1$, but $\rho(\boldsymbol{G})=0.949<1$. Hence, the condition in \cite{zhu2017network} is only sufficient.
\end{itemize}

\item[2.]
\noindent
\textit{Heterogeneous NAR process:}

    Consider an NAR$(1,1)$ process with $A_1=\begin{bmatrix}a_1&0\\0&a_2\end{bmatrix}$ and $B_1=\begin{bmatrix}b_1&0\\0&b_2\end{bmatrix}$, so that $\boldsymbol{G}=\begin{bmatrix}a_1&b_1\\b_2&a_2\end{bmatrix}.$ The stationary condition in Proposition \ref{stationary} becomes:
    \[A(z)=1-(a_1+a_2)z-(b_1b_2-a_1a_2)z^2\not=0,|z|\leq 1\]
    which holds if and only if the following three relationships are simultaneously satisfied \[|a_1+a_2|<2,\ |b_1b_2-a_1a_2|<1,\ b_1b_2-a_1a_2<1\pm(a_1+a_2)\]
    For a concrete numerical illustration, let $A_1=\begin{bmatrix}0.8&0\\0&0.5\end{bmatrix}$ and $B_1=\begin{bmatrix}0.1&0\\0&0.6\end{bmatrix}$, so that $\boldsymbol{G}=\begin{bmatrix}0.8&0.1\\0.6&0.5\end{bmatrix}.$ 
    In this case, the condition in \cite{zhu2018grouped} and \cite{knight2019generalised} requires $max\{|a_i|+|b_i|\}<1$ which is clearly violated. Nevertheless,  $\rho(\boldsymbol{G})=0.937<1$, which shows the sufficiency of the former condition. To visually illustrate the difference:
    \begin{itemize}
\item[(I)] Fix $a_2=0.1$ and $b_2=0.1$, then the region containing stationary solutions as a function of $a_1$ and $b_1$ is depicted in the left panel of Figure \ref{fig:stability-plot} (blue) together with the region implied by the condition in \cite{zhu2018grouped} and \cite{knight2019generalised}.
\item[(II)] Fix $a_2=0.5$ and $b_2=0.5$, and then the corresponding plot for the region of stationary solutions as a function of $a_2$ and $b_2$ is depicted in the right panel of Figure \ref{fig:stability-plot}.

\begin{figure} [H]
\centering
\subfigure[$a_2=0.1$ and $b_2=0.1$]{
\begin{minipage}[t]{6cm}
\centering
\includegraphics[scale=0.3]{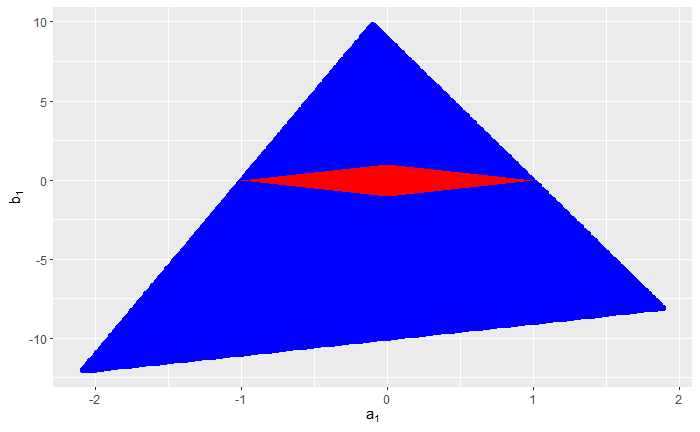}
\end{minipage}
}
\subfigure[$a_2=0.5$ and $b_2=0.5$]{
\begin{minipage}[t]{6cm}
\centering
\includegraphics[scale=0.3]{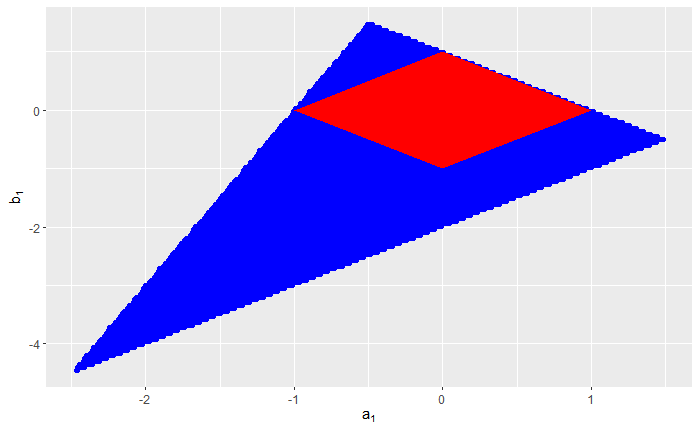}
\end{minipage}
}
\caption{Stability regions implied by the conditions in Proposition \ref{stationary} vs those used in the literature
\label{fig:stability-plot}}
\end{figure}

\end{itemize}
\end{itemize}

\end{remark5}

\section{Estimation Procedures for the NAR Model and their Asymptotic Properties} \label{sec:theory}

We consider the following estimators for the NAR model: (1) ordinary least squares (OLS), (2) generalized least squares (GLS), and (3) the empirical counterpart of (2) (EGLS).
In all subsequent developments, we consider the following two regimes for the network size: (I) the number of nodes $N$ remains fixed, while the number of time points $T$ diverges to infinity, and (II) both the network size $N$ and time points $T$ grow to infinity at appropriately defined rates.

\subsection{OLS Estimator} \label{sec:ols}
This estimator formally defined next, ignores the covariance structure of the error term: 

\begin{equation}\label{beta-ols}
\hat{\beta}_{OLS}=(\sum_{t=1}^{T}\mathbb{Z}_{t-1}^T\mathbb{Z}_{t-1})^{-1}\sum_{t=1}^{T}\mathbb{Z}_{t-1}^T\mathbb{X}_t
=\beta+(\sum_{t=1}^{T}\mathbb{Z}_{t-1}^T\mathbb{Z}_{t-1})^{-1}\sum_{t=1}^{T}\mathbb{Z}_{t-1}^T\epsilon_t.
\end{equation}

\medskip\noindent
\textbf{(I) Fixed network size $N$}:

\begin{p2}[Asymptotic Properties of the OLS Estimator]	\label{P:OLS} 
Suppose Assumptions \ref{A1}-\ref{A3} hold. For a stationary process $\mathbb{X}_t$ defined by the NAR$(q_1,q_2)$ model \eqref{compact}, i.e., $\mathbb{X}_{t}=\mathbb{Z}_{t-1}\beta+ \epsilon_{t}$, with finite network size $N$, $\hat{\beta}_{OLS}$ is a consistent estimator of $\beta$ and its asymptotic distribution is given by
\begin{equation}
\label{OLS}
\sqrt{T}(\hat{\beta}_{OLS}-\beta)\rightarrow_{d}N(0, P^{-1}QP^{-1})
\end{equation}
where $P:=
E(\mathbb{Z}_{t}^T\mathbb{Z}_{t})$, $Q:=
E(\mathbb{Z}_{t}^T\Sigma_{\epsilon}\mathbb{Z}_{t})$.
\end{p2}

\begin{proof}
The two main components in the proof of Proposition~\ref{P:OLS} are: \\ (a) verifying that $\frac{1}{T}\sum\limits_{t=1}^{T}\mathbb{Z}_{t-1}^T\mathbb{Z}_{t-1}\rightarrow_{p}P$ in which the stability (stationarity) of the process is leveraged as discussed in Lemma \ref{L:LLNOLS}; (b) proving that $\frac{1}{\sqrt{T}}\sum\limits_{t=1}^{T}\mathbb{Z}_{t-1}^T\epsilon_{t}\rightarrow_{d}N(0, Q )$ which leverages a central limit theorem for martingale differences (Theorem 5.3.4 in \cite{fuller2009introduction}), since the summand terms are indeed dependent (due to the presence of temporal dependence); details are provided in Lemma \ref{L:CLTOLS}.

The detailed proof is given in Appendix \ref{a4} and is based on Lemmas
\ref{L:LLNOLS}, \ref{L:BDD} and \ref{L:CLTOLS} provided in Appendices \ref{a1}-\ref{a3}.
\end{proof}

\begin{remark5}
In practice, the quantity $E(\mathbb{Z}_{t}^T\mathbb{Z}_{t})$ can be estimated by $\frac{1}{T}\sum\limits_{t=1}^T\mathbb{Z}_{t}^T\mathbb{Z}_{t}$, and $E(\mathbb{Z}_{t}^T\Sigma_{\epsilon}\mathbb{Z}_{t})$ by $\frac{1}{T}\sum\limits_{t=1}^T\mathbb{Z}_{t}^T\hat{\Sigma}_\epsilon\mathbb{Z}_{t}$, wherein $\hat{\Sigma}_\epsilon=\frac{1}{T}\sum\limits_{t=1}^T\hat\epsilon_t\hat\epsilon_t^T$. 

Table \ref{ttt} evaluates how $P^{-1}QP^{-1}$ improves for larger sample size $T$ by considering \[\text{RMSE}=||P^{-1}QP^{-1}-(\frac{1}{T}\sum\limits_{t=1}^T\mathbb{Z}_{t}^T\mathbb{Z}_{t})^{-1}\frac{1}{T}\sum\limits_{t=1}^T\mathbb{Z}_{t}^T\hat{\Sigma}_\epsilon\mathbb{Z}_{t}(\frac{1}{T}\sum\limits_{t=1}^T\mathbb{Z}_{t}^T\mathbb{Z}_{t})^{-1}||_F.\]
\end{remark5}
\begin{table}[H] 
\centering
\caption{NAR$(1,1)$ model with $N=100$, $A=0.3I_{100}$, $B=0.3I_{100}$ and $W$ and $\Phi$ being a banded matrix of ``width" 5. The error covariance matrix comes from a spatial autoregressive model with parameter $\rho=0.5$ (for details see 
Section \ref{sec:SAR}).}
\begin{tabular}{c|ccccc}
\hline
$T$ & 50 & 100 & 300 & 500 & 1000\\ \hline
RMSE & 11.36 & 8.78 & 7.23 & 6.52 & 5.87 \\ \hline
\end{tabular}
\label{ttt}
\end{table}

\noindent
\textbf{(II) Growing network size with $N\leq T$}:

\begin{p3}[Asymptotic Properties of the OLS Estimator]
\label{P:DOLS}
Suppose Assumptions \ref{A1}-\ref{A2} hold. Let $\mathbb{X}_t$ be a stationary process generated by the NAR$(q_1,q_2)$ model \eqref{compact}: $\mathbb{X}_{t}=\mathbb{Z}_{t-1}\beta+ \epsilon_{t}$ with growing network size $N$. Define $D\in \mathbb{R}^{k\times(2Nq+Np)}$ for any finite $k$. Further, assume:
\begin{itemize}
\item
$D$ has bounded row sums; i.e., for $i=1,\cdots,k$, there exists a finite constant $c$ such that $\sum\limits_{j=1}^{2Nq+Np}d_{i,j}\leq c$ where $d_{i,j}$ is the ij-th element of D.
\item
 $N\leq T$.
\end{itemize}
Then,
\begin{equation}
\label{DOLS}
\frac{1}{\sqrt{T}}D(\sum_{t=1}^{T}\mathbb{Z}_{t-1}^T\mathbb{Z}_{t-1})(\hat{\beta}_{OLS}-\beta)\rightarrow_{d}N(0, DQD^{T})
\end{equation}
where $Q:=
E(\mathbb{Z}_{t}^T\Sigma_{\epsilon}\mathbb{Z}_{t})$.
\end{p3}

\begin{proof}
The proof of Proposition \ref{P:DOLS} is given in Appendix \ref{a6} and uses Lemma \ref{L:CLTDOLS} that is also stated and proved in the same Appendix.
\end{proof}

\begin{remark6}
Next, we illustrate the nature of $D$ and that of the confidence regions implied by the Proposition. 
Consider an NAR$(1,1)$ model with $N=100$, $A=0.3I_{100}$, $B=0.3I_{100}$ and $W$ and $\Phi$ being a banded matrix of ``width" 5. The error covariance matrix comes from a spatial autoregressive model with parameter $\rho=0.5$ (for details see 
Section \ref{sec:SAR}). Set $D_{(2\times 200)}=\begin{bmatrix}\begin{array}{c|c} e_i^T&\boldsymbol{0}_{100}^T\\
\boldsymbol{0}_{100}^T&e_i^T\end{array}\end{bmatrix}$.  Define $K:=\frac{1}{\sqrt{T}}D(\sum_{t=1}^{T}\mathbb{Z}_{t-1}^T\mathbb{Z}_{t-1})(\hat{\beta}_{OLS}-\beta)$, then \[K^T(DQD^T)^{-1}K\sim \chi^2_2.\]
The resulting confidence regions for an arbitrary pair of $(a_i,b_i)$ for different values of the sample size $T$ are depicted in
Figure \ref{Fig:CR}.
\begin{figure} [H]
\centering
\subfigure[T=150]{
\begin{minipage}[t]{6cm}
\centering
\includegraphics[scale=0.3]{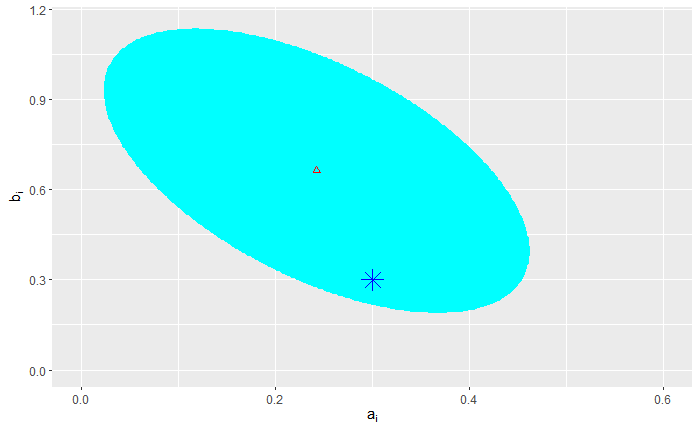}
\end{minipage}
}
\subfigure[T=300]{
\begin{minipage}[t]{6cm}
\centering
\includegraphics[scale=0.3]{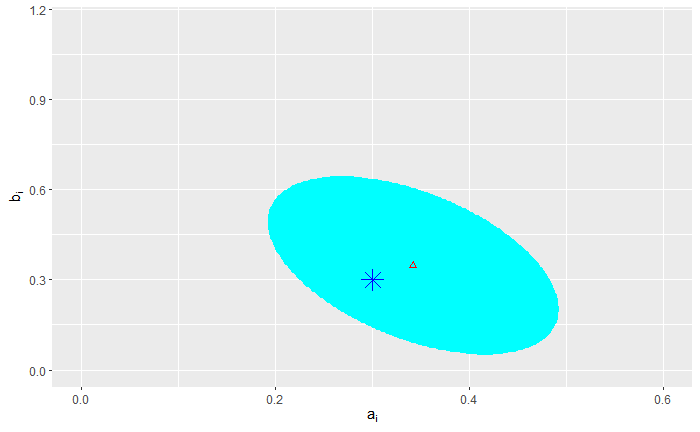}
\end{minipage}
} 
\subfigure[T=450]{
\begin{minipage}[t]{6cm}
\centering
\includegraphics[scale=0.3]{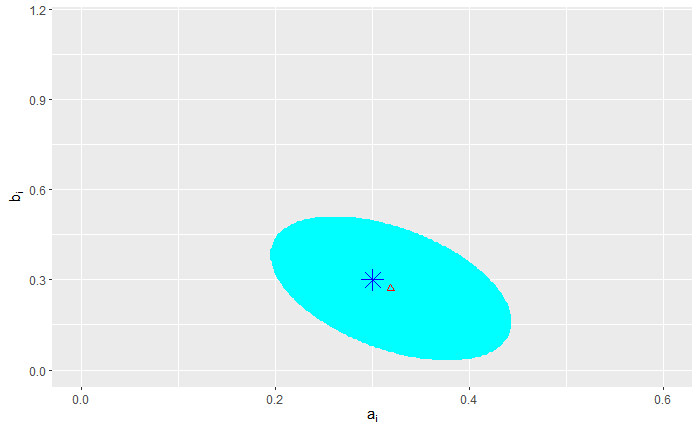}
\end{minipage}
} 
\subfigure[T=600]{
\begin{minipage}[t]{6cm}
\centering
\includegraphics[scale=0.3]{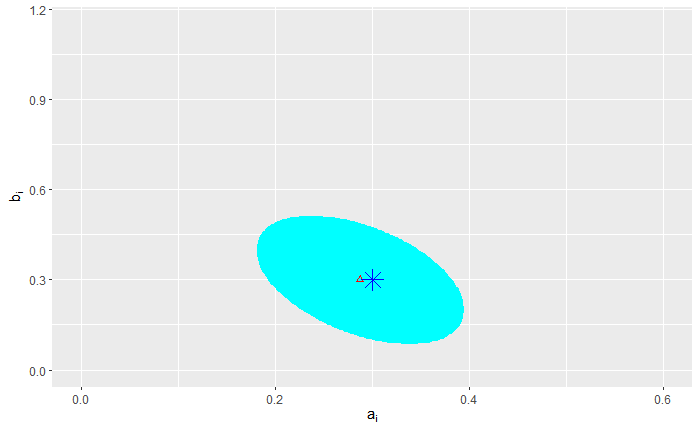}
\end{minipage}
} 
\caption{Confidence Regions for different sample sizes $T$. The true value of the parameters are denoted by $*$. \label{Fig:CR}}
\end{figure}
It can be seen that the volume of the confidence region shrinks for larger $T$ as expected, and also the center of the region gets closer to the true value of the parameters.
\end{remark6}

\textbf{(III) Growing network size with $N>T$}:

For $N>T$ case, we consider a \textit{ridge} regularized estimator defined as:
\begin{equation}\label{beta-ols-r}
\hat{\beta}_{ridge}=(\sum_{t=1}^{T}\mathbb{Z}_{t-1}^T\mathbb{Z}_{t-1}+T M)^{-1}\sum_{t=1}^{T}\mathbb{Z}_{t-1}^T\mathbb{X}_t,
\end{equation} 
where $M:= diag\{\lambda_1I_N,\lambda_2I_N,\cdots,\lambda_1I_N,\lambda_2I_N,\lambda_3I_{Np}\}$.
The ridge regularization proves beneficial both empirically, as shown in Section \ref{sec:performance} and technically, ensuring that $(\sum_{t=1}^{T}\mathbb{Z}_{t-1}^T\mathbb{Z}_{t-1}+T M)$ is invertible for $N>T$ case.

\begin{p9}[Asymptotic Properties of the Ridge Estimator]
\label{P:DOLSR}
Suppose Assumptions \ref{A1}-\ref{A2} hold. Let $\mathbb{X}_t$ be a stationary process generated by the NAR$(q_1,q_2)$ model \eqref{compact}: $\mathbb{X}_{t}=\mathbb{Z}_{t-1}\beta+ \epsilon_{t}$ with growing network size $N$. Define $D\in \mathbb{R}^{k\times(2Nq+Np)}$ for any finite $k$. Further, assume:
\begin{itemize}
\item
$D$ has bounded row sums; i.e., for $i=1,\cdots,k$, there exists a finite constant $c$ such that $\sum\limits_{j=1}^{2Nq+Np}d_{i,j}\leq c$ where $d_{i,j}$ is the ij-th element of D.
\item
$N>T$.
\item 
$\lambda_i=o(\frac{1}{\sqrt T})$ for $i=1,2,3$.
\end{itemize}
Then,
\begin{equation}
\label{DOLSR}
\frac{1}{\sqrt{T}}D(\sum_{t=1}^{T}\mathbb{Z}_{t-1}^T\mathbb{Z}_{t-1}+T M)(\hat{\beta}_{ridge}-\beta)\rightarrow_{d}N(0, DQD^{T}),
\end{equation}
where $Q:=
E(\mathbb{Z}_{t}^T\Sigma_{\epsilon}\mathbb{Z}_{t})$
\end{p9}
\begin{proof}
The proof is given in Appendix \ref{a9}.
\end{proof}

\subsection{GLS estimator}

The GLS estimator is defined next, and uses the covariance structure of the error term:
\begin{equation}\label{beta-gls}
\hat{\beta}_{GLS}=(\sum_{t=1}^{T}\mathbb{Z}_{t-1}^T\Sigma_{\epsilon}^{-1}\mathbb{Z}_{t-1})^{-1}\sum_{t=1}^{T}\mathbb{Z}_{t-1}^T\Sigma_{\epsilon}^{-1}\mathbb{X}_t\\
=\beta+(\sum_{t=1}^{T}\mathbb{Z}_{t-1}^T\Sigma_{\epsilon}^{-1}\mathbb{Z}_{t-1})^{-1}\sum_{t=1}^{T}\mathbb{Z}_{t-1}^T\Sigma_{\epsilon}^{-1}\epsilon_t,
\end{equation}
where $\Sigma_\epsilon:=E(\epsilon_t\epsilon_t^T)$.

As in the case of the OLS estimator, we consider two cases, with the network size $N$ being fixed and growing as a function of $T$.

\medskip\noindent
\textbf{(I) Fixed network size $N$}

\begin{p4}[Asymptotic Properties of the GLS Estimator]
\label{P:GLS}
Suppose Assumptions \ref{A1}-\ref{A3} hold. For a stationary process $\mathbb{X}_t$ defined by the NAR$(q_1,q_2)$ model \eqref{compact}, i.e., $\mathbb{X}_{t}=\mathbb{Z}_{t-1}\beta+ \epsilon_{t}$ with finite network size $N$, $\hat{\beta}_{GLS}$ is a consistent estimator of $\beta$ and its asymptotic distribution is given by 
\begin{equation}
\label{GLS}
\sqrt{T}(\hat{\beta}_{GLS}-\beta)\rightarrow_{d}N(0, Q^{-1})
\end{equation}
where $Q:=
E(\mathbb{Z}_{t}^T\Sigma_{\epsilon}^{-1}\mathbb{Z}_{t})$.
\end{p4}

\begin{proof}
The proof is given in Appendix \ref{b3} and leverages Lemmas
\ref{L:LLNGLS} and \ref{L:CLTGLS} also provided in Appendices
\ref{b1}-\ref{b2}.
\end{proof}

Note that the result of Proposition \ref{P:GLS} is a direct extension of that in Proposition \ref{P:OLS}, with only the form of the covariance matrix in the limiting distribution changing.

\medskip\noindent
\textbf{(II) Growing network size with $N\leq T$:}

\begin{p5}[Asymptotic Properties of the GLS Estimator]
\label{P:DGLS}
Suppose Assumptions \ref{A1}-\ref{A2} hold. Let $\mathbb{X}_t$ be a stationary process generated by the NAR$(q_1,q_2)$ model \eqref{compact}, i.e., $\mathbb{X}_{t}=\mathbb{Z}_{t-1}\beta+ \epsilon_{t}$ with growing network size $N$. Define $D\in \mathbb{R}^{k\times(2Nq+Np)}$ for any finite $k$. Further, assume:
\begin{itemize}
\item
$D$ has bounded row sums; i.e., for $i=1,\cdots,k$, there exists a finite constant $c$ such that $\sum\limits_{j=1}^{2Nq+Np}d_{i,j}\leq c$ where $d_{i,j}$ is the ij-th element of D.
\item
 $N<T$.
\end{itemize}
Then,
\begin{equation}
\label{DGLS}
\frac{1}{\sqrt{T}}D(\sum_{t=1}^{T}\mathbb{Z}_{t-1}^T\Sigma_{\epsilon}^{-1}\mathbb{Z}_{t-1})(\hat{\beta}_{ridge}-\beta)\rightarrow_{d}N(0, DQD^T)
\end{equation}
where $Q:=
E(\mathbb{Z}_{t}^T\Sigma_{\epsilon}^{-1}\mathbb{Z}_{t})$.
\end{p5}

\begin{proof}
The proof of Proposition \ref{P:DGLS} is given in Appendix \ref{b5} based on Lemma \ref{L:CLTDGLS} also provided therein.
\end{proof}

Note that the covariance matrix of the limiting distribution takes into consideration the dependence of the error term vis-a-vis its OLS counterpart.

\textbf{(III) Growing network size with $N> T$:}

The ridge regularized GLS estimator is defined next, and uses the covariance structure of the error term:
\begin{equation}\label{beta-gls-r}
\hat{\beta}_{ridge}=(\sum_{t=1}^{T}\mathbb{Z}_{t-1}^T\Sigma_{\epsilon}^{-1}\mathbb{Z}_{t-1}+T M)^{-1}\sum_{t=1}^{T}\mathbb{Z}_{t-1}^T\Sigma_{\epsilon}^{-1}\mathbb{X}_t,
\end{equation}
where $\Sigma_\epsilon:=E(\epsilon_t\epsilon_t^T)$ and $M:= diag\{\lambda_1I_N,\lambda_2I_N,\cdots,\lambda_1I_N,\lambda_2I_N,\lambda_3I_{Np}\}$.

\begin{p10}[Asymptotic Properties of the ridge regularized GLS Estimator ]
\label{P:DGLSR}
Suppose Assumptions \ref{A1}-\ref{A2} hold. Let $\mathbb{X}_t$ be a stationary process generated by the NAR$(q_1,q_2)$ model \eqref{compact}, i.e., $\mathbb{X}_{t}=\mathbb{Z}_{t-1}\beta+ \epsilon_{t}$ with growing network size $N$. Define $D\in \mathbb{R}^{k\times(2Nq+Np)}$ for any finite $k$. Further, assume:
\begin{itemize}
\item
$D$ has bounded row sums; i.e., for $i=1,\cdots,k$, there exists a finite constant $c$ such that $\sum\limits_{j=1}^{2Nq+Np}d_{i,j}\leq c$ where $d_{i,j}$ is the ij-th element of D.
\item
 $N>T$.
\item 
$\lambda_i=o(\frac{1}{\sqrt T})$ for $i=1,2,3$.
\end{itemize}
Then,
\begin{equation}
\label{DGLSR}
\frac{1}{\sqrt{T}}D(\sum_{t=1}^{T}\mathbb{Z}_{t-1}^T\Sigma_{\epsilon}^{-1}\mathbb{Z}_{t-1}+TM)(\hat{\beta}_{GLS}-\beta)\rightarrow_{d}N(0, DQD^T)
\end{equation}
where $Q:=
E(\mathbb{Z}_{t}^T\Sigma_{\epsilon}^{-1}\mathbb{Z}_{t})$.
\end{p10}

\begin{proof}
The proof is given in Appendix \ref{a10}.
\end{proof}

\subsection{EGLS Estimator}

For the EGLS estimator, we first establish its consistency and asymptotic normality under very general conditions:

\begin{p8}[Asymptotic Properties of EGLS Estimator]
\label{P:EGLS}
Suppose Assumptions \ref{A1}-\ref{A3} hold. For a stationary process $\mathbb{X}_t$ defined by the NAR$(q_1,q_2)$ model \eqref{compact}, i.e., $\mathbb{X}_{t}=\mathbb{Z}_{t-1}\beta+ \epsilon_{t}$ with finite network size $N$, and in addition assuming that there exists a consistent estimator $\hat{\Sigma}_{\epsilon}$ for  $\Sigma_{\epsilon}$. Then, the EGLS estimator $\hat{\beta}=(\sum_{t=1}^{T}\mathbb{Z}_{t-1}^T\hat{\Sigma}_{\epsilon}^{-1}\mathbb{Z}_{t-1})^{-1}\sum_{t=1}^{T}\mathbb{Z}_{t-1}^T\hat{\Sigma}_{\epsilon}^{-1}\mathbb{X}_t$ is asymptotically equivalent to the GLS estimator, and \[\sqrt{T}(\hat{\beta}-\beta)\rightarrow_{d}N(0, Q^{-1})\ for\ fixed\ N.\]

As before, define $D\in \mathbb{R}^{k\times(2Nq+Np)}$ for any finite $k$.
\begin{itemize}

\item[(i)] If Assumption \ref{A2} and the conditions in Proposition \ref{P:DOLS} hold, then
\[\frac{1}{\sqrt{T}}D(\sum_{t=1}^{T}\mathbb{Z}_{t-1}^T\hat{\Sigma}_{\epsilon}^{-1}\mathbb{Z}_{t-1})(\hat{\beta}-\beta)\rightarrow_{d}N(0, DQD^T)\ for\ diverging\ N\leq T;\]

\item[(ii)] If Assumption \ref{A2} and the conditions in Proposition \ref{P:DGLSR} hold, then
\[\frac{1}{\sqrt{T}}D(\sum_{t=1}^{T}\mathbb{Z}_{t-1}^T\Sigma_{\epsilon}^{-1}\mathbb{Z}_{t-1}+TM)(\hat{\beta}_{ridge}-\beta)\rightarrow_{d}N(0, DQD^T)\ for\ diverging\ N>T,\]
\end{itemize}
where $\hat{\beta}_{ridge}=(\sum_{t=1}^{T}\mathbb{Z}_{t-1}^T\Sigma_{\epsilon}^{-1}\mathbb{Z}_{t-1}+T M)^{-1}\sum_{t=1}^{T}\mathbb{Z}_{t-1}^T\Sigma_{\epsilon}^{-1}\mathbb{X}_t$ and $Q:=
E(\mathbb{Z}_{t}^T\Sigma_{\epsilon}^{-1}\mathbb{Z}_{t})$.
\end{p8}

\begin{proof}
 The proof of the Proposition is given in Appendix \ref{p3.5}.
\end{proof}

Next, we examine the EGLS estimator for the following two popular models for the structure of the covariance matrix of the error term for the NAR$(q_1,q_2)$ model.
The first corresponds to a spatial autoregressive structure for the error term, while the second considers a factor model.

\subsubsection{Spatial autoregressive covariance structure}
\label{sec:SAR}

The following model is considered for the error term of the NAR$(q_1,q_2)$ model:
\begin{equation}\label{SAR-error}
\epsilon_t=\rho \Phi\epsilon_t+u_t, \ \ \  u_t\sim F(0,\sigma_u^2I),
\end{equation}
wherein $\Phi\in \mathbb{R}^{N\times N}$ is a row-normalized matrix, and $u_t$ is drawn from a distribution $F$ with mean 0 and variance $\sigma_u^2$.
For this model, the covariance matrix of the error term
$\Sigma_\epsilon$ takes the form $\Sigma_{\epsilon}(\rho)=\sigma_u^2(I-\rho \Phi)^{-1}(I-\rho \Phi)^{-T}$.
To apply the result of Proposition \ref{P:EGLS}, we need to
obtain a consistent estimator for the parameter $\rho$, which in turn provides a consistent estimator for $\Sigma_\epsilon$.

\begin{remark}
Note that the weight matrix $W$ in model \eqref{eq:matrix} 
captures the \textit{autoregressive structure} amongst the nodes of the network, while the weight matrix $\Phi$ captures additional \textit{contemporaneous dependence} amongst them. In general, we expect that $W\neq \Phi$.
\end{remark}

Next, we discuss the estimation procedure for obtaining $\hat{\rho}$.

[Step 1:] Calculate the OLS estimator for the NAR model parameters:
\[\hat{\beta}_0=(\sum_{t=1}^{T}\mathbb{Z}_{t-1}^T\mathbb{Z}_{t-1})^{-1}\sum_{t=1}^{T}\mathbb{Z}_{t-1}^T\mathbb{X}_t.\]

[Step 2:] Calculate residuals obtained by $\hat{\epsilon}_{t}=\mathbb{X}_t-\mathbb{Z}_{t-1}\hat{\beta_0}$.

Then, define the quasi-loglikelihood function for $(\rho,\sigma^2_u)$ given by
\[\log L=-\frac{NT}{2}\log 2\pi-\frac{NT}{2}\log\sigma^2_u+T\log|S(\rho)|-\frac{1}{2\sigma^2_u}\sum\limits_{t=1}^{T}(\epsilon_t-\rho\Phi\epsilon_t)^T(\epsilon_t-\rho\Phi\epsilon_t)\]
with $S(\rho)=I-\rho \Phi$.

Given a value for $\rho$, the Quasi-Maximum Likelihood Estimator (QMLE) of $\sigma^2_u$ is defined as: 
\[\hat{\sigma}^2_u=\frac{1}{NT}\sum\limits_{t=1}^{T}(S(\rho)\hat{\epsilon}_t)^T(S(\rho)\hat{\epsilon}_t).\]

Thus, the profile log-likelihood function for $\rho$ is given by
\[\log L_\rho=-\frac{NT}{2}\log2\pi-\frac{NT}{2}+T\log|S(\rho)|-\frac{NT}{2}\log\hat{\sigma}^2_u,\]
\[\frac{\partial \log L_\rho}{\partial \rho}=-Ttr(S(\rho)^{-1}\Phi)+NT\frac{\sum\limits_{t=1}^{T}\epsilon_t^T\Phi^T(S(\rho)\epsilon_t)}{\sum\limits_{t=1}^{T}(S(\rho)\epsilon_t)^TS(\rho)\epsilon_t} \overset{\Delta}{=}0.\]
Then, $\hat{\rho}$ corresponds to the solution to $\frac{\partial lnl}{\partial \rho}=0$. Following along the lines in \cite{lee2004asymptotic,lee2010estimation}, Lemma \ref{L:SAR}, establishes its consistency.

[Step 3:] Update $\Sigma_{\epsilon}(\hat\rho)$ and obtain the EGLS estimator:
\[\hat{\beta}=(\sum_{t=1}^{T}\mathbb{Z}_{t-1}^T\Sigma(\hat\rho)^{-1}\mathbb{Z}_{t-1})^{-1}\sum_{t=1}^{T}\mathbb{Z}_{t-1}^T\Sigma(\hat\rho)^{-1}\mathbb{X}_t.\]

\begin{remark2}
Note that 
$\Sigma_{\epsilon}(\hat\rho)=\sigma_u^2(I-\hat{\rho} \Phi)^{-1}(I-\hat{\rho} \Phi)^{-T}$, so a consistent $\hat\rho$ translates to $\Sigma_{\epsilon}(\hat{\rho})$ being also consistent through a direct application of the Continuous Mapping Theorem \citep{van2000asymptotic}.
\end{remark2}

\begin{c1}[Asymptotic Properties of the EGLS Estimator]
Suppose Assumptions \ref{A1}-\ref{A3} hold. Then, for a stationary process $\mathbb{X}_t$ defined by the NAR$(q_1,q_2)$ model \eqref{compact}, i.e. $\mathbb{X}_{t}=\mathbb{Z}_{t-1}\beta+ \epsilon_{t}$ with finite network size $N$, a result in \cite{lutkepohl2005new} estbalishes that since $\Sigma_{\epsilon}(\hat{\rho})$ is consistent, the EGLS estimator $$\hat{\beta}=(\sum_{t=1}^{T}\mathbb{Z}_{t-1}^T\Sigma(\hat\rho)^{-1}\mathbb{Z}_{t-1})^{-1}\sum_{t=1}^{T}\mathbb{Z}_{t-1}^T\Sigma(\hat\rho)^{-1}\mathbb{X}_t$$ 
is asymptotically equivalent to the GLS estimator, and \[\sqrt{T}(\hat{\beta}-\beta)\rightarrow_{d}N(0, Q^{-1})\ for\ fixed\ N.\]
As before, define $D\in \mathbb{R}^{k\times(2Nq+Np)}$ for any finite $k$.
\begin{itemize}
\item[(i)] If Assumption \ref{A2} and the conditions in Proposition \ref{P:DOLS} hold, then
\[\frac{1}{\sqrt{T}}D(\sum_{t=1}^{T}\mathbb{Z}_{t-1}^T\Sigma_{\epsilon}(\hat{\rho})^{-1}\mathbb{Z}_{t-1})(\hat{\beta}-\beta)\rightarrow_{d}N(0, DQD^T)\ for\ diverging\ N\leq T;\]

\item[(ii)] If Assumption \ref{A2} and the conditions in Proposition \ref{P:DGLSR} hold, then
\[\frac{1}{\sqrt{T}}D(\sum_{t=1}^{T}\mathbb{Z}_{t-1}^T\Sigma_{\epsilon}(\hat{\rho})^{-1}\mathbb{Z}_{t-1}+TM)(\hat{\beta}_{ridge}-\beta)\rightarrow_{d}N(0, DQD^T)\ for\ diverging\ N>T,\]
\end{itemize}
where $\hat{\beta}_{ridge}=(\sum_{t=1}^{T}\mathbb{Z}_{t-1}^T\Sigma_{\epsilon}(\hat{\rho})^{-1}\mathbb{Z}_{t-1}+T M)^{-1}\sum_{t=1}^{T}\mathbb{Z}_{t-1}^T\Sigma_{\epsilon}(\hat{\rho})^{-1}\mathbb{X}_t$ and $Q:=
E(\mathbb{Z}_{t}^T\Sigma_{\epsilon}^{-1}\mathbb{Z}_{t})$.
\end{c1}
		
\subsubsection{A Factor model for the error covariance structure} \label{sec:factor}

Next, for the NAR($q_1,q_2$) model \eqref{compact}, $\mathbb{X}_{t}=\mathbb{Z}_{t-1}\beta+ \epsilon_{t}$, it is assumed that the error term is generated according to the following factor model: $\epsilon_{t}=\Lambda F_t+u_{t}$ and $u_{t}\sim F(0,\sigma^2I) $. where the $N\times k$ matrix $\Lambda$ contains fixed factor loadings satisfying ${\Lambda}^T\Lambda$ being diagonal, $F_t$ is a $k\times 1$ random factor satisfying ${F}^TF/T=I$.

Letting $E=\begin{Bmatrix}
\epsilon_{1}^T\\
\epsilon_{2}^T\\
\vdots\\
\epsilon_{T}^T
\end{Bmatrix}$, $F=\begin{Bmatrix}
F_1^T\\
F_2^T\\
\vdots\\
F_{T}^T
\end{Bmatrix}$, and $U=\begin{Bmatrix}
u_1^T\\
u_2^T\\
\vdots\\
u_{T}^T
\end{Bmatrix}$, we can then write:\[E=F\Lambda^T+U.\]

The EGLS estimator for this setting is obtained based on the following procedure:

[Step 1:] Calculate the OLS estimator:
\[\hat{\beta}_0=(\sum_{t=1}^{T}\mathbb{Z}_{t-1}^T\mathbb{Z}_{t-1})^{-1}\sum_{t=1}^{T}\mathbb{Z}_{t-1}^T\mathbb{X}_t.\]

[Step 2:] Calculate residuals given by $\hat{\epsilon}_{t}=\mathbb{X}_t-\mathbb{Z}_{t-1}\hat{\beta_0}$.

For large $N, T$ settings, we can estimate $\Lambda$ and $F$ by
\[\mathop{min}\limits_{\Lambda^k,F^k}S(k),\ with\ S(k)=(NT)^{-1}\sum\limits_{i=1}^{N}\sum\limits_{t=1}^{T}(\hat{\epsilon}_{it}-{\lambda_i^{k}}^TF_t^k)^2, \ \ k=1,\cdots,K,\]
subject to the normalization ${F^k}^TF^k/T=I_k$ and ${\Lambda^k}^T\Lambda^k$ being diagonal.

Let $g(N,T)$ be a penalty function (e.g., $g(N,T)=\frac{(N+T-k)}{NT}\log (NT)$). Define the information criterion
\[IC(k)=\ln(S(k))+kg(N,T),\]
thus, we can estimate the number of factors $k$ by:
\[\hat{k}_{IC}=\mathop{argmin}_{0\leq k \leq k_{max}}IC(k).\]
		
Integrating out $\Lambda^k$, the problem is equivalent to maximizing $tr(F^{k^T}(\hat{E}^T\hat{E})F^k)$. The estimated factor matrix $\hat{F}^k$ is $\sqrt{T}$ times the eigenvectors corresponding to the $k$ largest eigenvalues of the $T\times T$ matrix $\hat{E}^T\hat{E}$ and $\hat{\Lambda}^{k^T}=\hat{F}^{k^T}\hat{E}/T$. Further, $\hat\sigma^2=\frac{1}{NT-k(T+N-k)}\sum\limits_{i=1}^{N}\sum\limits_{t=1}^{T}\hat{u}_{it}^2$. From Lemma \ref{L:FM}, $\hat{\Sigma}_\epsilon$ is a consistent estimator of $\Sigma_\epsilon$.

[Step 3:] Update $\hat{\Sigma}_{\epsilon}$ and obtain the EGLS estimator:
\[\hat{\beta}=(\sum_{t=1}^{T}\mathbb{Z}_{t-1}^T\hat{\Sigma}_{\epsilon}^{-1}\mathbb{Z}_{t-1})^{-1}\sum_{t=1}^{T}\mathbb{Z}_{t-1}^T\hat{\Sigma}_{\epsilon}^{-1}\mathbb{X}_t.\]

\begin{c2}[Asymptotic Properties of the EGLS Estimator]
Suppose Assumptions \ref{A1}-\ref{A3} hold. For a stationary process $\mathbb{X}_t$ defined by the NAR$(q_1,q_2)$ model \eqref{compact}, i.e. $\mathbb{X}_{t}=\mathbb{Z}_{t-1}\beta+ \epsilon_{t}$ with finite network size N, a result in \cite{lutkepohl2005new} shows that since $\hat{\Sigma}_{\epsilon}$ is consistent, the EGLS estimator $$\hat{\beta}=(\sum_{t=1}^{T}\mathbb{Z}_{t-1}^T\hat{\Sigma}_{\epsilon}^{-1}\mathbb{Z}_{t-1})^{-1}\sum_{t=1}^{T}\mathbb{Z}_{t-1}^T\hat{\Sigma}_{\epsilon}^{-1}\mathbb{X}_t$$ 
is asymptotically equivalent to the GLS estimator, and \[\sqrt{T}(\hat{\beta}-\beta)\rightarrow_{d}N(0, Q^{-1})\ for\ fixed\ N.\]
As before, define $D\in \mathbb{R}^{k\times(2Nq+Np)}$ for any finite $k$.
\begin{itemize}

\item[(i)] If Assumption \ref{A2} and the conditions in Proposition \ref{P:DOLS} hold, then
\[\frac{1}{\sqrt{T}}D(\sum_{t=1}^{T}\mathbb{Z}_{t-1}^T\hat{\Sigma}_{\epsilon}^{-1}\mathbb{Z}_{t-1})(\hat{\beta}-\beta)\rightarrow_{d}N(0, DQD^T)\ for\ diverging\ N\leq T;\]

\item[(ii)] If Assumption \ref{A2} and the conditions in Proposition \ref{P:DGLSR} hold, then
\[\frac{1}{\sqrt{T}}D(\sum_{t=1}^{T}\mathbb{Z}_{t-1}^T\Sigma_{\epsilon}^{-1}\mathbb{Z}_{t-1}+TM)(\hat{\beta}_{ridge}-\beta)\rightarrow_{d}N(0, DQD^T)\ for\ diverging\ N>T,\]
\end{itemize}
where $\hat{\beta}_{ridge}=(\sum_{t=1}^{T}\mathbb{Z}_{t-1}^T\Sigma_{\epsilon}^{-1}\mathbb{Z}_{t-1}+T M)^{-1}\sum_{t=1}^{T}\mathbb{Z}_{t-1}^T\Sigma_{\epsilon}^{-1}\mathbb{X}_t$ and $Q:=
E(\mathbb{Z}_{t}^T\Sigma_{\epsilon}^{-1}\mathbb{Z}_{t})$.
\end{c2}

\subsection{Misspecification of the Weight Matrix W}

In the NAR model, the weight matrix $W$ is assumed to be a prior specified and known. However, in many applications it is reasonable to assume that it may be empirically defined and hence exhibit misspecification. \cite{knight2019generalised} discuss data driven procedures of selecting $W$. In the sequel, we present the impact of misspecifying the weight matrix $W$ on the asymptotic distribution of the OLS and GLS estimators.

We assume that the weight matrix of the $NAR(q_1,q_2)$ model can be decomposed as $W_T^M=W+\pi_T$, with $W$ being the true weight matrix and $\pi_T$ the misspecified component; the subscript $T$ emphasizes the dependence of the misspecifciation on the sample size. Then, the misspecified design matrix $\mathbb{Z}_{t-1}^M$ can be written as: \[\mathbb{Z}_{t-1}^M=\mathbb{Z}_{t-1}+\begin{bmatrix}\boldsymbol{0}_{N\times N} &\mathop{diag}\{\pi_{T}\mathbb{X}_{t-1}\}&\cdots&\boldsymbol{0}_{N\times N} &\mathop{diag}\{\pi_{T}\mathbb{X}_{t-q}\}&\boldsymbol{0}_{N\times Np}\end{bmatrix}.\]

\begin{p11}
\label{P:MWM}
Consider an $NAR(q_1,q_2)$ model, with misspecified weight matrix $W_T^M$, and corresponding design matrix $\mathbb{Z}_{t-1}^M$.
Define \[P_{t-1}:=\begin{bmatrix}\boldsymbol{0}_{N\times N} &\mathop{diag}\{\pi_{T}\mathbb{X}_{t-1}\}&\cdots&\boldsymbol{0}_{N\times N} &\mathop{diag}\{\pi_{T}\mathbb{X}_{t-q}\}&\boldsymbol{0}_{N\times Np}\end{bmatrix}.\]
Let $\hat{\beta}^M$ denote the OLS/GLS estimator based on a misspecified weight matrix $W_T^M$. Then, the following hold:
\begin{itemize}
\item[(a)]
If $||\pi_T||_\infty=o(1)$, 
\[\hat{\beta}^M_{OLS}-\beta\rightarrow_p 0,\text{ for fixed} \ N,\] 
\[\frac{1}{T}D(\sum_{t=1}^{T}(\mathbb{Z}_{t-1}^M)^T\mathbb{Z}_{t-1}^M)(\hat{\beta}_{OLS}^M-\beta)\rightarrow_p 0,\text{ for diverging}  \ N,\]
\[\hat{\beta}^M_{GLS}-\beta\rightarrow_p 0,\text{ for fixed} \ N,\] 
and
\[\frac{1}{T}D(\sum_{t=1}^{T}(\mathbb{Z}_{t-1}^M)^T\Sigma_{\epsilon}^{-1}\mathbb{Z}_{t-1}^M)(\hat{\beta}_{GLS}^M-\beta)\rightarrow_p 0,\text{ for diverging} \ N.\]
\item[(b)]
If $||\pi_T||_\infty=o(\frac{1}{\sqrt{T}})$, 
\[\sqrt{T}(\hat{\beta}_{OLS}^M-\beta)\rightarrow_{d}N(0, P^{-1}QP^{-1})\text{ for fixed} \ N,\]
\[\frac{1}{\sqrt{T}}D(\sum_{t=1}^{T}(\mathbb{Z}_{t-1}^M)^T\mathbb{Z}_{t-1}^M)(\hat{\beta}^M_{OLS}-\beta)\rightarrow_{d}N(0, DQD^T),\text{ for diverging} \ N,\]
where $P:=
E(\mathbb{Z}_{t}^T\mathbb{Z}_{t})$, $Q:=
E(\mathbb{Z}_{t}^T\Sigma_{\epsilon}\mathbb{Z}_{t})$.
\[\sqrt{T}(\hat{\beta}_{GLS}^M-\beta)\rightarrow_{d}N(0, Q^{-1}),\text{ for fixed} \ N,\]
and
\[\frac{1}{\sqrt{T}}D(\sum_{t=1}^{T}(\mathbb{Z}_{t-1}^M)^T\Sigma_{\epsilon}^{-1}\mathbb{Z}_{t-1}^M)(\hat{\beta}^M_{GLS}-\beta)\rightarrow_{d}N(0, DQD^T),\text{ for diverging} \ N,\]
where $Q:=
E(\mathbb{Z}_{t}^T\Sigma_{\epsilon}^{-1}\mathbb{Z}_{t})$.
\end{itemize}
\end{p11}
\begin{proof}
 The proof of the Proposition is given in Appendix \ref{pos4}.
\end{proof}

\begin{remark8}
The result shows that the misspecification of the weight matrix $W$ needs to vanish at a $1/\sqrt{T}$ rate for the respective asymptotic distributions of the OLS/GLS estimators not to be impacted.

Next, we illustrate how the difference $||\hat{\beta}^M-\beta||_{F}$ and the coverage probability for confidence intervals constructed based on the asymptotic distribution results in Proposition \ref{P:MWM} behave as $||\pi_T||_\infty$ decreases as a function of the sample size $T$.

Consider an NAR$(1,1)$ model with $N=10$ and $A=diag(rep(0.5,10))$, $B=diag(rep(c(0.8,-0.8),5))$ and $W$ and $\Phi$ being a banded matrix of ``width" 1 and $\Sigma_\epsilon=I_N$. The experiment is replicated 100 times.

Figure \ref{heatmap2} depicts the heatmap of the misspecification $\pi_T=W_T^M-W$ as a function of the sample size $T$ for $T=100$ (left panel), $T=1000$ (middle panel) and $T=10000$ (right panel).
\begin{figure} [H]
\centering
\subfigure[Heatmaps of $\pi_T=W_T^M-W$ for $||\pi_T||_\infty=(\frac{1}{T})^{\frac{1}{2}}$]{
\begin{minipage}[t]{0.315\textwidth}
\centering
\includegraphics[width=5.0cm]{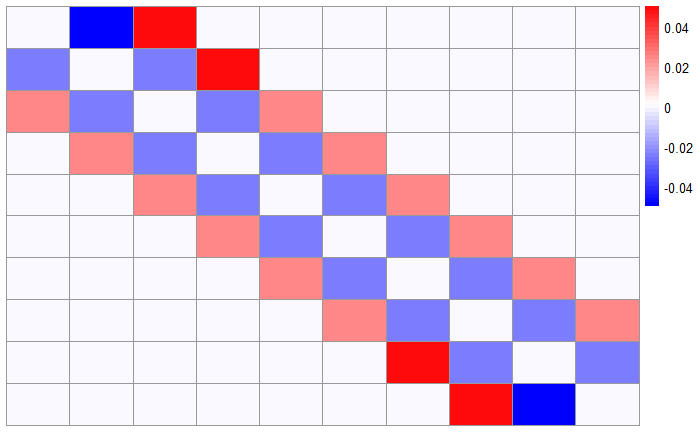}

\end{minipage}

\begin{minipage}[t]{0.315\textwidth}
\centering
\includegraphics[width=5.0cm]{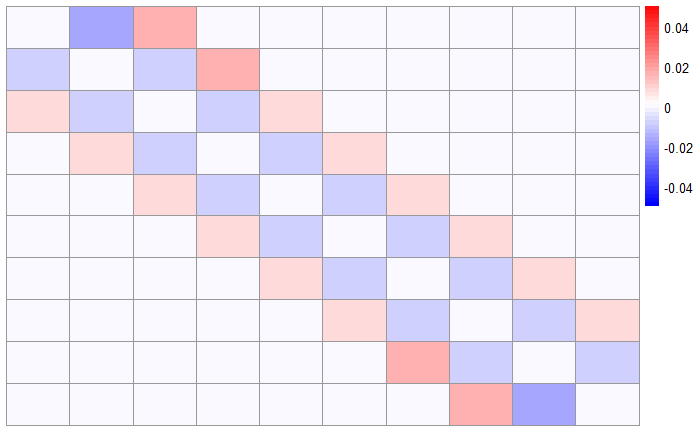}

\end{minipage}

\begin{minipage}[t]{0.315\textwidth}
\centering
\includegraphics[width=5.0cm]{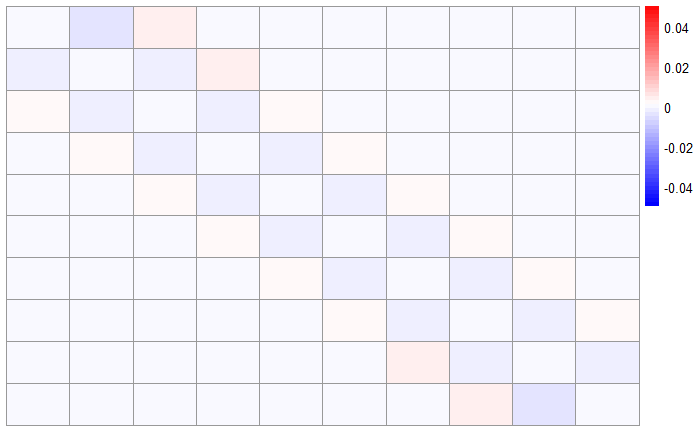}

\end{minipage}

}

\subfigure[Heatmaps of $\pi_T=W_T^M-W$ for $||\pi_T||_\infty=(\frac{1}{T})^{\frac{2}{3}}$]{
\begin{minipage}[t]{0.315\textwidth}
\centering
\includegraphics[width=5.0cm]{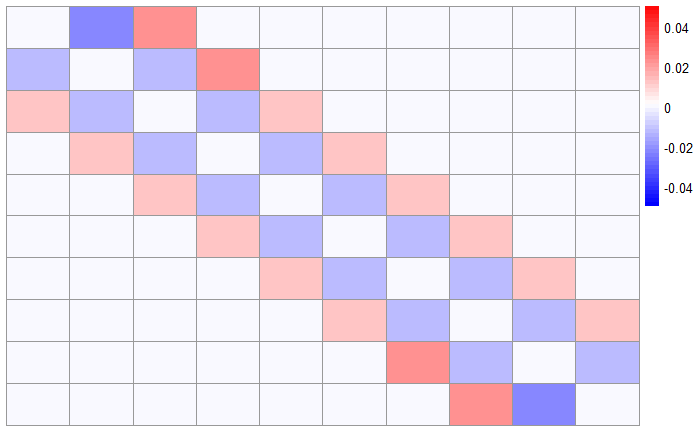}

\end{minipage}

\begin{minipage}[t]{0.315\textwidth}
\centering
\includegraphics[width=5.0cm]{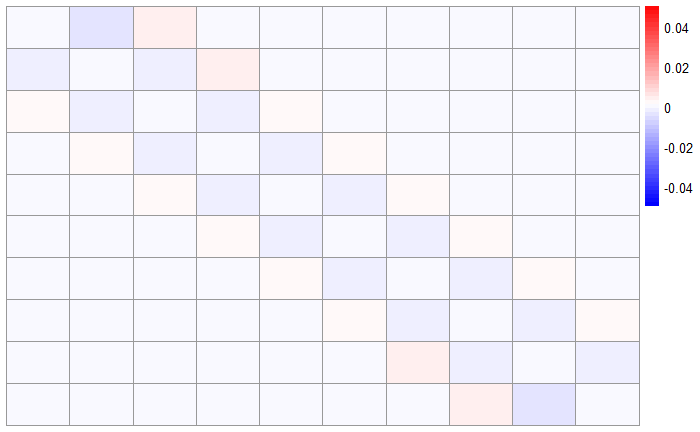}

\end{minipage}

\begin{minipage}[t]{0.315\textwidth}
\centering
\includegraphics[width=5.0cm]{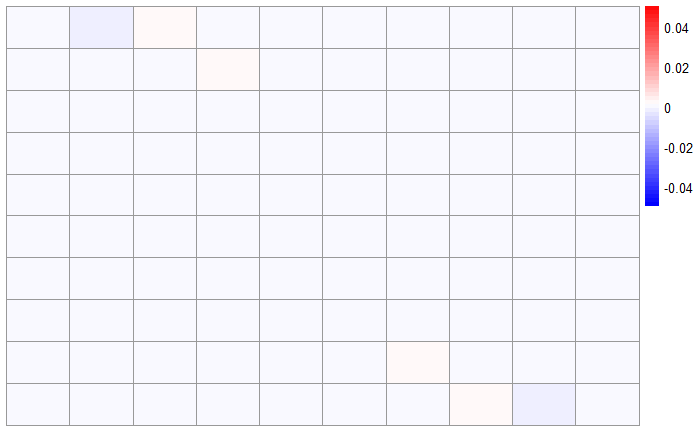}

\end{minipage}
} 
\caption{Heatmaps of $\pi_T=W_T^M-W$ for $||\pi_T||_\infty$ vanishing at different rates.\label{heatmap2} \\$T=100$ (left panel), $T=1000$ (middle panel) and $T=10000$ (right panel).}
\end{figure}

Figure \ref{rmse} depicts how both the estimation error and the coverage probability of $\hat{\beta}$ (red) and $\hat{\beta}^M$ (blue) change as $||\pi_T||_\infty$ decreases at different rates. 

\begin{figure} [H]
\centering
\subfigure[Estimation error for $||\pi_T||_\infty=(\frac{1}{T})^{\frac{1}{2}}$]{
\begin{minipage}[t]{6cm}
\centering
\includegraphics[scale=0.35]{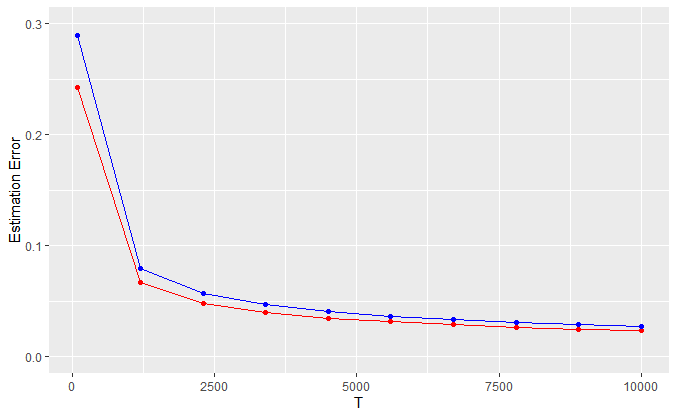}
\end{minipage}
}
\subfigure[Estimation error for $||\pi_T||_\infty=(\frac{1}{T})^{\frac{2}{3}}$]{
\begin{minipage}[t]{6cm}
\centering
\includegraphics[scale=0.35]{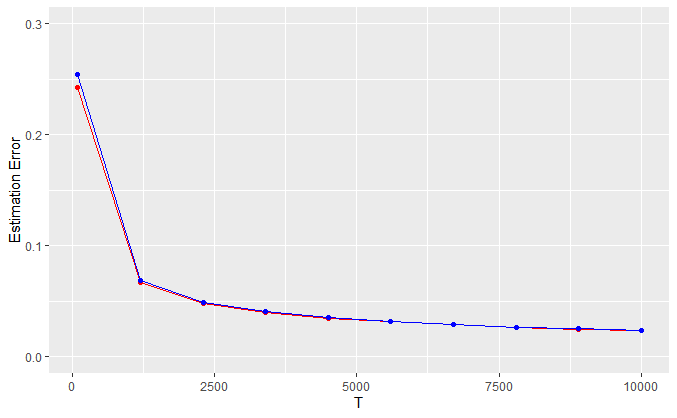}
\end{minipage}
} 
\subfigure[Coverage probability for $||\pi_T||_\infty=(\frac{1}{T})^{\frac{1}{2}}$]{
\begin{minipage}[t]{6cm}
\centering
\includegraphics[scale=0.35]{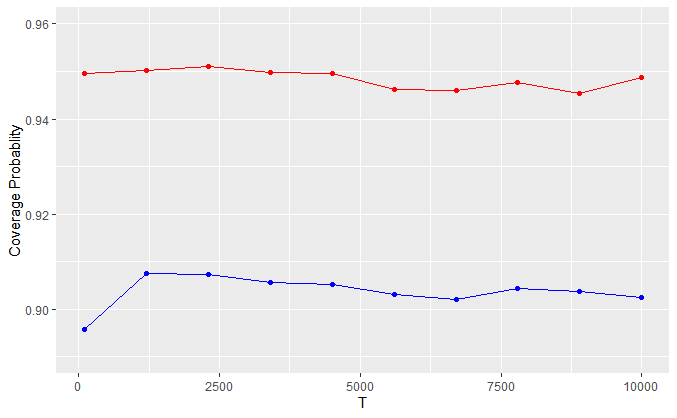}
\end{minipage}
} 
\subfigure[Coverage probability for $||\pi_T||_\infty=(\frac{1}{T})^{\frac{2}{3}}$]{
\begin{minipage}[t]{6cm}
\centering
\includegraphics[scale=0.35]{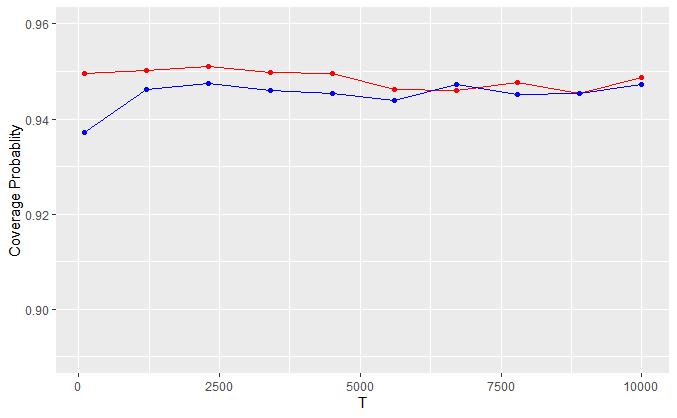}
\end{minipage}
} 
\caption{Estimation error and coverage probability of $\hat{\beta}$ (red) and $\hat{\beta}^M$ (blue), as $||\pi_T||_\infty$ decreases at different rates\label{rmse}}
\end{figure}

It can be seen that in general $||\hat{\beta}^M-\beta||_{F}$ decreases for smaller $||\pi_T||_\infty$ as expected. Further, comparing Figures \ref{rmse}(c) and (d), it can be concluded that for $||\pi_T||_\infty=O(\frac{1}{\sqrt{T}})$, the asymptotic distribution of $\hat{\beta}^M$ is not equivalent to $\hat{\beta}$, but for $||\pi_T||_\infty=(\frac{1}{T})^{\frac{2}{3}}=o(\frac{1}{\sqrt{T}})$, the asymptotic distribution of $\hat\beta^M$ is equivalent to $\hat\beta$.
\end{remark8}

\section{Performance Evaluation}
\label{sec:performance}

Several factors influence the performance of the various estimators proposed, including the sample size $T$, the number of nodes $N$, the structure of the weight matrix $W$, the lag orders $(q_1,q_2)$ and the parameterization of the error covariance matrix $\Sigma_\epsilon$ (SAR vs factor structure). 

The performance metrics considered include the root-mean-square error (RMSE) of the model parameters, together with the coverage probability of the constructed confidence intervals and their average length.

Next, we describe the data generating mechanism and the settings considered. Each experiment is based on 500 replications of data generated from the NAR$(q_1,q_2)$ model 
$\mathbb{X}_{t}=A \mathbb{X}_{t-1} + B W \mathbb{X}_{t-1} +\mathbb{Y}_{t-1}\gamma+\epsilon_{t}$ where $\mathbb{Y}_{t-1}:=\begin{bmatrix}\mathbb{Y}_{1,(t-1)}&\mathbb{Y}_{2,(t-1)}&\cdots&\mathbb{Y}_{p,(t-1)}\end{bmatrix}\in \mathbb{R}^{N\times p}$ and $\gamma:=\begin{bmatrix}\gamma_1&\gamma_2&\cdots&\gamma_p\end{bmatrix}^T$. We fix the network size to $N=100$. Further, the error terms are generated either through a SAR structure $[\epsilon_t\sim N(0,(I-\rho \Phi)^{-1}(I-\rho \Phi)^{-T})]$ or through a factor structure $\epsilon_t=\Lambda F_t+u_t$ with $u_t\sim N(0,I)$ for factor structure. Finally, the exogenous covariates are generated according to $Y_{ik,(t-1)}\sim N(0, 1)$, for all $i,k,t$.

The specific form of the primary model parameters $(A,B,\gamma)$
together with $W$, $\Phi$ and $\Lambda$ are specified in subsequent sections.


\subsection{Estimation Accuracy}

We focus on the performance of the EGLS estimator. We examine the influence of $T, W$, the structure of the error term, and the lag order $q=q_1=q_2$ on the RMSE metric. The latter is defined as follows for the three sets of model parameters: self-lags $||\hat{A}-A||_F/||A||_F$, network lags $||\hat{B}-B||_F/||B||_F$ and exogenous covariates $||\hat{\gamma}-\gamma||_F/||\gamma||_F$. We set $A=diag\{0.1I_{25},0.2I_{25},0.3I_{25},0.4I_{25}\}$, $B=diag\{0.4I_{25},0.3I_{25},0.2I_{25},0.1I_{25}\}$ and $\gamma=(-0.8\boldsymbol{1}_3^T,-0.4\boldsymbol{1}_2^T,0.4\boldsymbol{1}_2^T,0.8\boldsymbol{1}_3^T)^T$ in tables \ref{t5}, and \ref{t6}-\ref{t13}. In table \ref{t30}, we fix lag order $q_1=q_2=2$, $A_1=0.3I$, $A_2=0.3I$, $B_1=0.15I$, $B_2=0.15I$ and $\gamma=0.5\boldsymbol{1}_{10}$. The experiment is replicated 500 times, and we calculate the average over 500 estimates. For convenience, we set weight matrix $W$ to be a row-normalized banded matrix with different of different 
bandwidths.

We design 6 sets of numerical experiments, that examine the influence of various factors, as outlined next.

\begin{itemize}
\item In Table \ref{t5}, the influence of the sample size $T$ is examined. We fix $q_1=q_2=1$, $\rho=0.5$, $W$ and $\Phi$ as  banded matrices of width 5.
\item In Table \ref{t6}, the influence of the weight matrix $W$ is examined. We fix $q_1=q_2=1$, $\rho=0.5$, $T=400$, $\Phi$ as a banded matrix of width 5.
\item In Table \ref{t7}, the influence of $\rho$ is examined. We fix $q_1=q_2=1$, $T=400$, $W$ and $\Phi$ as banded matrices of width 5.
\item In Table \ref{t12}, the influence of the number of factors $k$ in the factor covariance model is examined. We fix $q_1=q_2=1$, $T=400$, $W$ as a banded matrix of width 5 and investigate how RMSE is influenced by the number of factors $k$ in the model $\epsilon_{t}=\Lambda F_t+u_{t}$ where $F_t\sim N(0,I)$ and $\lambda_i \sim U(0,1)$.
\item In Table \ref{t13}, the influence of the weight matrix $\Phi$ in the SAR model is examined. We fix $q_1=q_2=1$, $T=400$, $\rho=0.5$ and $W$ as a banded matrix of width 5.
\item In table \ref{t30}, we set $q_1=q_2=2$, $\rho=0.5$ and $W$ and $\Phi$ as banded matrices of width 5 and compare the estimates and their RMSE with $T$ increasing.
\end{itemize}

Due to space considerations, we show the results for the influence of the sample size $T$, and present all other Tables in Appendix \ref{tf}.

\begin{table}[H]
\centering
\caption{$N=100$, $q_1=q_2=1$, $\rho=0.5$, W and $\Phi$ band matrix of width 5. Performance evaluation for $T=150,300,450$.}

\begin{tabular}{c|c|c|c|c|c|c|c}

\hline
&Estimator&\multicolumn{2}{|c|}{T=150}&\multicolumn{2}{|c|}{T=300}&\multicolumn{2}{|c}{T=450}\\ \hline
&True value&Estimate&RMSE&Estimate&RMSE&Estimate&RMSE\\ \hline
$a_1\sim a_{25}$&0.1&0.090&0.613 &0.096&0.431 &0.098&0.355 \\ 
$a_{26}\sim a_{50}$&0.2&0.189&0.298 &0.195&0.208 &0.197&0.171 \\ 
$a_{51}\sim a_{75}$&0.3&0.289&0.189 &0.295&0.132 &0.296&0.108 \\ 
$a_{76}\sim a_{100}$&0.4&0.388&0.135 &0.394&0.093 &0.396&0.076 \\ 
$b_1\sim b_{25}$&0.4&0.400&0.305 &0.399&0.215 &0.399&0.172 \\ 
$b_{26}\sim b_{50}$&0.3&0.299&0.417 &0.299&0.293 &0.300&0.235 \\ 
$b_{51}\sim b_{75}$&0.2&0.199&0.628 &0.200&0.437 &0.200&0.359 \\ 
$b_{76}\sim b_{100}$&0.1&0.102&1.200 &0.100&0.825 &0.100&0.665 \\ 
$\gamma_1\sim \gamma_{3}$&-0.8&-0.811&0.042 &-0.805&0.030 &-0.803&0.023 \\ 
$\gamma_{4}\sim \gamma_{5}$&-0.4&-0.406&0.051 &-0.402&0.037 &-0.402&0.029 \\ 
$\gamma_{6}\sim \gamma_{7}$&0.4&0.405&0.079 &0.402&0.057 &0.401&0.046 \\ 
$\gamma_{8}\sim \gamma_{10}$&0.8&0.811&0.053 &0.805&0.036 &0.804&0.029 \\ \hline

\end{tabular}

\label{t5}
\end{table}

It can be seen that the accuracy of all model parameters -self-lags, network lags and regression coefficients of the exogenous covariates- increases as the sample size increases.

Next, a summary of the results of the remaining five scenarios (Tables in Appendix \ref{tf}) is presented. The results in Table~\ref{t6} (Table~\ref{t13}) confirm the robustness of the EGLS estimates over weight matrices $W$ ($\Phi$) with different bandwidths, while the RMSE exhibits a slight increase for network effect parameters when the band width increases. Table~\ref{t7} summarizes the performance of the EGLS estimates for different values of $\rho$. It can be seen that the estimates themselves are stable,  while the RMSE slightly decreases when $\rho$ increases (from close to zero to close to one). Further, Table~\ref{t12} shows that the number of factors $k$ has no significant effect on EGLS estimation, while the RMSE for the network parameters decrease slightly when $k$ increases. Finally, Table \ref{t30} shows that that EGLS performance improves for larger $T$, for NAR processes with higher temporal lags. An overall conclusion of the various simulation scenarios is that the performance of the EGLS estimator for the network parameters ($B$) is more sensitive to changes in $W$, $\Phi$ and $\rho$ than the autoregressive parameters ($A$).

\subsection{Coverage Probability and Length of Confidence Intervals}

For NAR$(1,1)$, we fix $A=B=0.4I_N$ and $\gamma=0.4\times\boldsymbol{1}_{10}$. For NAR$(2,2)$, $A_1=B_1=0.3I_N$, $A_2=B_2=0.15I_N$ and $\gamma=0.5\times\boldsymbol{1}_{10}$. We explore how different factors influence the coverage probability (CP) and length of confidence intervals (CI) of the various estimators. The results are based on 500 replicates. We set the network size $N=100$, and $\epsilon_t$ follows either a spatial autoregressive model with parameter $\rho$, or a $k$ factor model. For each model parameter (100 $\alpha_i$'s, 100 $\beta_i$'s, 10 $\gamma_i$'s), we calculate its CI and the corresponding CP and length length. The $95\%$ CI is calculated using $CI_i=(\hat{\beta}_i-z_{0.975}SE(\hat{\beta}),\hat{\beta}_i+z_{0.975}SE(\hat{\beta}))$, where $SE(\hat{\beta})=(\sum\limits_{t}\mathbb{Z}_t^T\mathbb{Z}_t)^{-1}(\sum\limits_{t}\mathbb{Z}_t^T\Sigma_{\epsilon}\mathbb{Z}_t)(\sum\limits_{t}\mathbb{Z}_t^T\mathbb{Z}_t)^{-1}$ for the OLS estimator and $SE(\hat{\beta})=(\sum\limits_{t}\mathbb{Z}_t^T\Sigma_{\epsilon}^{-1}\mathbb{Z}_t)^{-1}$ for the GLS and EGLS estimators.

We design the following 6 sets of numerical experiments:
\begin{itemize}
\item In Table \ref{t31}, the influence of the lag order $q$ of the NAR model is investigated. We set $q_1=q_2=2$, $\rho=0.5$ and $W$ as a banded matrix of width 5 and compare the CP and the length of the CI as $T$ increases.
\item In Table \ref{t1}, the influence of the sample size $T$ is investigated. We fix $\epsilon_{t}=\rho W\epsilon_{t}+u_{t}$ with $\rho=0.5$, $W$ as a banded matrix of width 5 and explore how the CP and length of the CI are influenced by $T$.
\item In Table \ref{t2}, the influence of the weight matrix $W$ is investigated. We fix $\epsilon_{t}=\rho W\epsilon_{t}+u_{t}$ with $\rho=0.5$, $T=400$ and explore how the bandwidth of $W$ influences the CP and length of the CI.
\item In Table \ref{t3}, the influence of $\rho$ is investigated for the spatial autoregressive model $\epsilon_{t}=\rho W\epsilon_{t}+u_{t}$. We fix $T=400$, and $W$ as a banded matrix of width 5.
\item In Table \ref{t4}, the influence of the number of factors $k$ in the factor model for the error term $\epsilon_{t}=\Lambda F_t+u_{t}$, with $F_t\sim N(0,1)$ and $\lambda_i \sim U(0,1)$. is investigated. We fix $T=400$ and $W$ as a banded matrix of width 5.
\item In Table \ref{t10}, the influence of the weight matrix $\Phi$ in the SAR model is investigated. We fix $T=400$, $\rho=0.5$ and $W$ as a banded matrix of width 5.
\end{itemize}

The results in the corresponding tables are averaged over the respective set of parameters (e.g., $\alpha_i$'s) and over the 500 replicates.

\begin{table}[H] 
\centering
\caption{$N=100$, $q_1=q_2=2$, $\rho=0.5$ and W and $\Phi$ band matrix of width 5. Length of confidence interval and coverage probability for $T=150,300,450$.}
\begin{tabular}{c|c|c|c|c|c|c|c|c|c}
 
\hline
&Estimator&\multicolumn{2}{|c|}{T=150}&\multicolumn{2}{|c|}{T=300}&\multicolumn{2}{|c|}{T=450}&\multicolumn{2}{|c}{T=600}\\ \hline
&&CI&CP&CI&CP&CI&CP&CI&CP\\ \hline
&OLS& 0.183 & 0.950 & 0.128 & 0.951 & 0.104 & 0.950 & 0.090 & 0.950 \\ 
   $A_1$ &GLS & 0.169 & 0.951 & 0.119 & 0.951 & 0.097 & 0.949 & 0.084 & 0.950 \\ 
   &EGLS & 0.169 & 0.951 & 0.119 & 0.951 & 0.097 & 0.949 & 0.084 & 0.950 \\ \hline
   & OLS & 0.183 & 0.949 & 0.128 & 0.950 & 0.104 & 0.950 & 0.090 & 0.951 \\ 
 $A_2$ &GLS & 0.169 & 0.950 & 0.119 & 0.949 & 0.097 & 0.949 & 0.084 & 0.950 \\ 
  &EGLS  & 0.169 & 0.950 & 0.119 & 0.949 & 0.097 & 0.949 & 0.084 & 0.950 \\ \hline
 &OLS & 0.460 & 0.950 & 0.322 & 0.950 & 0.262 & 0.950 & 0.227 & 0.949 \\ 
     $B_1$ &GLS & 0.439 & 0.952 & 0.307 & 0.950 & 0.250 & 0.950 & 0.216 & 0.949 \\ 
   &EGLS & 0.439 & 0.952 & 0.307 & 0.950 & 0.250 & 0.950 & 0.216 & 0.949 \\ \hline
   & OLS & 0.460 & 0.949 & 0.322 & 0.949 & 0.262 & 0.948 & 0.227 & 0.950 \\ 
  $B_2$ &GLS & 0.439 & 0.951 & 0.307 & 0.949 & 0.250 & 0.949 & 0.216 & 0.951 \\ 
  &EGLS & 0.439 & 0.951 & 0.307 & 0.949 & 0.250 & 0.949 & 0.216 & 0.951 \\ \hline
  & OLS & 0.035 & 0.951 & 0.024 & 0.948 & 0.020 & 0.948 & 0.017 & 0.949 \\ 
  $\gamma$ &GLS & 0.032 & 0.953 & 0.023 & 0.949 & 0.018 & 0.948 & 0.016 & 0.948 \\ 
   &EGLS  & 0.032 & 0.953 & 0.023 & 0.949 & 0.018 & 0.948 & 0.016 & 0.948 \\ \hline
 
\end{tabular}
\label{t31}
\end{table}

Due to space considerations, we show the results for the influence of the sample size $T$ in the first scenario, and present all other Tables in Appendix \ref{tf}. Table \ref{t31} shows that the length of the confidence intervals decreases for larger sample sizes $T$, while the coverage probabilities improve (get closer to the nominal $95\%$ level). These results are in accordance with the theoretical developments. Next, a summary of the results of the remaining five scenarios (Tables in Appendix \ref{tf}) is presented. Table \ref{t1} indicates that the length of  confidence intervals (95\% nominal level) for $A$, $B$ and $\gamma$ decreases, as $T$ increases for NAR processes with lag one. Table \ref{t2} shows that as the band width of $W$ increases,the  length of the $95\%$ confidence intervals of the network effect parameters $B$ increases, while the corresponding coverage probabilities are robust. From Table \ref{t3}, it can be seen that both the length of confidence intervals and coverage probabilities are robust for both the GLS and EGLS estimates with respect to changes in $\rho$ (except for the network effect parameters $B$ for which the CI length decreases slightly), while the length of CIs for the OLS estimates increases for all model parameters. Table \ref{t4} shows that as the number of factors $k$ increases, the length of the $95\%$ confidence intervals for $A$, $B$ and $\gamma$ increases for the OLS estimator, while for the GLS and EGLS estimators, the corresponding length of the CIs decreases. Finally, results summarized in Table \ref{t10} confirm that the length of CIs and coverage probabilities are robust with respect to changes on the band width of matrix $\Phi$. Overall, the coverage probabilities in all simulation scenarios are close to the nominal level ($95\%$), which implies that the estimators are unbiased and the estimated variances are close to the true variances.

\subsection{Influence of the Distribution of the Error Term}
\label{sec:influence-error-term}

We consider distributions with heavier tails than Gaussian. Specifically, we generate errors from a $t-$distribution; $\epsilon_t\sim t_v(0,(I-0.5 \Phi)^{-1}(I-0.5 \Phi)^{-T})$ and $Y_{ik,t}\sim t_v(0,I_{10})$, with $v=4,8,16$ degrees of freedom and compare the results on a NAR model with a SAR error covariance matrix and normally distributed errors.
In Table \ref{t14}, we set $N=100$, $q_1=q_2=1$, $T=400$, $\rho=0.5$, $W$ and $\Phi$ as banded matrices of width 5. Further, $A=diag\{0.1I_{25},0.2I_{25},0.3I_{25},0.4I_{25}\}$, $B=diag\{0.4I_{25},0.3I_{25},0.2I_{25},0.1I_{25}\}$, $\gamma=(-0.8\boldsymbol{1}_3^T,-0.4\boldsymbol{1}_2^T,0.4\boldsymbol{1}_2^T,0.8\boldsymbol{1}_3^T)^T$ and investigate how the RMSE of the EGLS estimator is influenced by the distribution of the error term. 
In Table \ref{t11}, we set $T=400$, $\rho=0.5$, $W$ and $\Phi$ as banedd matrix of width 5, $A=B=0.4I_{100}$, $\gamma=0.4\times\boldsymbol{1}_{10}$ and explore how the CP and the length of the CI are influenced by the distribution of the error term.

\begin{table}[H] 
\centering
\caption{Estimation performance evaluation for the NAR model with errors following a $t$-distribution with  4, 8, 16 degrees of freedom and also normal distribution. Setting considered: $N=100$, $q_1=q_2=1$, $T=400$, $\rho=0.5$ and $W$ and $\Phi$ being banded matrices of width 5. }

\begin{tabular}{c|c|c|c|c|c|c|c|c|c}
 
\hline
&Estimator&\multicolumn{2}{|c|}{DF=4}&\multicolumn{2}{|c|}{DF=8}&\multicolumn{2}{|c|}{DF=16}&\multicolumn{2}{|c}{Normal Dist.}\\ \hline
&True value&Estimate&RMSE&Estimate&RMSE&Estimate&RMSE&Estimate&RMSE\\ \hline
$a_1\sim a_{25}$&0.1&0.096&0.388 &0.096&0.381 &0.096&0.379 &0.097&0.373 \\ 
$a_{26}\sim a_{50}$&0.2&0.196&0.195 &0.197&0.190 &0.196&0.189 &0.196&0.181 \\ 
$a_{51}\sim a_{75}$&0.3&0.300&0.133 &0.300&0.132 &0.300&0.134 &0.295&0.114 \\ 
$a_{76}\sim a_{100}$&0.4&0.393&0.092 &0.393&0.092 &0.392&0.092 &0.396&0.080 \\ \hline
$b_1\sim b_{25}$&0.4&0.400&0.183 &0.400&0.185 &0.400&0.186 &0.399&0.185 \\ 
$b_{26}\sim b_{50}$&0.3&0.299&0.253 &0.299&0.257 &0.299&0.258 &0.301&0.255 \\ 
$b_{51}\sim b_{75}$&0.2&0.200&0.380 &0.20&0.388 &0.199&0.391 &0.198&0.376 \\ 
$b_{76}\sim b_{100}$&0.1&0.099&0.719 &0.100&0.727 &0.100&0.735 &0.100&0.713 \\\hline
$\gamma_1\sim \gamma_{3}$&-0.8&-0.803&0.027 &-0.804&0.025 &-0.804&0.025 &-0.804&0.024 \\ 
$\gamma_{4}\sim \gamma_{5}$&-0.4&-0.402&0.035 &-0.402&0.035 &-0.402&0.035 &-0.401&0.031 \\ 
$\gamma_{6}\sim \gamma_{7}$&0.4&0.402&0.054 &0.402&0.051 &0.402&0.051 &0.402&0.047 \\ 
$\gamma_{8}\sim \gamma_{10}$&0.8&0.804&0.034 &0.804&0.032 &0.804&0.032 &0.804&0.032 \\ \hline
 
\end{tabular}

\label{t14}
\end{table}

\begin{table}[H] 
\centering
\caption{CP and length of CI performance evaluation for the NAR model with errors following a $t$-distribution with  4, 8, 16 degrees of freedom and also normal distribution. Setting considered: $N=100$, $q_1=q_2=1$, $T=400$, $\rho=0.5$ and $W$ and $\Phi$ being banded matrices of width 5. }
\begin{tabular}{c|c|c|c|c|c|c|c|c|c}
 
\hline
&Estimator&\multicolumn{2}{|c|}{DF=4}&\multicolumn{2}{|c|}{DF=8}&\multicolumn{2}{|c|}{DF=16}&\multicolumn{2}{|c}{Normal Dist.} \\ \hline
&&CI&CP&CI&CP&CI&CP&CI&CP\\ \hline
 & OLS& 0.085 & 0.838 & 0.104 & 0.911 & 0.112 & 0.933& 0.119 & 0.952 \\ 
  $a_i$ &GLS&  0.078 & 0.840 & 0.096 & 0.912 & 0.103 & 0.934& 0.110 & 0.953 \\ 
   &EGLS& 0.078 & 0.840 & 0.096 & 0.912 & 0.103 & 0.934& 0.110 & 0.953 \\\hline 
   & OLS& 0.166 & 0.837 & 0.203 & 0.909 & 0.219 & 0.932& 0.234 & 0.951 \\ 
  $b_i$ &GLS& 0.159 & 0.837 & 0.194 & 0.911 & 0.209 & 0.932 & 0.224 & 0.950\\ 
   &EGLS& 0.159 & 0.837 & 0.194 & 0.911 & 0.209 & 0.932 & 0.224 & 0.950\\ \hline
   & OLS& 0.015 & 0.822 & 0.018 & 0.913 & 0.020 & 0.931& 0.021 & 0.949  \\ 
 $\gamma_i$ &GLS & 0.014 & 0.824 & 0.017 & 0.908 & 0.018 & 0.934 & 0.019 & 0.952\\ 
   &EGLS& 0.014 & 0.823 & 0.017 & 0.909 & 0.018 & 0.934& 0.019 & 0.952 \\ \hline
 
\end{tabular}
\label{t11}
\end{table}

The results in Table \ref{t14} show that non-normal errors have almost no impact on the quality of the obtained estimates. On the other hand, the results in Table \ref{t11} show that the obtained CIs are short and their CP below the nominal level, especially for 4 and 8 degrees of freedom. Further, for 16 degrees of freedom the difference to the CIs obtained from normal errors becomes fairly small and vanishes for the $\gamma$ parameters. 

Hence, if practitioners suspect that in their applications the error term exhibits tails heavier than Gaussian, the use of the residual bootstrap \citealp{rilstone1996using,kim1999asymptotic,lutkepohl2000bootstrapping} is recommended. The key steps of the residual bootstrap as it pertains to the NAR model and its theoretical justification are summarized in Appendix \ref{Appendix-bootstrap}.

We apply the stated algorithm to obtain $95\%$ CIs, wherein the setting is as follows: $T=400$, $\rho=0.5$, $W$ and $\Phi$ banded matrices of width 5, $A=B=0.4I_{100}$, $\gamma=0.4\times\boldsymbol{1}_{10}$ and $\epsilon_t$ follows a $t$-distribution with 4 degrees of freedom. The results are based on 5000 replicates. A comparison between CIs obtained from the theoretical results and those from the residual bootstrap are presented in Table  \ref{tt11}.

\begin{table}[H] 
\centering
\caption{Comparison of CIs based on the asymptotic distribution of the respective estimator and that obtained by the residual bootstrap. The setting is: $T=400$, $\rho=0.5$, $W$ and $\Phi$ banded matrices of width 5, $A=B=0.4I_{100}$, $\gamma=0.4\times\boldsymbol{1}_{10}$ and $\epsilon_t$ follows a t-distribution with 4 degrees of freedom. }

\begin{tabular}{c|c|c|c|c|c}
 
\hline
&Estimator& Asymptitc CI & Asymptotic CP & Bootstrap CI & Bootstrap CP\\ \hline
$a_i$ & OLS& $(0.356,0.441)$ & 0.838&$(0.336,0.456)$&0.949  \\ 
   &EGLS& $(0.359,0.437)$ & 0.840&$(0.341,0.451)$& 0.947  \\\hline 
   $b_i$ & OLS& $(0.315,0.481)$ & 0.837&$(0.276,0.511)$&0.950  \\ 
   &EGLS& $(0.319,0.478)$ & 0.837&$(0.286,0.509)$&0.950 \\ \hline
  $\gamma_i$ & OLS& $(0.392,0.407)$ & 0.822 &$(0.377,0.422)$&0.949  \\ 
   &EGLS& $(0.393,0.407)$ & 0.823&$(0.386,0.414)$&0.948  \\ \hline
 
\end{tabular}

\label{tt11}
\end{table}

The results show that the CP of the residual bootstrap based CIs attains the nominal level.

\section{Application to Air Quality Index Data}
\label{sec:application}

\begin{figure}[H] 
\centering
\subfigure[Locations of the 319 stations]{
\begin{minipage}[t]{4.5cm}
\centering
\includegraphics[scale=0.33]{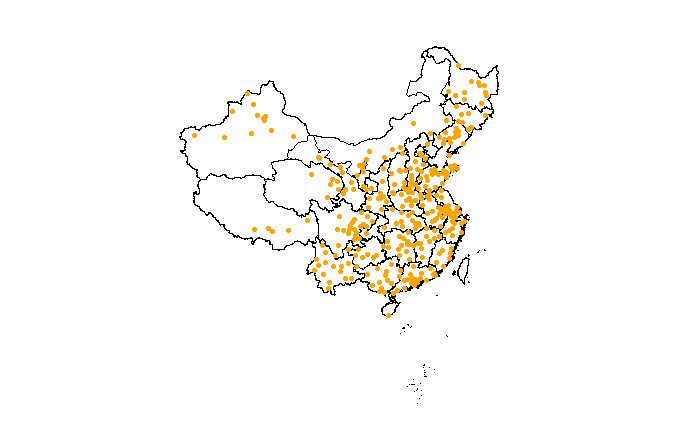}
\label{RDA1}
\end{minipage}
}
\subfigure[Box plot of Monthly AQI values]{
\begin{minipage}[t]{6cm}
\centering
\includegraphics[scale=0.33]{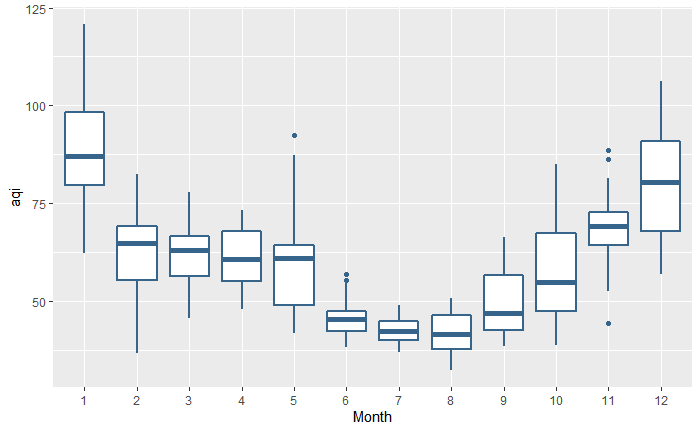}
\label{MAQI}
\end{minipage}
}
\subfigure[Average AQI for each station]{
\begin{minipage}[t]{3cm}
\centering
\includegraphics[scale=0.33]{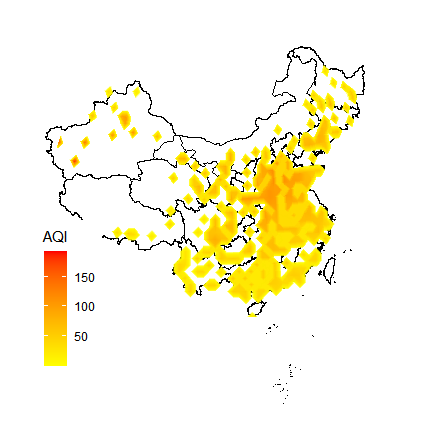}
\end{minipage}
\label{AAQI}
}
\caption{Spatial distribution of AQI monitoring stations and their average values, together with monthly variability in the AQI values \label{AQI}}
\end{figure}

We employ the proposed NAR($q_1,q_2$) model to analyze Air Quality Index (AQI) data together with relevant weather condition covariates, collected from $N=319$ stations across China for the period from March 20th, 2019 to March 19th, 2020, for a total of $T=366$ observations. The AQI data are obtained from the China National Environmental Monitoring Centre, while the weather covariates from the National Centers for Environmental Information\footnote{\texttt{https://www.ncdc.noaa.gov/isd}}. The locations of the stations (left panel), boxplots of the AQI for each month across all stations (middle panel) and the average AQI for each station across all observations (right panel) are depicted in Figure \ref{AQI}. We consider the log transformed AQI as the response variable. It can be seen from the middle panel of Figure \ref{AQI} that the average AQI reaches its peak in winter, while the pollution level is relatively low in summer; hence, we fit \textit{separate} models for each season. Exogenous covariates $\mathbb{Y}_t$ included in the NAR($q,q$) model include air temperature, relative humidity, wind speed rate and sky condition total coverage. 

Note also that the right panel of Figure \ref{AQI} indicates substantial heterogeneity, with the north-northwest regions of the country exhibiting higher AQI levels.

 The model is fitted with both a SAR covariance structure and a factor model one. To construct the network for the SAR version, both the weight matrix $W$ and $\Phi$ correspond to a row-normalized adjacent matrix obtained as follows: let $D_{ij}$ be the spatial distance between two stations and $\sigma^2$ be the variance of all distances, then the $ij$-th element of $\Phi$ is defined as:
 \begin{equation*}
\phi_{ij}:=\left\{
 \begin{array}{ll}
 \frac{D_{ij}^{-1}}{\sum\limits_j D_{ij}^{-1}} & \text{if} \ i\not=j \\
 0 & \text{if} \ i=j.
\end{array}\right.
\end{equation*}
Recall that $\Phi$ aims to capture any additional spatial dependence not reflected in the struuture of the NAR model.

Further, the $ij$-th element of $W$ is defined as:
 \begin{equation*}
w_{ij}:=\left\{
 \begin{array}{ll}
 \frac{D_{ij}^{-1}}{\sum\limits_j D_{ij}^{-1}} & \text{if} \ i\not=j\ and\ D_{ij}\leq 500 \text{ km}  \\
 0 & \text{if} \ i=j.
\end{array}\right.
\end{equation*}

Based on the following BIC criterion
\[BIC(q)=\log |\hat{\Sigma}_\epsilon|+\frac{(2Nq+p)\log T}{T},\]
an NAR(1,1) model was selected for each season. A plot of the partial autocorrelation function for the AQI variable (not shown) corroborates this choice for the temporal autoregressive and network lags.

To select the number of factors in the corresponding model, we employed the following information criterion:
\[IC(k)=\log(S(k))+\frac{k(N+T-k)}{NT}\log (NT).\]
It resulted in selecting a single factor ($k=1$) for each season's NAR model.

Given the large number of stations (319) and limited sample size for each season, the autoregressive $a_i$ and network lag $b_i$ coefficients, together with those of the external covariates were obtained based on regularized (ridge) EGLS and depicted in Figures \ref{Fig:lEGLS1a} and \ref{Fig:lEGLS1b} and tabulated in Table \ref{Tab:lEGLSCI1}-\ref{Tab:lEGLSCI4} (in Appendix \ref{tf}), respectively. The NAR model estimates show great variation amongst different regions and different seasons. 

\begin{figure} [H]
\centering
\subfigure[Spring]{
\begin{minipage}[t]{6cm}
\centering
\includegraphics[scale=0.4]{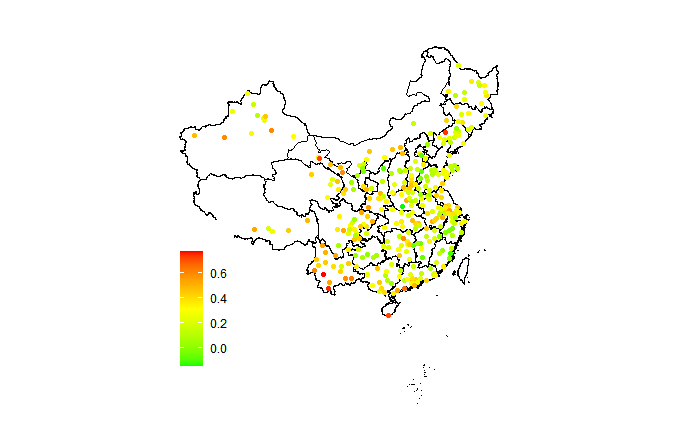}
\end{minipage}
}
\subfigure[Summer]{
\begin{minipage}[t]{6cm}
\centering
\includegraphics[scale=0.4]{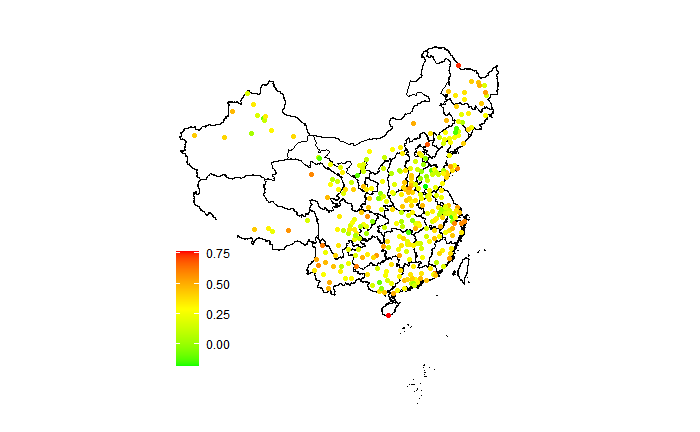}
\end{minipage}
} 
\subfigure[Fall]{
\begin{minipage}[t]{6cm}
\centering
\includegraphics[scale=0.4]{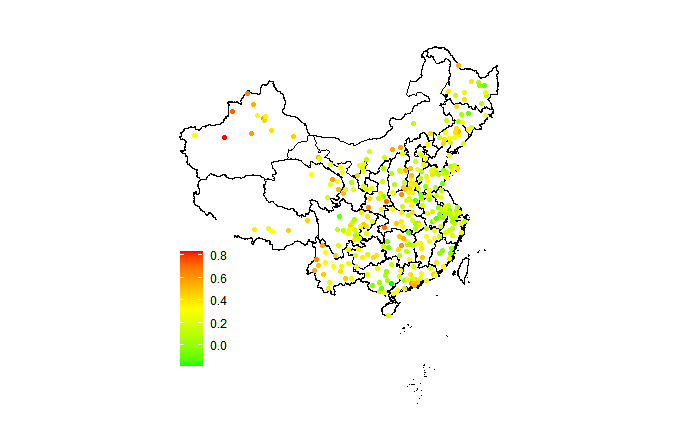}
\end{minipage}
} 
\subfigure[Winter]{
\begin{minipage}[t]{6cm}
\centering
\includegraphics[scale=0.4]{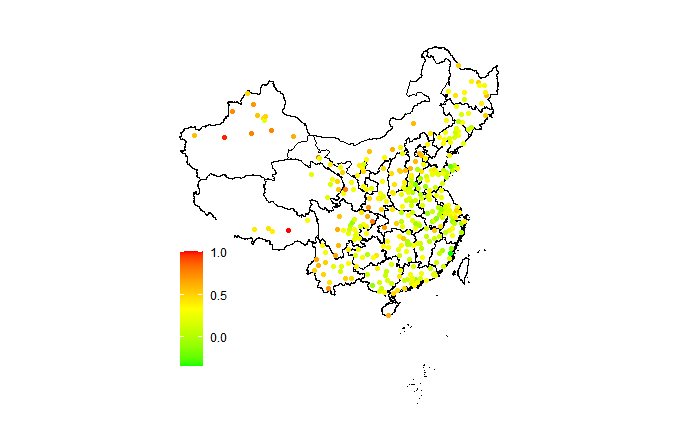}
\end{minipage}
} 
\caption{Autoregressive coefficients $a_i$\label{Fig:lEGLS1a}}
\end{figure}

\begin{figure} [H]
\centering
\subfigure[Spring]{
\begin{minipage}[t]{6cm}
\centering
\includegraphics[scale=0.4]{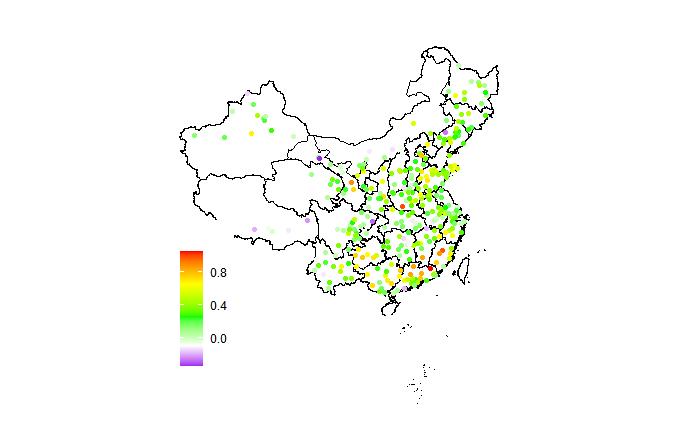}
\end{minipage}
}
\subfigure[Summer]{
\begin{minipage}[t]{6cm}
\centering
\includegraphics[scale=0.4]{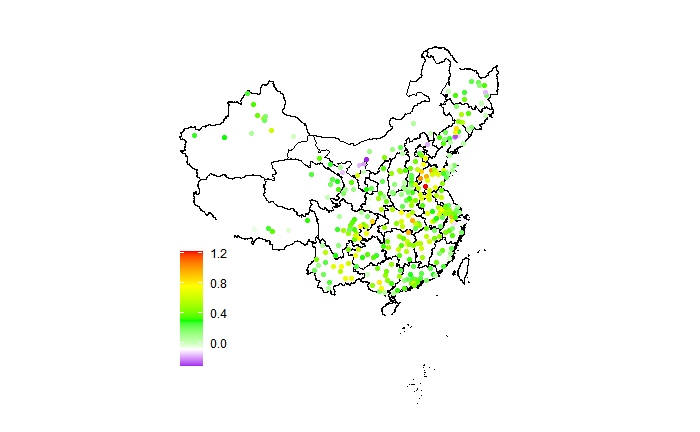}
\end{minipage}
} 
\subfigure[Fall]{
\begin{minipage}[t]{6cm}
\centering
\includegraphics[scale=0.4]{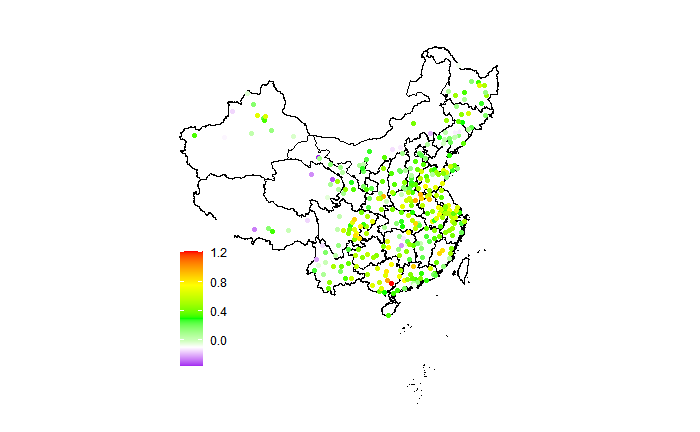}
\end{minipage}
} 
\subfigure[Winter]{
\begin{minipage}[t]{6cm}
\centering
\includegraphics[scale=0.4]{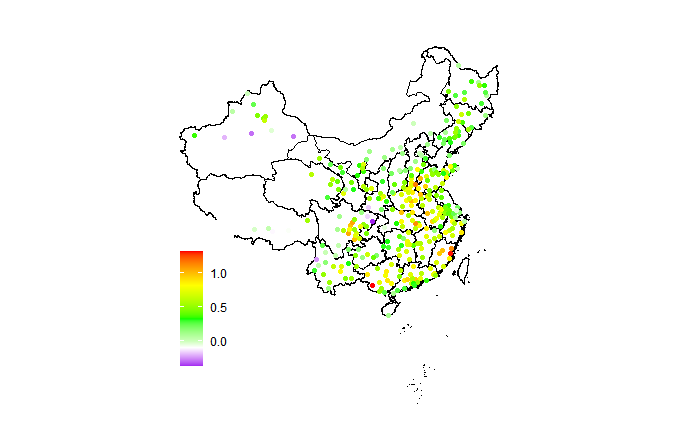}
\end{minipage}
} 
\caption{Network lag coefficients $b_i$\label{Fig:lEGLS1b}}
\end{figure}

The results in Table \ref{Tab:lEGLSCI1}-\ref{Tab:lEGLSCI4} show that all covariates employed are statistically significant. For relative humidity, its magnitude remains constant across the four seasons and its impact is positive in reducing air pollution (negative sign of the regression coefficient). Analogously, the impact of the wind speed is fairly similar across the four seasons and positive for air quality. The impact of the air temperature is positive and similar during the Summer and Fall seasons; further, it exhibits a bigger positive impact in Winter and a small negative impact in the Spring.

\begin{figure} [H]
\centering
\subfigure[Spring]{
\begin{minipage}[t]{6cm}
\centering
\includegraphics[scale=0.35]{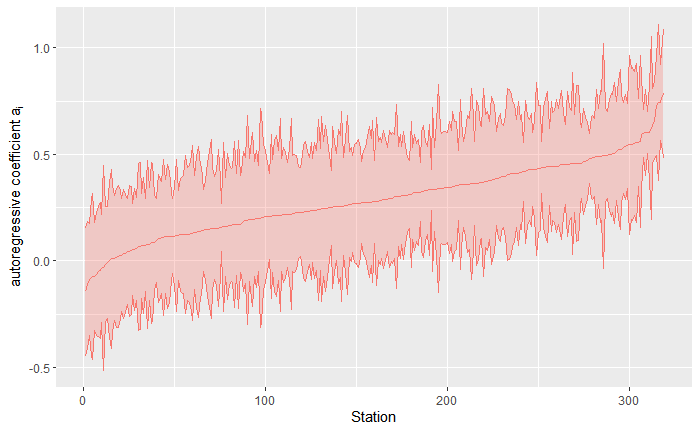}
\end{minipage}
}
\subfigure[Summer]{
\begin{minipage}[t]{6cm}
\centering
\includegraphics[scale=0.35]{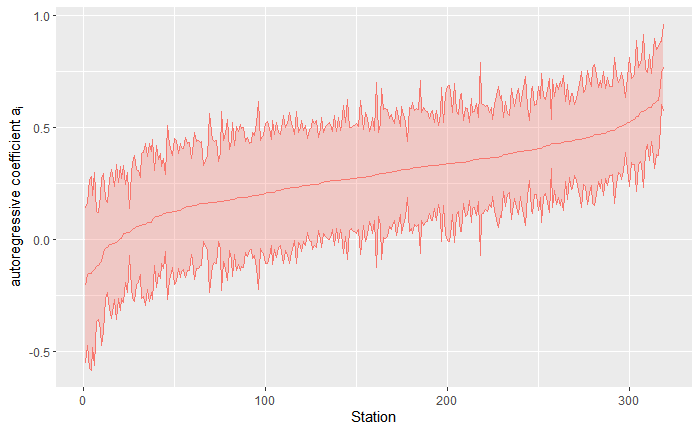}
\end{minipage}
} 
\subfigure[Fall]{
\begin{minipage}[t]{6cm}
\centering
\includegraphics[scale=0.35]{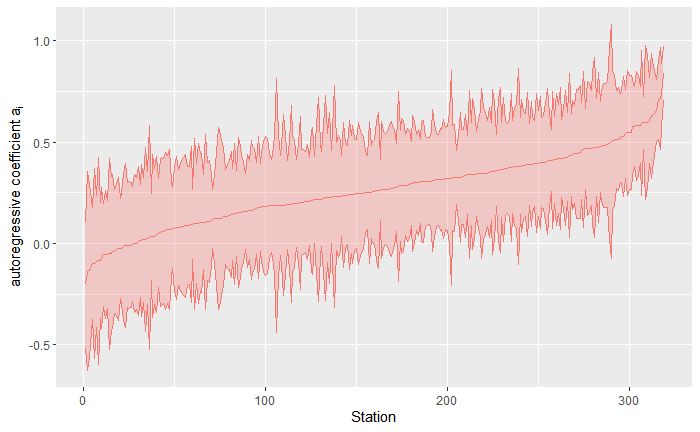}
\end{minipage}
} 
\subfigure[Winter]{
\begin{minipage}[t]{6cm}
\centering
\includegraphics[scale=0.35]{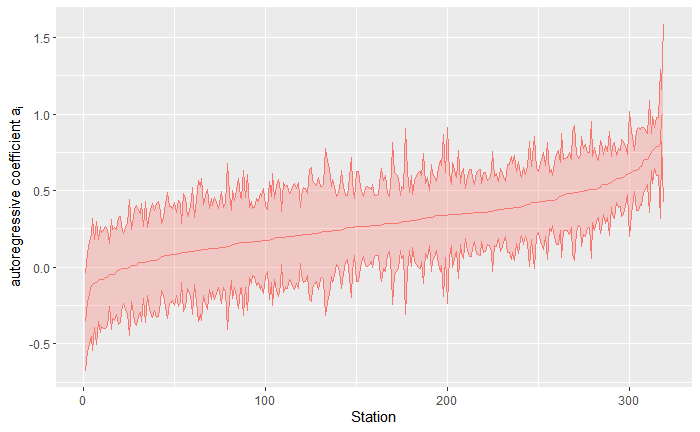}
\end{minipage}
} 
\caption{Estimates and corresponding confidence intervals of autoregressive effects ($a_i$) across different seasons\label{Fig:lEGLSCI2}}
\end{figure}

\begin{figure} [H]
\centering
\subfigure[Spring]{
\begin{minipage}[t]{6cm}
\centering
\includegraphics[scale=0.35]{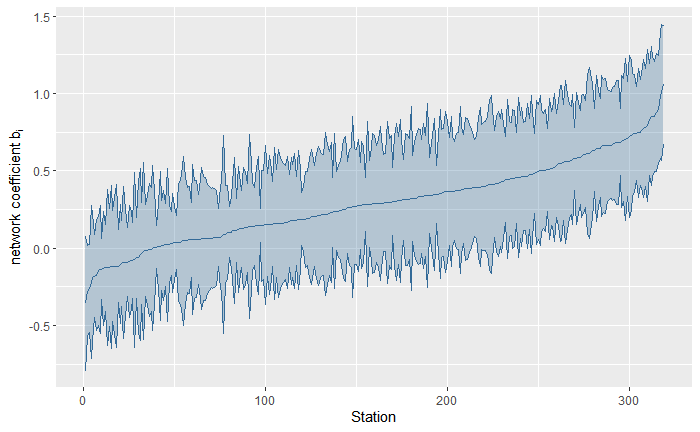}
\end{minipage}
}
\subfigure[Summer]{
\begin{minipage}[t]{6cm}
\centering
\includegraphics[scale=0.35]{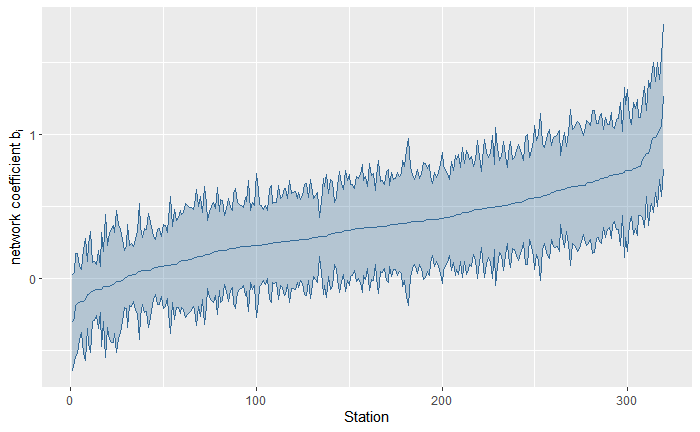}
\end{minipage}
} 
\subfigure[Fall]{
\begin{minipage}[t]{6cm}
\centering
\includegraphics[scale=0.35]{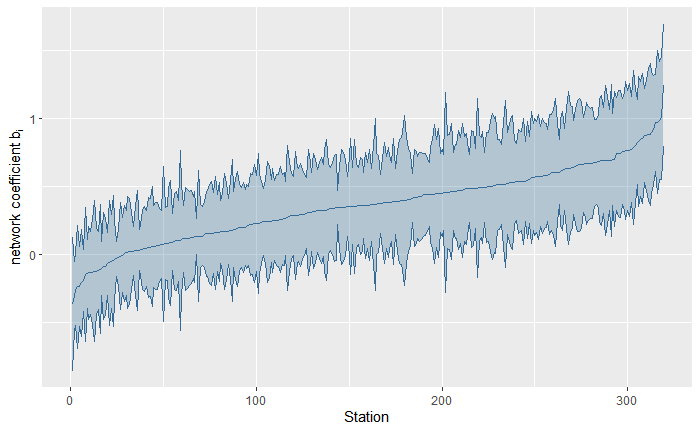}
\end{minipage}
} 
\subfigure[Winter]{
\begin{minipage}[t]{6cm}
\centering
\includegraphics[scale=0.35]{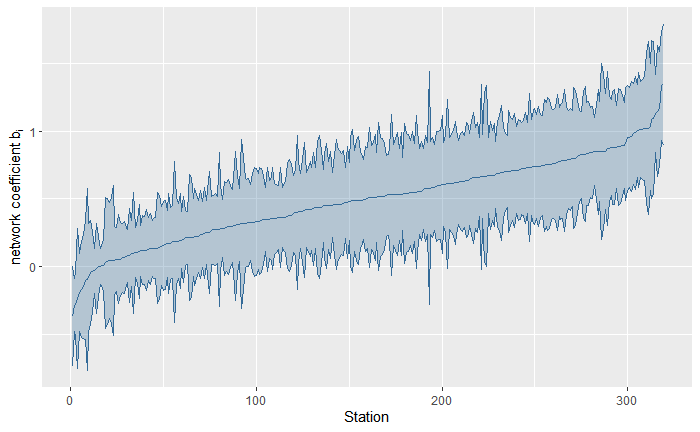}
\end{minipage}
} 
\caption{Estimates and confidence intervals of network effects ($b_i$) across different seasons\label{Fig:lEGLSCI3}}
\end{figure}

To aid interpretation, Table \ref{tab:provinces} presents the average autoregressive and network lag coefficients for all the provinces and selected big cities. Current air pollution in China is caused by multiple pollutants, with great variations among different regions and different seasons. Future studies should focus on improving the understanding of the associations between air quality and meteorological conditions, variations of emissions in different regions, and transport and transformation of pollutants in both intra- and inter-regional contexts.

It can be seen that provinces with the largest autoregressive coefficients are Hainan, Yunnan and Tibet, while provinces with the highest network coefficients are Guangxi, Fujian and Anhui. The topography of the province (island or plateau) may be related to the presence of such large autoregressive coefficients (e.g., Yunnan–Guizhou Plateau, Tibetan Plateau and the Hainan Island). In contrast to regions with large autoregressive coefficients, provinces with the largest network coefficients are coastal areas (e.g., Fujian and Guangxi). During Winter months, northern regions like Beijing tend to have larger autoregressive coefficients and smaller network coefficients compared to other seasons, and temperature inversion may be the cause. During an inversion, warmer air is held above cooler air, so air pollution is trapped by it, which makes the air pollution hard to diffuse.

\begin{table}[H]
\caption{Average autoregressive coefficient, network coefficient for each province/city.\label{tab:provinces}}
\centering
\resizebox{\columnwidth}{!}{
\begin{tabular}{rlrrrrrrrr}
  \hline
 & Province & $a_i$ (Spring)& $b_i$ (Spring) & $a_i$ (Summer) & $b_i$ (Summer) & $a_i$ (Fall) & $b_i$ (Fall) & $a_i$ (Winter) & $b_i$ (Winter)\\ 
  \hline
1 & Beijing & 0.26 & 0.38 & 0.38 & 0.24 & 0.31 & 0.20 & 0.60 & 0.01 \\ 
  2 & Tianjin & 0.26 & 0.34 & -0.03 & 0.74 & 0.33 & 0.15 & 0.35 & 0.30 \\ 
  3 & Hebei & 0.20 & 0.33 & 0.29 & 0.37 & 0.33 & 0.24 & 0.36 & 0.36 \\ 
  4 & Shanxi & 0.24 & 0.24 & 0.26 & 0.26 & 0.25 & 0.28 & 0.42 & 0.28 \\ 
  5 & Inner Mongolia & 0.34 & 0.13 & 0.30 & 0.06 & 0.29 & 0.16 & 0.45 & 0.14 \\ 
  6 & Liaoning & 0.25 & 0.32 & 0.20 & 0.20 & 0.33 & 0.10 & 0.26 & 0.32 \\ 
  7 & Jilin & 0.30 & 0.34 & 0.28 & 0.29 & 0.20 & 0.30 & 0.23 & 0.44 \\ 
  8 & Heilongjiang & 0.26 & 0.24 & 0.41 & 0.17 & 0.21 & 0.29 & 0.38 & 0.30 \\ 
  9 & Shanghai & 0.34 & 0.06 & 0.46 & 0.11 & 0.32 & 0.32 & 0.35 & 0.16 \\ 
  10 & Jiangsu & 0.34 & 0.21 & 0.26 & 0.44 & 0.15 & 0.54 & 0.26 & 0.39 \\ 
  11 & Zhejiang & 0.13 & 0.34 & 0.35 & 0.30 & 0.15 & 0.46 & 0.16 & 0.58 \\ 
  12 & Anhui & 0.30 & 0.27 & 0.28 & 0.49 & 0.10 & 0.59 & 0.11 & 0.68 \\ 
  13 & Fujian & 0.12 & 0.62 & 0.32 & 0.22 & 0.14 & 0.46 & 0.02 & 0.94 \\ 
  14 & Jiangxi & 0.14 & 0.40 & 0.26 & 0.49 & 0.28 & 0.28 & 0.20 & 0.59 \\ 
  15 & Shandong & 0.18 & 0.47 & 0.21 & 0.53 & 0.18 & 0.45 & 0.16 & 0.54 \\ 
  16 & Henan & 0.27 & 0.29 & 0.36 & 0.34 & 0.30 & 0.45 & 0.13 & 0.82 \\ 
  17 & Hubei & 0.31 & 0.15 & 0.23 & 0.56 & 0.33 & 0.38 & 0.26 & 0.60 \\ 
  18 & Hunan & 0.30 & 0.15 & 0.26 & 0.52 & 0.25 & 0.30 & 0.22 & 0.55 \\ 
  19 & Guangdong & 0.32 & 0.44 & 0.35 & 0.27 & 0.33 & 0.34 & 0.25 & 0.59 \\ 
  20 & Guangxi & 0.27 & 0.49 & 0.24 & 0.40 & 0.10 & 0.67 & 0.19 & 0.70 \\ 
  21 & Hainan & 0.72 & -0.07 & 0.77 & -0.08 & 0.25 & 0.35 & 0.62 & 0.11 \\ 
  22 & Chongqing & 0.20 & 0.06 & 0.03 & 0.72 & 0.23 & 0.54 & 0.31 & 0.44 \\ 
  23 & Sichuan & 0.33 & 0.15 & 0.25 & 0.41 & 0.24 & 0.49 & 0.35 & 0.49 \\ 
  24 & Guizhou & 0.13 & 0.64 & 0.41 & 0.41 & 0.29 & 0.49 & 0.29 & 0.36 \\ 
  25 & Yunnan & 0.50 & 0.21 & 0.37 & 0.42 & 0.39 & 0.23 & 0.43 & 0.34 \\ 
  26 & Tibet & 0.39 & -0.12 & 0.35 & 0.18 & 0.38 & 0.02 & 0.52 & 0.07 \\ 
  27 & Shaanxi & 0.30 & 0.36 & 0.21 & 0.24 & 0.31 & 0.40 & 0.39 & 0.38 \\ 
  28 & Gansu & 0.38 & 0.23 & 0.23 & 0.17 & 0.29 & 0.17 & 0.41 & 0.38 \\ 
  29 & Qinghai & 0.34 & 0.21 & 0.32 & 0.23 & 0.37 & 0.01 & 0.32 & 0.48 \\ 
  30 & Ningxia & 0.12 & 0.40 & 0.17 & 0.07 & 0.31 & 0.13 & 0.34 & 0.52 \\ 
  31 & Xinjiang & 0.35 & 0.19 & 0.26 & 0.26 & 0.52 & 0.19 & 0.62 & 0.16 \\ 
   \hline
\end{tabular}
}
\end{table}
The results are broadly in accordance with findings in recent studies that have investigated spatial and temporal variations of air pollutants in China. \citep{wang2014spatial}.
In North East China, coal-based industries such as iron and steel manufacturing and coal-fired power plants are key drivers for increased AQI levels the main causes of air pollution.
In the Northern China Plain where Beijing, Tianjin, Hebei Province and Henan Province are located, the network effect is high compared to other regions. Emissions from fossil fuel combustion and biomass burning for home heating in the winter months result in high concentration of air pollutants. Surrounded by the Yan and Taihang Mountains, particulates brought by south easterly winds may accumulate in the region, whereas cold front from the north together with their winds are weakened by the mountains and hence result in increased pollution levels. Further, sandstorms from the Hunshandake, Ulan Buh and Ordos deserts also contribute to the network effects observed for Northern China Plain stations. A number of studies have discussed air pollution patterns in this region and potential drivers \citep{wang2017inter,xiao2020analysis}. For the Yangtze River Delta, where Shanghai and Jiangsu are located, the network coefficient is large compared to the autoregressive coefficient. Particulates brought by cold fronts from the Mongolian Plateau also contribute to the network effect.  Finally, it is worth noting that eastern regions exhibit on average larger network coefficients, while western regions have higher autoregressive coefficients.

Next, based on the obtained estimates, we empirically examine the stability/stationarity condition given in Proposition \ref{stationary}. We find that $\rho(\hat{G})<1$ by the NAR(1,1) model for all seasons. However, we note that for several stations the condition used in previous work in the literature 
\begin{equation}\label{eq:condition-stationarity}
\mathop{max}\limits_{1\leq i\leq N}\{\sum\limits_{l=1}^q(|a_i^{(l)}|+|b_i^{(l)}|)\}<1
\end{equation}
is violated; examples include a station in Fuxin for Spring with $a_i=0.838$ and $b_i=-0.313$, a station in Shaoxing for Summer with $a_i=0.668$ and $b_i=-0.362$ and so forth. This shows that the latter condition is restrictive and imposing it may lead to deteriorating predictive performance, as briefly illustrated in Table \ref{illustration-stationarity}. Specifically, the Predictive MSE (PMSE, defined in the sequel) is shown for each season's NAR(1,1) model with the EGLS estimated autoregressive and network lag coefficients satisfying (by truncating their values) condition \eqref{eq:condition-stationarity} and also without imposing it. 
%
\begin{table}[H] 
\centering
\caption{PMSE with and without condition \ref{eq:condition-stationarity} imposed\label{illustration-stationarity}}
\begin{tabular}{c|c|c|c|c}
   \hline
  & Spring&Summer&Fall&Winter \\ \hline
original EGLS estimates &  0.14024 & 0.07328&0.06691& 0.08542 \\ \hline
truncated EGLS estimates & 0.14247 & 0.07353& 0.07065& 0.09352 \\ \hline
\end{tabular}
\end{table}
It can be seen that imposing the standard condition in the literature for model stability has a deleterious effect to the predictive performance of the NAR model. As our results in Section \ref{sec:stability} show, this
condition is too stringent and can be replaced by the proposed condition \eqref{eq:condition-stationarity} in this work.

Next, we consider how the estimated NAR(1,1) model performs in terms of forecasts together with a number of competing models.
The parameters of the NAR(1,1) model are estimated by both OLS and EGLS, with SAR and factor covariance structures and a ridge penalty (since $N>T$).
For comparison purposes, we also consider an NAR(1,1) model with $A=aI$ and $B=bI$, a regularized VAR(1) model with ridge and lasso penalties and finally a simple AR(1) model, applied to each station's data.
The evaluation is based on the last 20 days (test data) of each season, that are used to calculate PMSEs for the different models, defined as: 
$$
(N|T_{test}|)^{-1}\sum\limits_{t\in T_{test}}||\mathbb{X}_t-\mathbb{Z}_{t-1}\hat{\beta}||_F^2.$$
The results are shown in Table \ref{t41}. 
\begin{table}[H] 
\centering
\caption{PMSE for different estimators across different seasons}

\begin{tabular}{c|c|c|c|c}
 
  \hline
  & Spring&Summer&Fall&Winter \\ \hline
OLS &  0.0656 & 0.0665&0.0999& 0.0889 \\ \hline
EGLS w/ spatial covariance & 0.0654 & 0.0656& 0.0990& 0.0880 \\ \hline
EGLS w/ factor structure & 0.0613 & 0.0660& 0.1030& 0.0884 \\ \hline
NAR with $A=aI$ and $B=bI$ & 0.0658 & 0.0661 & 0.1019 & 0.0925 \\ \hline
VAR(1) with ridge & 0.0927 & 0.0835 & 0.1466 & 0.1180 \\ \hline
VAR(1) with lasso & 0.0727 & 0.0791 & 0.1239 & 0.1068 \\ \hline
AR(1) & 0.0684 & 0.0693 & 0.1136 & 0.1054 \\ \hline
\end{tabular}
\label{t41}
\end{table}

It can be seen that the NAR based predictions clearly outperform the VAR and AR(1) ones, across all seasons. Further, the EGLS based one for the posited NAR model exhibits better performance than its OLS counterpart and also the predictions of the homogeneous NAR model. Differences are  minuscule for Summer, but around 5\% in magnitude for the other seasons.

\section{Conclusion and Discussion}
\label{sec:discussion}

The paper presented a general flexible framework for NAR processes that
can accommodate node-specific network effects, exogenous covariates, 
errors that can exhibit 
heavier than Gaussian tails and a variety of error covariance matrices.
It can also be regarded as a VAR model with specific structure in the transition matrices that reduces the number of parameters, and also aids in interpretability. The latter connection also enables us to provide a significantly weaker stability condition compared to those available in the literature for significantly simpler models, thus expanding the applicability of the framework, as also illustrated in the real data application. The parameter reduction requires a priori knowledge of the weight matrices. However, the results established show that the model parameters estimates are robust to a certain degree of misspecification of these matrices.

To that end, how to ``design" the weight matrices $W$ to optimize performance is a topic of future research. Further, in very high-dimensional settings, the use of sparsity inducing penalties (such as the lasso and its variants) is of interest, together with inference procedures based on ideas of debiasing the resulting parameter estimates.

\bibliographystyle{unsrtnat}


\begin{appendix}

\noindent
{\huge\textbf{Appendix for: \\
A General Framework for Network Autoregressive Processes}}

\section{Proof of Stability Result (Section \ref{sec:stability})}\label{pos}
\begin{proof}[Proof of Proposition \ref{stationary}]
The NAR$(q_1,q_2)$ model defined in \eqref{eq:matrix} can be rewritten in terms of a VAR($1$) process taking the form:
\begin{equation}
\boldsymbol{X}_t=\boldsymbol{G}\boldsymbol{X}_{t-1}+\mathcal{E}_t,
\end{equation}
where
\begin{equation}
\begin{split}
&\boldsymbol{X}_t:=\begin{bmatrix}\mathbb{X}_t\\\mathbb{X}_{t-1}\\\vdots\\\mathbb{X}_{t-q+1}\end{bmatrix},\ \mathcal{E}_t:=\begin{bmatrix}\Tilde{\epsilon}_t\\0\\\vdots\\0 \end{bmatrix}, \\
&\boldsymbol{G}:=\begin{bmatrix}G_1&G_2&\cdots&G_{q-1}&G_q\\I_N&0&\cdots&0&0\\0&I_N&\cdots&0&0\\\vdots&\vdots&\ddots&\vdots&\vdots\\0&0&\cdots&I_N&0 \end{bmatrix}.
\end{split}
\end{equation}
Hence, the proof of stationarity follows from the general theory for VAR processes (see e.g. Section~2 of \cite{lutkepohl2005new} and Chapter~10 of \cite{citeulike:105953}).
\end{proof}
\section{OLS results}
\subsection{LEMMA \ref{L:LLNOLS}}\label{a1}
\begin{l7}
\label{L:LLNOLS}
Under the conditions of Proposition~\ref{P:OLS}, the following holds
\begin{equation}
\label{LLN:OLS}
\frac{1}{T}\sum_{t=1}^{T}\mathbb{Z}_{t-1}^T\mathbb{Z}_{t-1}\rightarrow_{p}
E(\mathbb{Z}_{t}^T\mathbb{Z}_{t}).
\end{equation}
\end{l7}
\begin{proof}[Proof of LEMMA \ref{L:LLNOLS}]
\label{proof:l7}
Following the proof of Proposition 11.1 in \cite{citeulike:105953}, for stationary $\mathbb{X}_t$ defined in \eqref{varq}, \[\det(I_{N}-G_1z-\cdots-G_q z^q)\not=0,\ \text{for} \ |z|\leq 1\] ensures the $\text{MA}(\infty)$ representation for the process $\mathbb{X}_t$ is absolutely summable. Hence, $\mathbb{X}_t$ is ergodic for first moments with 
\[E(\mathbb{X}_t)=\begin{bmatrix}
I_N& \boldsymbol{0}_{N\times N(q-1)}
\end{bmatrix}\sum_{j=0}^{\infty}\boldsymbol{G}^jE(\mathcal{E}_{t-j})=0\]
from Proposition 10.2(b), 10.5(a) of \cite{citeulike:105953}, and is also ergodic for second moments with 
\[E(\mathbb{X}_t\mathbb{X}_t^T)=\begin{bmatrix}
I_N& \boldsymbol{0}_{N\times N(q-1)}
\end{bmatrix}\sum\limits_{i=0}^{\infty}\boldsymbol{G}^i\Sigma_{E}(\boldsymbol{G}^i)^T\begin{bmatrix}
I_N\\ \boldsymbol{0}_{N\times N(q-1)}
\end{bmatrix},\]
where $\Sigma_{E}:=Var(\mathcal{E}_t)=\begin{bmatrix}
\Sigma_{\Tilde{\epsilon}}&0\\
0&0
\end{bmatrix}$ from Proposition 10.2(d) of \cite{citeulike:105953}.

As a result, we have 

\[\frac{1}{T}\sum\limits_{t=1}^T\boldsymbol{X}_t\boldsymbol{X}_t^T\rightarrow_{p}E(\boldsymbol{X}_t\boldsymbol{X}_t^T).\]

Also, by Assumption \ref{A1}, $\{\mathbb{Y}_{t},t\in \mathbb{N}\}$ is a sequence of i.i.d. random vectors with $E(\mathbb{Y}_{t})=0$ and $E(\mathbb{Y}_{t}\mathbb{Y}_{t}^T)=\Sigma_{Y}$, thus $\frac{1}{T}\sum_{t=1}^T\mathbb{Y}_{t}\mathbb{Y}_{t}^T\rightarrow_pE(\mathbb{Y}_{t}\mathbb{Y}_{t}^T)$ by the law of large numbers.

Moreover, by Assumption \ref{A1}, $\{\epsilon_t,t\in \mathbb{N}\}$ is independent of $\{\mathbb{Y}_{t},t\in \mathbb{N}\}$. Next, note that (a) the row sums of $W$ is equal to 1; (b) the non-zero elements in $\mathbb{Z}_{t-1}^T\mathbb{Z}_{t-1}$ consist of elements in $\mathbb{Y}_{t}\mathbb{Y}_{t}^T$, elements in $\boldsymbol{X}_{t-1}^T\boldsymbol{X}_{t-1}$, and their weighed averages. Thus, we conclude that $E(\mathbb{Z}_{t}^T\mathbb{Z}_{t})$ exists, and:

\[\frac{1}{T} \sum_{t=1}^{T}\mathbb{Z}_{t-1}^T\mathbb{Z}_{t-1} \rightarrow_{p}
E(\mathbb{Z}_{t}^T\mathbb{Z}_{t}),\]
\end{proof}
which completes the proof.

\subsection{LEMMA \ref{L:BDD}}\label{a2} 
\begin{l8}
\label{L:BDD}
Suppose Assumptions \ref{A1}-\ref{A3} hold and $\mathbb{X}_t$ is the NAR$(q_1,q_2)$ model defined as the stationary solution of \eqref{compact}, i.e. $\mathbb{X}_{t}=\mathbb{Z}_{t-1}\beta+ \epsilon_{t}$. For a diverging $N$, we further impose Assumption~\ref{A2}. Then, $E(|X_{i,t_1}X_{l,t_2}X_{m,t_3}X_{n,t_4}|)\leq c_3<\infty$ for any $1\leq i,l,m,n\leq N$ wherein $c_3$ is large positive constant.
\end{l8}

\begin{proof}[Proof of LEMMA \ref{L:BDD}]

We consider the following two cases:
\begin{itemize}
    \item 
    $N$ is finite and bounded fourth moments for $\epsilon_t$ and $\mathbb{Y}_{t}$.
    \item
    $N$ is diverging and $\epsilon_t$ and $\mathbb{Y}_{t}$ are sub-Weibull.
\end{itemize}
Note that $X_{i,t}=\sum\limits_{k=0}^{\infty}\sum\limits_{j= 1}^{Nq}g^k_{ij}\Tilde{\epsilon}_{j,(t-k)}$, where $g^k_{ij}$ denotes the $ij$-th element of $\boldsymbol{G}^k$. Since $\rho(\boldsymbol{G})<1$, for any $1\leq i,l,m,n\leq N$:
\begin{equation}\nonumber
\begin{split}
&E(|X_{i,t_1}X_{l,t_2}X_{m,t_3}X_{n,t_4}|)\\ 
&= E(|\sum\limits_{k_1= 0}^{\infty}\sum\limits_{j_1= 1}^{Nq}g^{k_1}_{ij_1}\Tilde{\epsilon}_{j_1,(t_1-k_1)}||\sum\limits_{k_2= 0}^{\infty}\sum\limits_{j_2= 1}^{Nq}g^{k_2}_{lj_2}\Tilde{\epsilon}_{j_2,(t_2-k_2)}||\sum\limits_{k_3= 0}^{\infty}\sum\limits_{j_3= 1}^{Nq}g^{k_3}_{mj_3}\Tilde{\epsilon}_{j_3,(t_3-k_3)}||\sum\limits_{k_4= 0}^{\infty}\sum\limits_{j_4= 1}^{Nq}g^{k_4}_{nj_4}\Tilde{\epsilon}_{j_4,(t_4-k_4)}|)\\
&\leq E(\sum\limits_{k_1= 0}^{\infty}\sum\limits_{j_1= 1}^{Nq}\sum\limits_{k_2= 0}^{\infty}\sum\limits_{j_2= 1}^{Nq}\sum\limits_{k_3= 0}^{\infty}\sum\limits_{j_3= 1}^{Nq}\sum\limits_{k_4= 0}^{\infty}\sum\limits_{j_4= 1}^{Nq}|g^{k_1}_{ij_1}\Tilde{\epsilon}_{j_1,(t_1-k_1)}g^{k_2}_{lj_2}\Tilde{\epsilon}_{j_2,(t_2-k_2)}g^{k_3}_{mj_3}\Tilde{\epsilon}_{j_3,(t_3-k_3)}g^{k_4}_{nj_4}\Tilde{\epsilon}_{j_4,(t_4-k_4)}|)\\
\end{split}
\end{equation}
By Assumption \ref{A1} , $E(|\epsilon_{it}\epsilon_{lt}\epsilon_{mt}\epsilon_{nt}|)\leq c_1$ for some constant $c_1$, $E|Y_{i_1j_1,t}Y_{i_2j_2,t}Y_{i_3j_3,t}Y_{i_4j_4,t}|\leq c_2$ for some constant $c_2$ and $\epsilon_t$ is independent of $\mathbb{Y}_{t-1}$.
Thus
\begin{equation*}
    \begin{split}
      &E(|\Tilde{\epsilon}_{it_1}\Tilde{\epsilon}_{lt_2}\Tilde{\epsilon}_{mt_3}\Tilde{\epsilon}_{nt_4}|)\\
      &=E(|(\epsilon_{it_1}+\sum\limits_{k=1}^{p}c_{ik}Y_{ik,(t_1-1)})(\epsilon_{lt_2}+\sum\limits_{k=1}^{p}c_{lk}Y_{lk,(t_2-1)})(\epsilon_{mt_3}+\sum\limits_{k=1}^{p}c_{mk}Y_{mk,(t_3-1)})(\epsilon_{nt_4}+\sum\limits_{k=1}^{p}c_{nk}Y_{nk,(t_4-1)}|)\leq c_4  
    \end{split}
\end{equation*}
for some constant $c_4$.
Denote by $(I-\boldsymbol{G})^{-1}_{ij}$ the $ij$-th element of $(I-\boldsymbol{G})^{-1}$. We then obtain
\begin{equation}\nonumber
\begin{split}
&E(|X_{i,t_1}X_{l,t_2}X_{m,t_3}X_{n,t_4}|)\\ 
&=\sum\limits_{k_1= 0}^{\infty}\sum\limits_{j_1= 1}^{Nq}\sum\limits_{k_2= 0}^{\infty}\sum\limits_{j_2= 1}^{Nq}\sum\limits_{k_3= 0}^{\infty}\sum\limits_{j_3= 1}^{Nq}\sum\limits_{k_4= 0}^{\infty}\sum\limits_{j_4= 1}^{Nq}E(|g^{k_1}_{ij_1}g^{k_2}_{lj_2}g^{k_3}_{mj_3}g^{k_4}_{nj_4}\Tilde{\epsilon}_{j_1,(t_1-k_1)}\Tilde{\epsilon}_{j_2,(t_2-k_2)}\Tilde{\epsilon}_{j_3,(t_3-k_3)}\Tilde{\epsilon}_{j_4,(t_4-k_4)}|)\\
&\overset{(a)}{\leq} c_4 \sum\limits_{k_1= 0}^{\infty}\sum\limits_{j_1= 1}^{Nq}\sum\limits_{k_2= 0}^{\infty}\sum\limits_{j_2= 1}^{Nq}\sum\limits_{k_3= 0}^{\infty}\sum\limits_{j_3= 1}^{Nq}\sum\limits_{k_4= 0}^{\infty}\sum\limits_{j_4= 1}^{Nq}|g^{k_1}_{ij_1}g^{k_2}_{lj_2}g^{k_3}_{mj_3}g^{k_4}_{nj_4}|\\
&=c_4\sum\limits_{k_1= 0}^{\infty}\sum\limits_{j_1= 1}^{Nq}|g^{k_1}_{ij_1}|\sum\limits_{k_2= 0}^{\infty}\sum\limits_{j_2= 1}^{Nq}|g^{k_2}_{lj_2}|\sum\limits_{k_3= 0}^{\infty}\sum\limits_{j_3= 1}^{Nq}|g^{k_3}_{mj_3}|\sum\limits_{k_4= 0}^{\infty}\sum\limits_{j_4= 1}^{Nq}|g^{k_4}_{nj_4}|\\
&=c_4\sum\limits_{j_1= 1}^{Nq}|(I-\boldsymbol{G})^{-1}_{ij_1}|\sum\limits_{j_2= 1}^{Nq}|(I-\boldsymbol{G})^{-1}_{lj_2}|\sum\limits_{j_3= 1}^{Nq}|(I-\boldsymbol{G})^{-1}_{mj_3}|\sum\limits_{j_4= 1}^{Nq}|(I-\boldsymbol{G})^{-1}_{nj_4}|\\
&\overset{(b)}{\leq}c_3\\
&<\infty,
\end{split}
\end{equation}
where (a) follows from the Dominated Convergence Theorem. For case (i), (b) is obvious for fixed $N$.

For case (ii), since $\epsilon_t$ and $\mathbb{Y}_t$ are  sub-Weibull, $\Tilde{\epsilon}_{t}=\epsilon_{t}+
\sum\limits_{k=1}^{p}C_k\mathbb{Y}_{k,(t-1)}$ is also sub-Weibull. Consider the stationary NAR$(q_1,q_2)$ model \eqref{var1}, i.e. $\boldsymbol{X}_t=\boldsymbol{G}\boldsymbol{X}_{t-1}+\mathcal{E}_t$, and assume $\Tilde{\epsilon}_{t}$ is sub-Weibull($\lambda$).
Following \citep{wong2020lasso}, since $\rho(\boldsymbol{G})<1$ for stationary NAR$(q_1,q_2)$, by definition of the spectral radius,
\[\mathop{lim}\limits_{m\rightarrow\infty}||\boldsymbol{G}^m||^{1/m}=\rho(\boldsymbol{G})<1,\]
i.e., there exists a positive integer $k<\infty$ such that $||\boldsymbol{G}||^k<1$. Therefore,
\[||\boldsymbol{X}_t||_{\psi_\lambda}\leq ||\boldsymbol{G}^k||||\boldsymbol{X}_{t-k}||_{\psi_\lambda}+\sum\limits_{i=1}^{k}||\boldsymbol{G}^{k-i}||||\mathcal{E}_{t-k+i}||_{\psi_\lambda},\]
where $||\boldsymbol{X}_t||_{\psi_\lambda}:=\mathop{sup}_{p\geq 1}(E|\boldsymbol{X}_t|^p)^{1/p}p^{-1/\lambda}$ is a pointwise supremum of norms. By stationarity,
\[||\boldsymbol{X}_t||_{\psi_\lambda}\leq\frac{||\Tilde{\epsilon}_t||_{\psi_\lambda}}{1-||\boldsymbol{G}^k||}(\sum\limits_{i=1}^{k}||\boldsymbol{G}^{k-i}||)<\infty.\]
Thus, \[E|X_{i,t}|^4\leq 4^{4/\lambda}||\boldsymbol{X}_t||_{\psi_\lambda}^4<\infty\]
due to the properties of the sub-Weibull distribution.
\end{proof}

\subsection{LEMMA \ref{CLT} [Central Limit Theorem for martingale differences]}
In Lemmas \ref{L:CLTOLS}, \ref{L:CLTDOLS}, \ref{L:CLTGLS} and \ref{L:CLTDGLS}, we show that the martingale difference sequences satisfy the central limit theorem for martingale differences (see Theorem 5.3.4 of \citep{fuller2009introduction}).

For completeness of the exposition, we provide the central limit theorem for martingale differences next:

\begin{CLT}
\label{CLT}
Let $\{Z_{tn}: 1 \leq t \leq n, n \geq 1\}$ denote a triangular array of random variables defined on the probability space $(\Omega, \mathcal{A}, P)$, and let $\{\mathcal{A}_{tn}: 0 \leq t \leq n, n \geq 1\}$ be any triangular array of sub-sigma-fields of $\mathcal{A}$ such that for each $n$ and $1 \leq t \leq n$, $Z_{tn}$ is $\mathcal{A}_{tn}$-measurable and $\mathcal{A}_{t-1,n}$ is contained in $\mathcal{A}_{tn}$. For $1 \leq k \leq n, 1 \leq j \leq n$, and
$n \geq 1$, let
\[S_{kn}=\sum\limits_{t=1}^kZ_{tn},\]
\[\delta_{tn}^2=E(Z_{tn}^2|\mathcal{A}_{t-1,n}),\]
\[V_{jn}^2=\sum\limits_{t=1}^j\delta_{tn}^2,\]
and
\[s_{nn}^2=E(V^2_{nn}).\]
Assume
\begin{itemize}
    \item 
    $E(Z_{tn}|\mathcal{A}_{t-1,n})=0$ a.s. for $1\leq t\leq n$,
    \item
    $V_{nn}^2s_{nn}^{-2}\rightarrow_p1,$
    \item
    $\mathop{lim}\limits_{n\rightarrow\infty}s_{nn}^{-2}\sum\limits_{j=1}^nE(Z_{jn}^2I(|Z_{jn}|\geq \epsilon s_{nn}))=0$ for all $\epsilon>0,$
\end{itemize}
where $I(A)$ denotes the indicator function of a set $A$. Then, as $n\rightarrow\infty$,
\[s_{nn}^{-1}S_{nn}\rightarrow_dN(0,1).\]
\end{CLT}
\subsection{LEMMA \ref{L:CLTOLS}}\label{a3}

\begin{l1}
\label{L:CLTOLS}
Suppose Assumptions \ref{A1}-\ref{A3} hold. For a stationary process $\mathbb{X}_t$ defined by the NAR$(q_1,q_2)$ model \eqref{compact}, i.e. $\mathbb{X}_{t}=\mathbb{Z}_{t-1}\beta+ \epsilon_{t}$, with finite network size $N$, the following hold:
\begin{equation}
\label{CLT:OLS}
\frac{1}{\sqrt{T}}\sum\limits_{t=1}^{T}\mathbb{Z}_{t-1}^T\epsilon_{t}\rightarrow_{d}N(0, Q )
\end{equation}
where $Q:=
E(\mathbb{Z}_{t}^T\Sigma_{\epsilon}\mathbb{Z}_{t}).$
\end{l1}

\begin{proof}[Proof of LEMMA \ref{L:CLTOLS}]
		
Let $\eta_{(2Nq+Np)\times 1}$ be a column vector of arbitrary real numbers such that $\eta^T \eta \not=0$. Let
\[\frac{1}{\sqrt{T}}\eta^T\sum\limits_{t=1}^{T}\mathbb{Z}_{t-1}^T\epsilon_{t}=\sum\limits_{t=1}^{T}Z_{tT}=S_{TT}\]
where $Z_{tT}=\frac{1}{\sqrt{T}}\eta^T\mathbb{Z}_{t-1}^T\epsilon_{t}$. Since $E(\epsilon_t|\mathcal{A}_{t-1})=0$ where $\mathcal{A}_{t-1}$ is the sigma-field generated by $\{\epsilon_{j}:j\leq t-1\}$, we have $E(Z_{tT}|\mathcal{A}_{t-1})=0$, so condition 1 of Lemma \ref{CLT} is satisfied.

Define
\[\delta_{tT}^2:=E(Z_{tT}^2|\mathcal{A}_{t-1})=E(\frac{1}{T}\eta^T\mathbb{Z}_{t-1}^T\epsilon_{t}\epsilon_{t}^T\mathbb{Z}_{t-1}\eta|\mathcal{A}_{t-1})=\frac{1}{T}\eta^T\mathbb{Z}_{t-1}^T\Sigma_{\epsilon}\mathbb{Z}_{t-1}\eta\]
and 
\[V_{TT}^2:=\frac{1}{T}\sum\limits_{t=1}^T\eta^T\mathbb{Z}_{t-1}^T\Sigma_{\epsilon}\mathbb{Z}_{t-1}\eta.\]
Hence, we have
\[V_{TT}^2\rightarrow_{p}\eta^TQ \eta\] by ergodicity of $\mathbb{X}_t$.
Further,
\[s_{TT}^2:=E(V_{TT}^2)=\sum\limits_{t=1}^T\frac{1}{T}\eta^TE(\mathbb{Z}_{t-1}^T\Sigma_{\epsilon}\mathbb{Z}_{t-1})\eta=\eta^TQ \eta,\]
thus
\[V_{TT}^2s_{TT}^{-2}\rightarrow_{p}1,\]
so condition 2 of Lemma \ref{CLT} is satisfied.
		
For condition 3 of Lemma \ref{CLT},
\begin{equation}\nonumber
\begin{split}
&s_{TT}^{-2}\sum\limits_{t=1}^TE(Z_{tT}^2I(|Z_{tT}|\geq\epsilon s_{TT}))\\&\leq s_{TT}^{-2}\sum\limits_{t=1}^TE((\epsilon s_{TT})^{-2}Z_{tT}^{4}I(|Z_{tT}|\geq\epsilon s_{TT}))\\&\leq s_{TT}^{-4}\epsilon^{-2}\sum\limits_{t=1}^TE((\frac{1}{\sqrt{T}}\eta^T\mathbb{Z}_{t-1}^T\epsilon_{t})^{4})\\
&= s_{TT}^{-4}\epsilon^{-2}\sum\limits_{t=1}^TE((\frac{1}{\sqrt{T}}\eta^T\mathbb{Z}_{t-1}^T\epsilon_{t})^{4}).\\
\end{split}
\end{equation}
Define $a_i:=\frac{1}{\sqrt{T}}\sum\limits_{k=1}^{q}(\eta_{2(k-1)N+i}X_{i,(t-k)}+\eta_{(2(k-1)N+N+i)}w_{i}^{T}\mathbb{X}_{t-k})+\frac{1}{\sqrt{T}}\sum\limits_{j=1}^p\eta_{2Nq+(j-1)N+i}Y_{ij,(t-1)}$, then for $E((\eta^T\frac{1}{\sqrt{T}}\mathbb{Z}_{t-1}^T\epsilon_{t})^{4})$,  we have:
\begin{equation}\nonumber
\begin{split}
&E((\eta^T\frac{1}{\sqrt{T}}\mathbb{Z}_{t-1}^T\epsilon_{t})^{4})\\
&=E((\frac{1}{\sqrt{T}}\sum\limits_{i=1}^N\sum\limits_{k=1}^{q}(\eta_{(2(k-1)N+i)}X_{i,(t-k)}+\eta_{(2(k-1)N+N+i)}w_{i}^{T}\mathbb{X}_{t-k})\epsilon_{it}\\
&+\frac{1}{\sqrt{T}}\sum\limits_{i=1}^N\sum\limits_{j=1}^p\eta_{2Nq+(j-1)N+i}Y_{ij,(t-1)}\epsilon_{it})^{4})\\
&= E((\sum\limits_{i=1}^N a_i\epsilon_{it})^{4})\\
&=E(E((\sum\limits_{i=1}^N a_i\epsilon_{it})^{4}|\mathcal{A}_{t-1}))\\
&=E(E(\sum\limits_{i,j,k,m}a_ia_ja_ka_m\epsilon_{it}\epsilon_{jt}\epsilon_{kt}\epsilon_{mt}|\mathcal{A}_{t-1}))\\
&=E(\sum\limits_{i,j,k,m}a_ia_ja_ka_mE(\epsilon_{it}\epsilon_{jt}\epsilon_{kt}\epsilon_{mt}|\mathcal{A}_{t-1}))\\
&\leq E(\sum\limits_{i,j,k,m}|a_ia_ja_ka_m|E(|\epsilon_{it}\epsilon_{jt}\epsilon_{kt}\epsilon_{mt}||\mathcal{A}_{t-1}))\\
&\overset{(a)}{\leq}  c_1E((\sum\limits_{i=1}^N|a_i|)^4)
\end{split}
\end{equation}
where (a) follows from $E(|\epsilon_{it}\epsilon_{jt}\epsilon_{kt}\epsilon_{mt}||\mathcal{A}_{t-1})\leq c_1<\infty$.

Next, we need to find an upper bound for
\begin{equation*}
    \begin{split}
        E((\sum\limits_{i=1}^N|a_i|)^4)&=E((\sum\limits_{i=1}^N|\frac{1}{\sqrt{T}}\sum\limits_{k=1}^{q}(\eta_{(2(k-1)N+i)}X_{i,(t-k)}+\eta_{(2(k-1)N+N+i)}w_{i}^{T}\mathbb{X}_{t-k})\\
        &+\frac{1}{\sqrt{T}}\sum\limits_{j=1}^p\eta_{2Nq+(j-1)N+i}Y_{ij,(t-1)})^4).
    \end{split}
\end{equation*}

Denote by $X_{2(k-1)N+i}'=X_{i,(t-k)}$ and $X_{2(k-1)N+N+i}'=w_i^T\mathbb{X}_{t-k}$ for $i=1,\cdots,N$. By Lemma \ref{L:BDD}, $E(|X_{i,t_1}X_{l,t_2}X_{m,t_3}X_{n,t_4}|)\leq c_3<\infty$ for any $1 \leq i,l,m,n\leq N$. Since row sum of W is 1, $X_{2(k-1)N+N+i}'$ can be seen as weighed average of $\mathbb{X}_{t-k}$, $E(|X_i'X_l'X_m'X_n'|)\leq c_5<\infty$ for any $1 \leq i,l,m,n\leq 2Nq$.
\begin{equation}\nonumber
\begin{split}
&E((\sum\limits_{i=1}^N|a_i|)^4)\\
&\leq 8E((\frac{1}{\sqrt{T}}\sum\limits_{i=1}^{2Nq}|\eta_iX_i'|)^4)+8E((\frac{1}{\sqrt{T}}\sum\limits_{i=1}^N|\sum\limits_{j=1}^p\eta_{2Nq+(j-1)N+i}Y_{ij,(t-1)}|)^{4})\\
&\overset{(b)}{\leq} \frac{8}{T^2}\sum\limits_{1 \leq i,l,m,n\leq 2Nq}|\eta_i\eta_l\eta_m\eta_n|E(|X_i'X_l'X_m'X_n'|)+\frac{8c_2}{T^2}(\sum\limits_{i=1}^N\sum\limits_{j=1}^p|\eta_{2Nq+(j-1)N+i}|)^4\\
&\overset{(c)}{\leq} \frac{8c_5}{T^2}\sum\limits_{1 \leq i,l,m,n\leq 2Nq}|\eta_i\eta_l\eta_m\eta_n|+\frac{8c_2}{T^2}(\sum\limits_{i=1}^N\sum\limits_{j=1}^p|\eta_{2Nq+(j-1)N+i}|)^4\\
&=O(\frac{1}{T^2})
\end{split}
\end{equation}
where (b) follows from \begin{multline}
E|Y_{i_1j_1,t}Y_{i_2j_2,t}Y_{i_3j_3,t}Y_{i_4j_4,t}|\leq c_2, \\ \text{for}\ i_1,i_2,i_3,i_4=1,\cdots,N,\ j_1,j_2,j_3,j_4=1,\cdots,p \text{ and all}\ t;
\end{multline} and (c) follows from $E(|X_i'X_l'X_m'X_n'|)\leq c_5<\infty$ for any $1 \leq i,l,m,n\leq 2Nq$.

So $s_{TT}^{-2}\sum\limits_{t=1}^TE(Z_{tT}^2I(|Z_{tT}|\geq\epsilon s_{TT}))=O(\frac{1}{T})\rightarrow0$.
		
Thus, we obtain   \[\frac{1}{\sqrt{T}}\sum\limits_{t=1}^{T}\mathbb{Z}_{t-1}^T\epsilon_{t}\rightarrow_{d}N(0, Q ).\]
\end{proof}	

\subsection{Proof of PROPOSITION \ref{P:OLS}}\label{a4}

\begin{proof}[Proof of PROPOSITION \ref{P:OLS}]
Under the conditions of proposition \ref{P:OLS}, we have:
\[\frac{1}{T}\sum\limits_{t=1}^{T}\mathbb{Z}_{t-1}^T\mathbb{Z}_{t-1}\rightarrow_{p}P\] by Lemma \ref{L:LLNOLS}, and
\[\frac{1}{\sqrt{T}}\sum\limits_{t=1}^{T}\mathbb{Z}_{t-1}^T\epsilon_{t}\rightarrow_{d}N(0, Q )\]
by Lemma \ref{L:CLTOLS}. By Equation \ref{beta-ols}, $\hat{\beta}_{OLS}$ can be written as $\hat{\beta}_{OLS}=\beta+(\sum\limits_{t=1}^{T}\mathbb{Z}_{t-1}^T\mathbb{Z}_{t-1})^{-1}\sum\limits_{t=1}^{T}\mathbb{Z}_{t-1}^T\epsilon_t$, so we have:
\begin{equation}\nonumber
\begin{split}
\sqrt{T}(\hat{\beta}-\beta)=&\sqrt{T}((\sum\limits_{t=1}^{T}\mathbb{Z}_{t-1}^T\mathbb{Z}_{t-1})^{-1}\sum\limits_{t=1}^{T}\mathbb{Z}_{t-1}^T\mathbb{X}_t-\beta)\\
&=\sqrt{T}((\sum\limits_{t=1}^{T}\mathbb{Z}_{t-1}^T\mathbb{Z}_{t-1})^{-1}\sum\limits_{t=1}^{T}\mathbb{Z}_{t-1}^T(\mathbb{Z}_{t-1}\beta+\epsilon_t)-\beta)\\
&=\sqrt{T}(\sum\limits_{t=1}^{T}\mathbb{Z}_{t-1}^T\mathbb{Z}_{t-1})^{-1}\sum\limits_{t=1}^{T}\mathbb{Z}_{t-1}^T\epsilon_t,
\end{split}
\end{equation}
so	\[\sqrt{T}(\hat{\beta}-\beta)\rightarrow_{d}N(0, P^{-1}QP^{-1})\]
where $P:=
E(\mathbb{Z}_{t}^T\mathbb{Z}_{t})$, $Q:=
E(\mathbb{Z}_{t}^T\Sigma_{\epsilon}\mathbb{Z}_{t})$.
\end{proof}	
\subsection{LEMMA \ref{L:CLTDOLS}}\label{a5}

\begin{l2}
\label{L:CLTDOLS}
Suppose Assumptions \ref{A1}-\ref{A3} hold. Let $\mathbb{X}_t$ be a stationary process generated by the NAR$(q_1,q_2)$ model \eqref{compact}: $\mathbb{X}_{t}=\mathbb{Z}_{t-1}\beta+ \epsilon_{t}$ with growing network size $N$. Define as before $D\in \mathbb{R}^{k\times (2Nq+Np)}$ for any finite $k$. Further, suppose that Assumption \ref{A2} and the additional conditions posited in Proposition \ref{P:DOLS} hold for $D$, as well as $N<T$.
Then, 
\begin{equation}
\label{CLT:DOLS}
\frac{1}{\sqrt{T}}D\sum\limits_{t=1}^{T}\mathbb{Z}_{t-1}^T\epsilon_{t}\rightarrow_{d}N(0, DQD^T),
\end{equation}
where $Q:=
E(\mathbb{Z}_{t}^T\Sigma_{\epsilon}\mathbb{Z}_{t})$.
\end{l2}

\begin{proof}[Proof of LEMMA \ref{L:CLTDOLS}]

Let $\eta_{k\times 1}$ be a column vector of arbitrary real numbers such that $\eta^T \eta \not=0$. Let
\[\frac{1}{\sqrt{T}}\eta^T\sum\limits_{t=1}^{T}D\mathbb{Z}_{t-1}^T\epsilon_{t}=\sum\limits_{t=1}^{T}Z_{tT}=S_{TT}\]
where $Z_{tT}=\frac{1}{\sqrt{T}}\eta^TD\mathbb{Z}_{t-1}^T\epsilon_{t}$. Since $E(\epsilon_t|\mathcal{A}_{t-1})=0$, where $\mathcal{A}_{t-1}$ is the sigma-field generated by $\{\epsilon_{j}:j\leq t-1\}$, we have $E(Z_{tT}|\mathcal{A}_{t-1})=0\ a.s.$, so condition 1 of Lemma \ref{CLT} is satisfied.

Define
\[\delta_{tT}^2:=E(Z_{tT}^2|\mathcal{A}_{t-1})=E(\frac{1}{T}\eta^TD\mathbb{Z}_{t-1}^T\epsilon_{t}\epsilon_{t}^T\mathbb{Z}_{t-1}D^{T}\eta|\mathcal{A}_{t-1})=\frac{1}{T}\eta^TD\mathbb{Z}_{t-1}^T\Sigma_{\epsilon}\mathbb{Z}_{t-1}D^{T}\eta\]
and 
\[V_{TT}^2:=\frac{1}{T}\sum\limits_{t=1}^T\eta^TD\mathbb{Z}_{t-1}^T\Sigma_{\epsilon}\mathbb{Z}_{t-1}D^T\eta.\]
Further, we obtain 
\[V_{TT}^2\rightarrow_{p}\eta^TDQD^T \eta.\] by ergodicity of $\mathbb{X}_t$.
We also have
\[s_{TT}^2:=E(V_{TT}^2)=\frac{1}{T}\sum\limits_{t=1}^T\eta^TDE(\mathbb{Z}_{t-1}^T\Sigma_{\epsilon}\mathbb{Z}_{t-1})D^T\eta=\eta^TDQD^T \eta,\]
thus
\[V_{TT}^2s_{TT}^{-2}\rightarrow_{p}1,\]
so condition 2 of Lemma \ref{CLT} is satisfied.

For condition 3 of Lemma \ref{CLT},
\begin{equation}\nonumber
\begin{split}
&s_{TT}^{-2}\sum\limits_{t=1}^TE(Z_{tT}^2I(|Z_{tT}|\geq\epsilon s_{TT}))\\&\leq s_{TT}^{-2}\sum\limits_{t=1}^TE((\epsilon s_{TT})^{-2}Z_{tT}^{4}I(|Z_{tT}|\geq\epsilon s_{TT}))\\&\leq s_{TT}^{-4}\epsilon^{-2}\sum\limits_{t=1}^TE((\frac{1}{\sqrt{T}}\eta^TD\mathbb{Z}_{t-1}^T\epsilon_{t})^{4})\\
&= s_{TT}^{-4}\epsilon^{-2}\sum\limits_{t=1}^TE((\frac{1}{\sqrt{T}}\eta^TD\mathbb{Z}_{t-1}^T\epsilon_{t})^{4})\\
\end{split}
\end{equation}

Define 
\begin{equation}\nonumber
\begin{split}
a_j&:=\frac{1}{\sqrt{T}}\sum\limits_{i=1}^k\sum\limits_{l=1}^q(\eta_id_{i,(2(l-1)N+j)}X_{j,(t-l)}+\eta_{i}d_{i,(2(l-1)N+N+j)}w_{j}^{T}\mathbb{X}_{t-l})\\
&+\frac{1}{\sqrt{T}}\sum\limits_{i=1}^k\sum\limits_{l=1}^p\eta_{i}d_{i,(2Nq+j+(l-1)N)}Y_{jl,(t-1)}.
\end{split}
\end{equation} 
For $E((\frac{1}{\sqrt{T}}\eta^TD\mathbb{Z}_{t-1}^T\epsilon_{t})^{4})$, we have:
\begin{equation}\nonumber
\begin{split}
&E((\frac{1}{\sqrt{T}}\eta^TD\mathbb{Z}_{t-1}^T\epsilon_{t})^{4})\\
&=E((\frac{1}{\sqrt{T}}\sum\limits_{i=1}^k\sum\limits_{j=1}^N\sum\limits_{l=1}^q(\eta_id_{i,(2(l-1)N+j)}X_{j,(t-l)}+\eta_{i}d_{i,(2(l-1)N+N+j)}w_{j}^{T}\mathbb{X}_{t-l})\epsilon_{jt}\\
&+\frac{1}{\sqrt{T}}\sum\limits_{i=1}^k\sum\limits_{l=1}^p\sum\limits_{j=1}^N\eta_{i}d_{i,(2Nq+j+(l-1)N)}Y_{jl,(t-1)}\epsilon_{jt})^{4})\\
&= E((\sum\limits_{j=1}^N a_j\epsilon_{jt})^{4})\\
&=E(E((\sum\limits_{j=1}^N a_i\epsilon_{jt})^{4}|\mathcal{A}_{t-1}))\\
&=E(E(\sum\limits_{i,j,k,m}a_ia_ja_ka_m\epsilon_{it}\epsilon_{jt}\epsilon_{kt}\epsilon_{mt}|\mathcal{A}_{t-1}))\\
&=E(\sum\limits_{i,j,k,m}a_ia_ja_ka_mE(\epsilon_{it}\epsilon_{jt}\epsilon_{kt}\epsilon_{mt}|\mathcal{A}_{t-1}))\\
&\leq E(\sum\limits_{i,j,k,m}|a_ia_ja_ka_m|E(|\epsilon_{it}\epsilon_{jt}\epsilon_{kt}\epsilon_{mt}||\mathcal{A}_{t-1}))\\
&\overset{(a)}{\leq}  c_1E((\sum\limits_{j=1}^N|a_j|)^4)
\end{split}
\end{equation}
where (a) follows from $E(|\epsilon_{it}\epsilon_{jt}\epsilon_{kt}\epsilon_{mt}||\mathcal{A}_{t-1})\leq c_1<\infty$.
Next, we need to find an upper bound for 
\begin{equation*}
\begin{split}
&E((\sum\limits_{j=1}^N|a_j|)^4)\\
&=E((\sum\limits_{j=1}^N|\frac{1}{\sqrt{T}}\sum\limits_{i=1}^k\sum\limits_{l=1}^q(\eta_id_{i,(2(l-1)N+j)}X_{j,(t-l)}+\eta_{i}d_{i,(2(l-1)N+N+j)}w_{j}^{T}\mathbb{X}_{t-l})+\\
&\frac{1}{\sqrt{T}}\sum\limits_{i=1}^k\sum\limits_{l=1}^p\eta_{i}d_{i,(2Nq+j+(l-1)N)}Y_{jl,(t-1)}|)^4).    
\end{split}
\end{equation*}

Denote by $X_{2(l-1)N+j}'=X_{j,(t-l)}$ and $X_{2(l-1)N+N+j}'=w_j^T\mathbb{X}_{t-l}$ for $i=1,\cdots,N$. By Lemma \ref{L:BDD}, $E(|X_{i,t_1}X_{l,t_2}X_{m,t_3}X_{n,t_4}|)\leq c_3<\infty$ for any $1 \leq i,l,m,n\leq 2Nq$. Since row sum of W is 1, $X_{2(k-1)N+N+i}'$ can be seen as weighed average of $\mathbb{X}_{t-k}$, $E(|X_i'X_l'X_m'X_n'|)\leq c_5<\infty$ for any $1 \leq i,l,m,n\leq 2Nq$.
\begin{equation}\nonumber
\begin{split}
&E((\sum\limits_{j=1}^N|a_j|)^4)\\
&\leq 8E((\frac{1}{\sqrt{T}}\sum\limits_{j=1}^{2Nq}|\sum\limits_{i=1}^k\eta_id_{i,j}X_j'|)^4)+8E((\frac{1}{\sqrt{T}}\sum\limits_{j=1}^N|\sum\limits_{i=1}^k\sum\limits_{l=1}^p\eta_{i}d_{i,(2Nq+j+(l-1)N)}Y_{jl,(t-1)}|)^{4})\\
&\leq 8c_5\frac{1}{T^2}(\sum\limits_{i=1}^k\sum\limits_{j=1}^{2Nq}\eta_id_{i,j})^4+8E((\frac{1}{\sqrt{T}}\sum\limits_{j=1}^N|\sum\limits_{i=1}^k\sum\limits_{l=1}^p\eta_{i}d_{i,(2Nq+j+(l-1)N)}Y_{jl,(t-1)}|)^{4}\\
&\leq 8c_5\frac{1}{T^2}(\sum\limits_{i=1}^k\sum\limits_{j=1}^{2Nq}\eta_id_{i,j})^4+8c_2\frac{1}{T^2}(\sum\limits_{j=1}^N\sum\limits_{i=1}^k\sum\limits_{l=1}^p\eta_{i}d_{i,(2Nq+j+(l-1)N)})^4\\
&\leq 8c_5\frac{1}{T^2}(\sum\limits_{i=1}^k\eta_i(\sum\limits_{j=1}^N\sum\limits_{j=1}^{2Nq}d_{i,j}))^4+8c_2\frac{1}{T^2}(\sum\limits_{i=1}^k\eta_{i}(\sum\limits_{l=1}^p\sum\limits_{j=1}^Nd_{i,(2Nq+j+(l-1)N)}))^4\\
&=O(\frac{1}{T^2}),
\end{split}
\end{equation}
since the row sum of $D$ is bounded.

Therefore, $s_{TT}^{-2}\sum\limits_{t=1}^TE(Z_{tT}^2I(|Z_{tT}|\geq\epsilon s_{TT}))=O(\frac{1}{T})\rightarrow0$.

Thus, we obtain   \[D\frac{1}{\sqrt{T}}\sum\limits_{t=1}^{T}\mathbb{Z}_{t-1}^T\epsilon_{t}\rightarrow_{d}N(0, DQD^T ).\]
\end{proof}

\subsection{Proof of PROPOSITION \ref{P:DOLS}}\label{a6}
\begin{proof}[Proof of PROPOSITION \ref{P:DOLS}]
 
By Equation \ref{beta-ols}, $\hat{\beta}_{OLS}$ can be written as $\hat{\beta}_{OLS}=\beta+(\sum\limits_{t=1}^{T}\mathbb{Z}_{t-1}^T\mathbb{Z}_{t-1})^{-1}\sum\limits_{t=1}^{T}\mathbb{Z}_{t-1}^T\epsilon_t$, so we have:

\begin{equation*}
\begin{split}
 \frac{1}{\sqrt{T}}D(\sum_{t=1}^{T}\mathbb{Z}_{t-1}^T\mathbb{Z}_{t-1})(\hat{\beta}_{OLS}-\beta)&=\frac{1}{\sqrt{T}}D(\sum_{t=1}^{T}\mathbb{Z}_{t-1}^T\mathbb{Z}_{t-1})((\sum\limits_{t=1}^{T}\mathbb{Z}_{t-1}^T\mathbb{Z}_{t-1})^{-1}\sum\limits_{t=1}^{T}\mathbb{Z}_{t-1}^T\mathbb{X}_t-\beta)\\
 &=\frac{1}{\sqrt{T}}D(\sum_{t=1}^{T}\mathbb{Z}_{t-1}^T\mathbb{Z}_{t-1})((\sum\limits_{t=1}^{T}\mathbb{Z}_{t-1}^T\mathbb{Z}_{t-1})^{-1}\sum\limits_{t=1}^{T}\mathbb{Z}_{t-1}^T(\mathbb{Z}_{t-1}\beta+\epsilon_t)-\beta)\\
 &=\frac{1}{\sqrt{T}}D(\sum_{t=1}^{T}\mathbb{Z}_{t-1}^T\epsilon_{t}).
 \end{split}
 \end{equation*}
 By Lemma \ref{L:CLTDOLS}, 
\[ \frac{1}{\sqrt{T}}D\sum\limits_{t=1}^{T}\mathbb{Z}_{t-1}^T\epsilon_{t}\rightarrow_{d}N(0, DQD^T ),\] so
 \[\frac{1}{\sqrt{T}}D(\sum_{t=1}^{T}\mathbb{Z}_{t-1}^T\mathbb{Z}_{t-1})(\hat{\beta}_{OLS}-\beta)\rightarrow_{d}N(0, DQD^T ).\]
\end{proof}

\subsection{Proof of PROPOSITION \ref{P:DOLSR}}\label{a9}
\begin{proof}[Proof of PROPOSITION \ref{P:DOLSR}]
By the definition of the ridge estimator in Equation \ref{beta-ols-r}, we get
\begin{equation*}
\begin{aligned}
\hat{\beta}_{ridge}&=argmin\frac{1}{T}\sum_{t=1}^{T}(\mathbb{X}_{t}-\mathbb{Z}_{t-1}\beta)^2+||M\beta||^2\\
&=(\sum_{t=1}^{T}\mathbb{Z}_{t-1}^T\mathbb{Z}_{t-1}+T M)^{-1}\sum_{t=1}^{T}\mathbb{Z}_{t-1}^T\mathbb{X}_t\\
&=(\sum_{t=1}^{T}\mathbb{Z}_{t-1}^T\mathbb{Z}_{t-1}+T  M)^{-1}\sum_{t=1}^{T}\mathbb{Z}_{t-1}^T\mathbb{X}_t\\
&=(\sum_{t=1}^{T}\mathbb{Z}_{t-1}^T\mathbb{Z}_{t-1}+T  M)^{-1}\sum_{t=1}^{T}(\mathbb{Z}_{t-1}^T\mathbb{Z}_{t-1}\beta+\mathbb{Z}_{t-1}^T\epsilon_t)\\
&=(\sum_{t=1}^{T}\mathbb{Z}_{t-1}^T\mathbb{Z}_{t-1}+T  M)^{-1}(\sum_{t=1}^{T}\mathbb{Z}_{t-1}^T\mathbb{Z}_{t-1}+T  M)\beta+(\sum_{t=1}^{T}\mathbb{Z}_{t-1}^T\mathbb{Z}_{t-1}+T  M)^{-1}\sum_{t=1}^{T}\mathbb{Z}_{t-1}^T\epsilon_t\\
&-(\sum_{t=1}^{T}\mathbb{Z}_{t-1}^T\mathbb{Z}_{t-1}+T  M)^{-1}T M\beta\\
&=\beta+(\sum_{t=1}^{T}\mathbb{Z}_{t-1}^T\mathbb{Z}_{t-1}+T  M)^{-1}\sum_{t=1}^{T}\mathbb{Z}_{t-1}^T\epsilon_t-(\sum_{t=1}^{T}\mathbb{Z}_{t-1}^T\mathbb{Z}_{t-1}+T  M)^{-1}T M\beta
\end{aligned}
\end{equation*} 

Since,
\[\frac{1}{T}(\sum_{t=1}^{T}\mathbb{Z}_{t-1}^T\mathbb{Z}_{t-1}+T  M)\rightarrow_{p}P+M,\]
we thus obtain:
\[\frac{1}{\sqrt{T}}D(\sum_{t=1}^{T}\mathbb{Z}_{t-1}^T\mathbb{Z}_{t-1}+TM)(\hat{\beta}_{ridge}-\beta) = \frac{1}{\sqrt{T}}D(\sum_{t=1}^{T}\mathbb{Z}_{t-1}^T\epsilon_t-TM\beta).\]

If $\lambda_1=o(\frac{1}{\sqrt T})$, $\lambda_2=o(\frac{1}{\sqrt T})$ and $\lambda_3=o(\frac{1}{\sqrt{T}})$,
\[\frac{1}{\sqrt{T}}D(\sum_{t=1}^{T}\mathbb{Z}_{t-1}^T\mathbb{Z}_{t-1}+TM)(\hat{\beta}_{ridge}-\beta)\rightarrow_{d}N(0, DQD^T ).\]
\end{proof}

\section{GLS results}

\subsection{LEMMA \ref{L:LLNGLS}}\label{b1}

\begin{l9}
\label{L:LLNGLS}
Under the conditions of proposition \ref{P:GLS},
\begin{equation}
\label{LLN:GLS}
\frac{1}{T}(\sum_{t=1}^{T}\mathbb{Z}_{t-1}^T\Sigma_{\epsilon}^{-1}\mathbb{Z}_{t-1})\rightarrow_{p}
E(\mathbb{Z}_{t}^T\Sigma_{\epsilon}^{-1}\mathbb{Z}_{t}).
\end{equation}
\end{l9}

\begin{proof}[Proof of LEMMA \ref{L:LLNGLS}]
The proof parallels that of LEMMA \ref{L:LLNOLS} and hence is omitted.
\end{proof}

\subsection{LEMMA \ref{L:CLTGLS}}\label{b2}

\begin{l3}
\label{L:CLTGLS}
Suppose Assumptions \ref{A1}-\ref{A3} hold. For a stationary process $\mathbb{X}_t$ defined by the NAR$(q_1,q_2)$ model \eqref{compact}, i.e. $\mathbb{X}_{t}=\mathbb{Z}_{t-1}\beta+ \epsilon_{t}$ with finite network size $N$, the following hold:
\begin{equation}
\label{CLT:GLS}
\frac{1}{\sqrt{T}}\sum\limits_{t=1}^{T}\mathbb{Z}_{t-1}^T\Sigma_{\epsilon}^{-1}\epsilon_{t}\rightarrow_{d}N(0, Q ),
\end{equation}
where $Q:=
E(\mathbb{Z}_{t}^T\Sigma_{\epsilon}^{-1}\mathbb{Z}_{t})$.
\end{l3}

\begin{proof}[Proof of LEMMA \ref{L:CLTGLS}]

Let $\eta_{(2Nq+Np)\times 1}$ be a column vector of arbitrary real numbers such that $\eta^T \eta \not=0$. Let
\[\frac{1}{\sqrt{T}}\eta^T\sum\limits_{t=1}^{T}\mathbb{Z}_{t-1}^T\Sigma_{\epsilon}^{-1}\epsilon_{t}=\sum\limits_{t=1}^{T}Z_{tT}=S_{TT},\]
where $Z_{tT}=\frac{1}{\sqrt{T}}\eta^T\mathbb{Z}_{t-1}^T\Sigma_{\epsilon}^{-1}\epsilon_{t}$. Since $E(\epsilon_t|\mathcal{A}_{t-1})=0$ where $\mathcal{A}_{t-1}$ is the sigma-field generated by $\{\epsilon_{j}:j\leq t-1\}$, we have $E(Z_{tT}|\mathcal{A}_{t-1})=0\ a.s.$, so condition 1 of Lemma \ref{CLT} is satisfied.

Define
\[\delta_{tT}^2:=E(Z_{tT}^2|\mathcal{A}_{t-1})=E(\frac{1}{T}\eta^T\mathbb{Z}_{t-1}^T\Sigma_{\epsilon}^{-1}\epsilon_{t}\epsilon_{t}^T\Sigma_{\epsilon}^{-1}\mathbb{Z}_{t-1}\eta|\mathcal{A}_{t-1})=\frac{1}{T}\eta^T\mathbb{Z}_{t-1}^T\Sigma_{\epsilon}^{-1}\mathbb{Z}_{t-1}\eta\]
and 
\[V_{TT}^2:=\frac{1}{T}\sum\limits_{t=1}^T\eta^T\mathbb{Z}_{t-1}^T\Sigma_{\epsilon}^{-1}\mathbb{Z}_{t-1}\eta.\]
Hence, we have
\[V_{TT}^2\rightarrow_{p}\eta^TQ \eta\] by ergodicity of $\mathbb{X}_t$.
Further,
\[s_{TT}^2:=E(V_{TT}^2)=\frac{1}{T}\sum\limits_{t=1}^T\eta^TE(\mathbb{Z}_{t-1}^T\Sigma_{\epsilon}^{-1}\mathbb{Z}_{t-1})\eta=\eta^TQ \eta,\]
thus
\[V_{TT}^2s_{TT}^{-2}\rightarrow_{p}1,\]
so that condition 2 of Lemma \ref{CLT} is satisfied.

For condition 3 of Lemma \ref{CLT},
\begin{equation}\nonumber
\begin{split}
&s_{TT}^{-2}\sum\limits_{t=1}^TE(Z_{tT}^2I(|Z_{tT}|\geq\epsilon s_{TT}))\\&\leq s_{TT}^{-2}\sum\limits_{t=1}^TE((\epsilon s_{TT})^{-2}Z_{tT}^{4}I(|Z_{tT}|\geq\epsilon s_{TT}))\\&\leq s_{TT}^{-4}\epsilon^{-2}\sum\limits_{t=1}^TE(|\frac{1}{\sqrt{T}}\eta^T\mathbb{Z}_{t-1}^T\Sigma_{\epsilon}^{-1}\epsilon_{t}|^4)\\
&\leq s_{TT}^{-4}\epsilon^{-2}\sum\limits_{t=1}^TE((\frac{1}{\sqrt{T}}\eta^T\mathbb{Z}_{t-1}^T\Sigma_{\epsilon}^{-1}\epsilon_{t})^4).\\
\end{split}
\end{equation}
Define 
\begin{equation}\nonumber
\begin{split}
a_i&:=\frac{1}{\sqrt{T}}\sum\limits_{j=1}^N\sum\limits_{k=1}^q(\eta_{2(k-1)N+j} X_{j,(t-k)}\alpha_{ij}+\eta_{(2(k-1)N+N+j)}w_{j}^{T}\mathbb{X}_{t-k}\alpha_{ij})\\
&+\frac{1}{\sqrt{T}}\sum\limits_{j=1}^p\sum\limits_{k=1}^N\eta_{2Nq+(j-1)N+k} Y_{kj,(t-1)}\alpha_{ki},
\end{split}
\end{equation}
where $\alpha_{ik}:=(\Sigma_{\epsilon}^{-1})_{ik}$,
then for $E((\frac{1}{\sqrt{T}}\eta^T\mathbb{Z}_{t-1}^T\Sigma_{\epsilon}^{-1}\epsilon_{t})^4)$, we have:
\begin{equation}\nonumber
\begin{split}
&E((\frac{1}{\sqrt{T}}\eta^T\mathbb{Z}_{t-1}^T\Sigma_{\epsilon}^{-1}\epsilon_{t})^4)\\
&=E((\frac{1}{\sqrt{T}}\sum\limits_{i=1}^N\sum\limits_{j=1}^N\sum\limits_{k=1}^q(\eta_{2(k-1)N+j} X_{j,(t-k)}\alpha_{ij}+\eta_{(2(k-1)N+N+j)}w_{j}^{T}\mathbb{X}_{t-k}\alpha_{ij})\epsilon_{it}\\
&+\frac{1}{\sqrt{T}}\sum\limits_{i=1}^N\sum\limits_{j=1}^p\sum\limits_{k=1}^N\eta_{2Nq+(j-1)N+k} Y_{kj,(t-1)}\alpha_{ki}\epsilon_{it})^4)\\
&= E((\sum\limits_{i=1}^N a_i\epsilon_{it})^{4})\\
&=E(E((\sum\limits_{i=1}^N a_i\epsilon_{it})^{4}|\mathcal{A}_{t-1}))\\
&=E(E(\sum\limits_{i,j,k,m}a_ia_ja_ka_m\epsilon_{it}\epsilon_{jt}\epsilon_{kt}\epsilon_{mt}|\mathcal{A}_{t-1}))\\
&=E(\sum\limits_{i,j,k,m}a_ia_ja_ka_mE(\epsilon_{it}\epsilon_{jt}\epsilon_{kt}\epsilon_{mt}|\mathcal{A}_{t-1}))\\
&\leq E(\sum\limits_{i,j,k,m}|a_ia_ja_ka_m|E(|\epsilon_{it}\epsilon_{jt}\epsilon_{kt}\epsilon_{mt}||\mathcal{A}_{t-1}))\\
&\overset{(a)}{\leq}  c_1E((\sum\limits_{i=1}^N|a_i|)^4),
\end{split}
\end{equation}
where (a) follows from $E(|\epsilon_{it}\epsilon_{jt}\epsilon_{kt}\epsilon_{mt}||\mathcal{A}_{t-1})\leq c_1<\infty$.

Next, we need to find an upper bound for 
\begin{equation}\nonumber
\begin{split}
&E((\sum\limits_{i=1}^N|a_i|)^4)\\
&=E((\sum\limits_{i=1}^N|\frac{1}{\sqrt{T}}\sum\limits_{j=1}^N\sum\limits_{k=1}^q(\eta_{2(k-1)N+j} X_{j,(t-k)}\alpha_{ij}\\
&+\eta_{(2(k-1)N+N+j)}w_{j}^{T}\mathbb{X}_{t-k}\alpha_{ij})+\frac{1}{\sqrt{T}}\sum\limits_{j=1}^p\sum\limits_{k=1}^N\eta_{2Nq+(j-1)N+k} Y_{kj,(t-1)}\alpha_{ki}|)^4).
\end{split}
\end{equation}

Denote by $X_{2(k-1)N+j}'=X_{j,(t-k)}$ and $X_{2(k-1)N+N+j}'=w_j^T\mathbb{X}_{t-k}$ for $j=1,\cdots,N$. By Lemma \ref{L:BDD}, $E(|X_{i,t_1}X_{l,t_2}X_{m,t_3}X_{n,t_4}|)\leq c_3<\infty$ for any $1 \leq i,l,m,n\leq 2Nq$. Since row sum of W is 1, $X_{2(k-1)N+N+i}'$ can be seen as weighed average of $\mathbb{X}_{t-k}$, $E(|X_i'X_l'X_m'X_n'|)\leq c_5<\infty$ for any $1 \leq i,l,m,n\leq 2Nq$. If the row sum of $\Sigma_{\epsilon}^{-1}$ is bounded by $c_0$:
\begin{equation}\nonumber
\begin{split}
&E((\sum\limits_{i=1}^N|a_i|)^4)\\
&\leq 8E((\frac{1}{\sqrt{T}}\sum\limits_{j=1}^{2Nq}|c_0\eta_jX_j'|)^4)+8E((\frac{c_0}{\sqrt{T}}\sum\limits_{i=1}^N\sum\limits_{j=1}^{p}\sum\limits_{k=1}^{N}|\eta_{2Nq+(j-1)N+k}Y_{kj,(t-1)}|)^4)\\
&\leq 8\frac{c_0^4}{T^2}\sum\limits_{1\leq i,j,l,m\leq 2Nq}|\eta_i\eta_l\eta_m\eta_n|E(|X_i'X_l'X_m'X_n'|)+8c_0^4c_2\frac{1}{T^2}\sum\limits_{2Nq+1\leq i,j,l,m\leq 2Nq+Np}|\eta_i\eta_l\eta_m\eta_n|\\
&\leq 8c_0^4c_5\frac{1}{T^2} \sum\limits_{1\leq i,j,l,m\leq 2Nq}|\eta_i\eta_l\eta_m\eta_n|+8c_0^4c_2\frac{1}{T^2}\sum\limits_{2Nq+1\leq i,j,l,m\leq 2Nq+Np}|\eta_i\eta_l\eta_m\eta_n|\\
&=O(\frac{1}{T^2}).
\end{split}
\end{equation}
		
Thus, $s_{TT}^{-2}\sum\limits_{t=1}^TE(Z_{tT}^2I(|Z_{tT}|\geq\epsilon s_{TT}))=O(\frac{1}{T})\rightarrow0$.
Thus we obtain
\[\frac{1}{\sqrt{T}}\sum\limits_{t=1}^{T}\mathbb{Z}_{t-1}^T\Sigma_{\epsilon}^{-1}\epsilon_{t}\rightarrow_{d}N(0, Q ).\]
\end{proof}
\subsection{Proof of PROPOSITION \ref{P:GLS}}\label{b3}
\begin{proof}[Proof of PROPOSITION \ref{P:GLS}]
Under the conditions of proposition \ref{P:GLS}, \[\frac{1}{T}\sum\limits_{t=1}^{T}\mathbb{Z}_{t-1}^T\Sigma_{\epsilon}^{-1}\mathbb{Z}_{t-1}\rightarrow_{p}Q\]
by Lemma \ref{L:LLNGLS}, and
\[\frac{1}{\sqrt{T}}\sum\limits_{t=1}^{T}\mathbb{Z}_{t-1}^T\Sigma_{\epsilon}^{-1}\epsilon_{t}\rightarrow_{d}N(0, Q )\]
by Lemma \ref{L:CLTGLS}.
We thus obtain
\[\sqrt{T}(\hat{\beta}-\beta)\rightarrow_{d}N(0, Q^{-1}).\]

By equation \ref{beta-gls}, $\hat\beta_{GLS}$ can be written as: $
\hat{\beta}_{GLS}=\beta+(\sum_{t=1}^{T}\mathbb{Z}_{t-1}^T\Sigma_{\epsilon}^{-1}\mathbb{Z}_{t-1})^{-1}\sum_{t=1}^{T}\mathbb{Z}_{t-1}^T\Sigma_{\epsilon}^{-1}\epsilon_t$, so we have:
\begin{equation}\nonumber
\begin{split}
\sqrt{T}(\hat{\beta}-\beta)=&\sqrt{T}((\sum\limits_{t=1}^{T}\mathbb{Z}_{t-1}^T\Sigma_{\epsilon}^{-1}\mathbb{Z}_{t-1})^{-1}\sum\limits_{t=1}^{T}\mathbb{Z}_{t-1}^T\Sigma_{\epsilon}^{-1}\mathbb{X}_t-\beta)\\
&=\sqrt{T}((\sum\limits_{t=1}^{T}\mathbb{Z}_{t-1}^T\Sigma_{\epsilon}^{-1}\mathbb{Z}_{t-1})^{-1}\sum\limits_{t=1}^{T}\mathbb{Z}_{t-1}^T\Sigma_{\epsilon}^{-1}(\mathbb{Z}_{t-1}\beta+\epsilon_t)-\beta)\\
&=\sqrt{T}(\sum\limits_{t=1}^{T}\mathbb{Z}_{t-1}^T\Sigma_{\epsilon}^{-1}\mathbb{Z}_{t-1})^{-1}\sum\limits_{t=1}^{T}\mathbb{Z}_{t-1}^T\Sigma_{\epsilon}^{-1}\epsilon_t.
\end{split}
\end{equation}
As a result,
\begin{equation*}
\sqrt{T}(\hat{\beta}_{GLS}-\beta)\rightarrow_{d}N(0, Q^{-1}),
\end{equation*}
where $Q:=
E(\mathbb{Z}_{t}^T\Sigma_{\epsilon}^{-1}\mathbb{Z}_{t})$.

\end{proof}	

\subsection{LEMMA \ref{L:CLTDGLS}}\label{b4}

\begin{l4}
\label{L:CLTDGLS}
Suppose Assumptions \ref{A1}-\ref{A3} hold. Let $\mathbb{X}_t$ be A stationary process generated by the NAR$(q_1,q_2)$ model \eqref{compact}, i.e., $\mathbb{X}_{t}=\mathbb{Z}_{t-1}\beta+ \epsilon_{t}$ with growing network size $N$. Define as before $D\in \mathbb{R}^{k\times (2Nq+Np)}$ for any finite $k$. Further, suppose that Assumption \ref{A2} and the additional conditions posited in Proposition \ref{P:DOLS} hold for $D$, as well as $N<T$.
Then, 
\begin{equation}
\label{CLT:DGLS}
\frac{1}{\sqrt{T}}D\sum\limits_{t=1}^{T}\mathbb{Z}_{t-1}^T\Sigma_{\epsilon}^{-1}\epsilon_{t}\rightarrow_{d}N(0, Q )
\end{equation}
where $Q:=
E(\mathbb{Z}_{t}^T\Sigma_{\epsilon}^{-1}\mathbb{Z}_{t})$.
		
\end{l4}

\begin{proof}[Proof of LEMMA \ref{L:CLTDGLS}]
Let $\eta_{k\times 1}$ be a column vector of arbitrary real numbers such that $\eta^T \eta \not=0$. Let
\[\frac{1}{\sqrt{T}}\eta^TD\sum\limits_{t=1}^{T}\mathbb{Z}_{t-1}^T\Sigma_{\epsilon}^{-1}\epsilon_{t}=\sum\limits_{t=1}^{T}Z_{tT}=S_{TT},\]
where $Z_{tT}=\frac{1}{\sqrt{T}}\eta^TD\mathbb{Z}_{t-1}^T\Sigma_{\epsilon}^{-1}\epsilon_{t}$. Since $E(\epsilon_t|\mathcal{A}_{t-1})=0$ where $\mathcal{A}_{t-1}$ is the sigma-field generated by $\{\epsilon_{j}:j\leq t-1\}$, we have $E(Z_{tT}|\mathcal{A}_{t-1})=0\ a.s.$, so condition 1 of Lemma \ref{CLT} is satisfied.

Define
\begin{equation*}
\begin{split}
\delta_{tT}^2&:=E(Z_{tT}^2|\mathcal{A}_{t-1})\\
&=E(\frac{1}{T}\eta^TD\mathbb{Z}_{t-1}^T\Sigma_{\epsilon}^{-1}\epsilon_{t}\epsilon_{t}^T\Sigma_{\epsilon}^{-1}\mathbb{Z}_{t-1}D^T\eta|\mathcal{A}_{t-1})\\
&=\frac{1}{T}\eta^TD\mathbb{Z}_{t-1}^T\Sigma_{\epsilon}^{-1}\mathbb{Z}_{t-1}D^T\eta    
\end{split}    
\end{equation*}

and 
\[V_{TT}^2:=\frac{1}{T}\sum\limits_{t=1}^T\eta^TD\mathbb{Z}_{t-1}^T\Sigma_{\epsilon}^{-1}\mathbb{Z}_{t-1}D^T\eta.\]
Further, we obtain
\[V_{TT}^2\rightarrow_{p}\eta^TDQD^T \eta.\]by ergodicity of $\mathbb{X}_t$.
We also have
\[s_{TT}^2:=E(V_{TT}^2)=\frac{1}{T}\sum\limits_{t=1}^T\eta^TDE(\mathbb{Z}_{t-1}^T\Sigma_{\epsilon}^{-1}\mathbb{Z}_{t-1})D^T\eta=\eta^TDQD^T \eta,\]
and thus
\[V_{TT}^2s_{TT}^{-2}\rightarrow_{p}1,\]
so that condition 2 of Lemma \ref{CLT} is satisfied.
	
For condition 3 of Lemma \ref{CLT},
\begin{equation}\nonumber
\begin{split}
&s_{TT}^{-2}\sum\limits_{t=1}^TE(Z_{tT}^2I(|Z_{tT}|\geq\epsilon s_{TT}))\\&\leq s_{TT}^{-2}\sum\limits_{t=1}^TE((\epsilon s_{TT})^{-2}Z_{tT}^{4}I(|Z_{tT}|\geq\epsilon s_{TT}))\\&\leq s_{TT}^{-4}\epsilon^{-2}\sum\limits_{t=1}^TE(|\frac{1}{\sqrt{T}}\eta^TD\mathbb{Z}_{t-1}^T\Sigma_{\epsilon}^{-1}\epsilon_{t}|^4)\\
&\leq s_{TT}^{-4}\epsilon^{-2}\sum\limits_{t=1}^TE((\frac{1}{\sqrt{T}}\eta^TD\mathbb{Z}_{t-1}^T\Sigma_{\epsilon}^{-1}\epsilon_{t})^4).\\
\end{split}
\end{equation}
Define $a_l:=\frac{1}{\sqrt{T}}\sum\limits_{i=1}^k\sum\limits_{j=1}^N\sum\limits_{m=1}^q(\eta_{i}d_{i,(2(m-1)N+j)} X_{j,(t-m)}\alpha_{lj}+\eta_{i}d_{i,(2(m-1)N+j+N)}w_{j}^{T}\mathbb{X}_{t-m}\alpha_{jl})\\+\frac{1}{\sqrt{T}}\sum\limits_{i=1}^k\sum\limits_{j=1}^p\sum\limits_{m=1}^N\eta_{i}d_{i,(2Nq+(j-1)N+m)} Y_{mj,(t-1)}\alpha_{ml}$, where $\alpha_{ik}:=(\Sigma_{\epsilon}^{-1})_{ik}$. For $E((\eta^TC_{T}^{-1}\mathbb{Z}_{t-1}^T\Sigma_{\epsilon}^{-1}\epsilon_{t})^4)$, we have:
\begin{equation}\nonumber
\begin{split}
&E((\eta^T\frac{1}{\sqrt{T}}\mathbb{Z}_{t-1}^T\Sigma_{\epsilon}^{-1}\epsilon_{t})^4)\\
&=E((\frac{1}{\sqrt{T}}\sum\limits_{i=1}^k\sum\limits_{j=1}^N\sum\limits_{l=1}^N\sum\limits_{m=1}^q(\eta_{i}d_{i,(2(m-1)N+j)} X_{j,(t-m)}\alpha_{lj}+\eta_{i}d_{i,(2(m-1)N+j+N)}w_{j}^{T}\mathbb{X}_{t-m}\alpha_{jl})\epsilon_{lt}\\
&+\frac{1}{\sqrt{T}}\sum\limits_{i=1}^k\sum\limits_{j=1}^p\sum\limits_{l=1}^N\sum\limits_{m=1}^N\eta_{i}d_{i,(2Nq+(j-1)N+m)} Y_{mj,(t-1)}\alpha_{ml}\epsilon_{lt})^4)\\
&=E((\sum\limits_{l=1}^N a_l\epsilon_{lt})^{4})\\
&=E(E((\sum\limits_{l=1}^N a_i\epsilon_{lt})^{4}|\mathcal{A}_{t-1}))\\
&=E(E(\sum\limits_{i,j,k,m}a_ia_ja_ka_m\epsilon_{it}\epsilon_{jt}\epsilon_{kt}\epsilon_{mt}|\mathcal{A}_{t-1}))\\
&=E(\sum\limits_{i,j,k,m}a_ia_ja_ka_mE(\epsilon_{it}\epsilon_{jt}\epsilon_{kt}\epsilon_{mt}|\mathcal{A}_{t-1}))\\
&\leq E(\sum\limits_{i,j,k,m}|a_ia_ja_ka_m|E(|\epsilon_{it}\epsilon_{jt}\epsilon_{kt}\epsilon_{mt}||\mathcal{A}_{t-1}))\\
&\overset{(a)}{\leq}  c_1E((\sum\limits_{l=1}^N|a_l|)^4),
\end{split}
\end{equation}
where (a) follows from $E(|\epsilon_{it}\epsilon_{jt}\epsilon_{kt}\epsilon_{mt}||\mathcal{A}_{t-1})\leq c_1<\infty$.
Next, we need to find an upper bound for
\begin{equation}\nonumber
\begin{split}
&E((\sum\limits_{l=1}^N|a_l|)^4)\\
&=E((\sum\limits_{l=1}^N|\frac{1}{\sqrt{T}}\sum\limits_{i=1}^k\sum\limits_{j=1}^N\sum\limits_{m=1}^q(\eta_{i}d_{i,(2(m-1)N+j)} X_{j,(t-m)}\alpha_{lj}+\eta_{i}d_{i,(2(m-1)N+j+N)}w_{j}^{T}\mathbb{X}_{t-m}\alpha_{lj})\\
&+\frac{1}{\sqrt{T}}\sum\limits_{i=1}^k\sum\limits_{j=1}^p\sum\limits_{m=1}^N\eta_{i}d_{i,(2Nq+(j-1)N+m)} Y_{mj,(t-1)}\alpha_{ml}|)^4).
\end{split}
\end{equation}
	
Denote by $X_{2(m-1)N+j}'=X_{j,(t-m)}$ and $X_{2(m-1)N+N+j}'=w_j^T\mathbb{X}_{t-m}$ for j=1,...,N. By Lemma \ref{L:BDD}, $E(|X_{i,t_1}X_{l,t_2}X_{m,t_3}X_{n,t_4}|)\leq c_3<\infty$ for any $1 \leq i,l,m,n\leq 2Nq$. Since row sum of W is 1, $X_{2(k-1)N+N+i}'$ can be seen as weighed average of $\mathbb{X}_{t-k}$, $E(|X_i'X_l'X_m'X_n'|)\leq c_5<\infty$ for any $1 \leq i,l,m,n\leq 2Nq$. If row sum of $\Sigma_{\epsilon}^{-1}$ is bounded by $c_0$:
\begin{equation}\nonumber
\begin{split}
&E((\sum\limits_{l=1}^N|a_l|)^4)\\
&\leq 8E((\frac{1}{\sqrt{T}}|c_0\sum\limits_{i=1}^k\sum\limits_{j=1}^{2Nq}\eta_id_{i,j}X_j'|)^4)+8E((\frac{1}{\sqrt{T}}|c_0\sum\limits_{i=1}^k\sum\limits_{j=1}^p\sum\limits_{m=1}^N\eta_{i}d_{i,(2Nq+(j-1)N+m)} Y_{mj,(t-1)}|)^4)\\
&\leq 8c_0^4c_5\frac{1}{T^2}(\sum\limits_{i=1}^k\sum\limits_{j=1}^{2Nq}\eta_id_{i,j})^4+8c_0^4c_2\frac{1}{T^2}(\sum\limits_{i=1}^k\eta_{i}(\sum\limits_{j=1}^p\sum\limits_{m=1}^Nd_{i,(2Nq+(j-1)N+m)}))^4\\
&= 8c_0^4c_5\frac{1}{T^2}(\sum\limits_{i=1}^k\eta_i(\sum\limits_{j=1}^{2Nq}d_{i,j}))^4+8c_0^4c_2\frac{1}{T^2}(\sum\limits_{i=1}^k\eta_{i}(\sum\limits_{j=1}^p\sum\limits_{m=1}^Nd_{i,(2Nq+(j-1)N+m)}))^4\\
&=O(\frac{1}{T^2}),
\end{split}
\end{equation}	
since the row sum of $D$ is bounded.

Therefore, $s_{TT}^{-2}\sum\limits_{t=1}^TE(Z_{tT}^2I(|Z_{tT}|\geq\epsilon s_{TT}))=O(\frac{1}{T})\rightarrow0$.
Thus, we obtain
\[\frac{1}{\sqrt{T}}D\sum\limits_{t=1}^{T}\mathbb{Z}_{t-1}^T\Sigma_{\epsilon}^{-1}\epsilon_{t}\rightarrow_{d}N(0, DQD^T ).\]
\end{proof}

\subsection{Proof of PROPOSITION \ref{P:DGLS}}\label{b5}
\begin{proof}[Proof of PROPOSITION \ref{P:DGLS}]

By equation \ref{beta-gls}, $\hat\beta_{GLS}$ can be written as: \[
\hat{\beta}_{GLS}=\beta+(\sum_{t=1}^{T}\mathbb{Z}_{t-1}^T\Sigma_{\epsilon}^{-1}\mathbb{Z}_{t-1})^{-1}\sum_{t=1}^{T}\mathbb{Z}_{t-1}^T\Sigma_{\epsilon}^{-1}\epsilon_t,\] so we have:
 \begin{equation*}
\begin{split}
 &\frac{1}{\sqrt{T}}D(\sum_{t=1}^{T}\mathbb{Z}_{t-1}^T\Sigma_{\epsilon}^{-1}\mathbb{Z}_{t-1})(\hat{\beta}_{GLS}-\beta)\\
 &=\frac{1}{\sqrt{T}}D(\sum_{t=1}^{T}\mathbb{Z}_{t-1}^T\Sigma_{\epsilon}^{-1}\mathbb{Z}_{t-1})((\sum\limits_{t=1}^{T}\mathbb{Z}_{t-1}^T\Sigma_{\epsilon}^{-1}\mathbb{Z}_{t-1})^{-1}\sum\limits_{t=1}^{T}\mathbb{Z}_{t-1}^T\Sigma_{\epsilon}^{-1}\mathbb{X}_t-\beta)\\
 &=\frac{1}{\sqrt{T}}D(\sum_{t=1}^{T}\mathbb{Z}_{t-1}^T\Sigma_{\epsilon}^{-1}\mathbb{Z}_{t-1})((\sum\limits_{t=1}^{T}\mathbb{Z}_{t-1}^T\Sigma_{\epsilon}^{-1}\mathbb{Z}_{t-1})^{-1}\sum\limits_{t=1}^{T}\mathbb{Z}_{t-1}^T\Sigma_{\epsilon}^{-1}(\mathbb{Z}_{t-1}\beta+\epsilon_t)-\beta)\\
 &=\frac{1}{\sqrt{T}}D(\sum_{t=1}^{T}\mathbb{Z}_{t-1}^T\Sigma_{\epsilon}^{-1}\epsilon_{t}).
 \end{split}
 \end{equation*}
 By Lemma \ref{L:CLTDGLS}, 
 \[\frac{1}{\sqrt{T}}D\sum\limits_{t=1}^{T}\mathbb{Z}_{t-1}^T\Sigma_{\epsilon}^{-1}\epsilon_{t}\rightarrow_{d}N(0, Q ),\]
 so that
 \[\frac{1}{\sqrt{T}}D(\sum_{t=1}^{T}\mathbb{Z}_{t-1}^T\Sigma_{\epsilon}^{-1}\mathbb{Z}_{t-1})(\hat{\beta}_{GLS}-\beta)\rightarrow_{d}N(0, DQD^T ).\]
\end{proof}

\subsection{Proof of PROPOSITION \ref{P:DGLSR}}\label{a10}
\begin{proof}[Proof of PROPOSITION \ref{P:DGLSR}]
By the definition of the ridge estimator in Equation \ref{beta-gls-r}, we have
\begin{equation*}
\begin{aligned}
\hat{\beta}_{ridge}&=argmin\frac{1}{T}\sum_{t=1}^{T}(\mathbb{X}_{t}-\mathbb{Z}_{t-1}\beta)^T\Sigma_{\epsilon}^{-1}(\mathbb{X}_{t}-\mathbb{Z}_{t-1}\beta)+||M\beta||^2\\
&=(\sum_{t=1}^{T}\mathbb{Z}_{t-1}^T\Sigma_{\epsilon}^{-1}\mathbb{Z}_{t-1}+T  M)^{-1}\sum_{t=1}^{T}\mathbb{Z}_{t-1}^T\Sigma_{\epsilon}^{-1}\mathbb{X}_t\\
&=(\sum_{t=1}^{T}\mathbb{Z}_{t-1}^T\Sigma_{\epsilon}^{-1}\mathbb{Z}_{t-1}+T  M)^{-1}\sum_{t=1}^{T}\mathbb{Z}_{t-1}^T\Sigma_{\epsilon}^{-1}\mathbb{X}_t\\
&=(\sum_{t=1}^{T}\mathbb{Z}_{t-1}^T\Sigma_{\epsilon}^{-1}\mathbb{Z}_{t-1}+T  M)^{-1}\sum_{t=1}^{T}(\mathbb{Z}_{t-1}^T\Sigma_{\epsilon}^{-1}\mathbb{Z}_{t-1}\beta+\mathbb{Z}_{t-1}^T\Sigma_{\epsilon}^{-1}\epsilon_t)\\
&=(\sum_{t=1}^{T}\mathbb{Z}_{t-1}^T\Sigma_{\epsilon}^{-1}\mathbb{Z}_{t-1}+T  M)^{-1}(\sum_{t=1}^{T}\mathbb{Z}_{t-1}^T\Sigma_{\epsilon}^{-1}\mathbb{Z}_{t-1}+T M)\beta\\
&+(\sum_{t=1}^{T}\mathbb{Z}_{t-1}^T\Sigma_{\epsilon}^{-1}\Sigma_{\epsilon}^{-1}\mathbb{Z}_{t-1}+T M)^{-1}\sum_{t=1}^{T}\mathbb{Z}_{t-1}^T\Sigma_{\epsilon}^{-1}\epsilon_t-(\sum_{t=1}^{T}\mathbb{Z}_{t-1}^T\Sigma_{\epsilon}^{-1}\mathbb{Z}_{t-1}+T  M)^{-1}T M\beta\\
&=\beta+(\sum_{t=1}^{T}\mathbb{Z}_{t-1}^T\Sigma_{\epsilon}^{-1}\mathbb{Z}_{t-1}+T  M)^{-1}\sum_{t=1}^{T}\mathbb{Z}_{t-1}^T\Sigma_{\epsilon}^{-1}\epsilon_t-(\sum_{t=1}^{T}\mathbb{Z}_{t-1}^T\Sigma_{\epsilon}^{-1}\mathbb{Z}_{t-1}+T  M)^{-1}T M\beta.
\end{aligned}
\end{equation*} 

Since
\[\frac{1}{T}(\sum_{t=1}^{T}\mathbb{Z}_{t-1}^T\Sigma_{\epsilon}^{-1}\mathbb{Z}_{t-1}+T M)\rightarrow_{p}Q+ M,\]
where
\[Q:=
E(\mathbb{Z}_{t}^T\Sigma_{\epsilon}^{-1}\mathbb{Z}_{t}),\]
we obtain:
\[\frac{1}{\sqrt{T}}D(\sum_{t=1}^{T}\mathbb{Z}_{t-1}^T\Sigma_{\epsilon}^{-1}\mathbb{Z}_{t-1}+TM)(\hat{\beta}_{ridge}-\beta)=\frac{1}{\sqrt{T}}D(\sum_{t=1}^{T}\mathbb{Z}_{t-1}^T\Sigma_{\epsilon}^{-1}\epsilon_t-TM\beta).\]
Thus, if $\lambda_1=o(\frac{1}{\sqrt{T}})$, $\lambda_2=o(\frac{1}{\sqrt{T}})$ and $\lambda_3=o(\frac{1}{\sqrt{T}})$,
\[\frac{1}{\sqrt{T}}D(\sum_{t=1}^{T}\mathbb{Z}_{t-1}^T\Sigma_{\epsilon}^{-1}\mathbb{Z}_{t-1}+TM)(\hat{\beta}_{ridge}-\beta)\rightarrow_{d}N(0, DQD^T ).\]
\end{proof}

\section{EGLS results}\label{p3.5}
\begin{proof}[Proof of Proposition \ref{P:EGLS}]
 By the Assumptions of Proposition \ref{P:EGLS}, $\hat{\Sigma}_\epsilon$ is a consistent estimator of $\Sigma_\epsilon$, so 
 \[\sum_{t=1}^{T}\mathbb{Z}_{t-1}^T\hat{\Sigma}_{\epsilon}^{-1}\mathbb{Z}_{t-1}\rightarrow_{p}\sum_{t=1}^{T}\mathbb{Z}_{t-1}^T\Sigma_{\epsilon}^{-1}\mathbb{Z}_{t-1}\]
 and
 \[\sum\limits_{t=1}^{T}\mathbb{Z}_{t-1}^T\hat{\Sigma}_{\epsilon}^{-1}\epsilon_{t}\rightarrow_p \sum\limits_{t=1}^{T}\mathbb{Z}_{t-1}^T\Sigma_{\epsilon}^{-1}\epsilon_{t}.\]
 Thus, 
 \[(\hat{\beta}_{EGLS}-\beta_{GLS})\rightarrow_p 0,\]
 $\hat{\beta}_{EGLS}$ is asymptotically equivalent to $\hat{\beta}_{GLS}$.
 Hence,
 \[\sqrt{T}(\hat{\beta}_{EGLS}-\beta)\rightarrow_{d}N(0, Q^{-1})\ for\ fixed\ N,\]
 and\[\frac{1}{\sqrt{T}}D(\sum_{t=1}^{T}\mathbb{Z}_{t-1}^T\hat{\Sigma}_{\epsilon}^{-1}\mathbb{Z}_{t-1})(\hat{\beta}_{EGLS}-\beta)\rightarrow_{d}N(0, DQD^T)\ for\ diverging\ N\leq T,\]
 and
 \[\frac{1}{\sqrt{T}}D(\sum_{t=1}^{T}\mathbb{Z}_{t-1}^T\hat\Sigma_{\epsilon}^{-1}\mathbb{Z}_{t-1}+TM)(\hat{\beta}_{ridge}-\beta)\rightarrow_{d}N(0, DQD^T)\ for\ diverging\ N>T.\]
 
\end{proof}

\begin{l5}[Consistency of $\hat{\rho}$]
\label{L:SAR}
We impose the following assumptions:
(i) $u_t\sim F(0,\sigma_u^2I)$ for some distribution $F$ and $E(|u|^{4+\gamma})$ for some $\gamma>0$ exists. (ii) $S(\rho_0)$ is nonsingular, where $\rho_0$ denotes the true model parameter, (iii) ${S^{-1}(\rho_0)}$ has bounded row and column sums, and (iv) $\{{S^{-1}(\rho)}\}$ is uniformly bounded in either row or column sums, uniformly in $\rho$ in a compact parameter space $P$. The true $\rho_0$ is in the interior of $P$.

Under these four conditions, $\theta:=(\rho_0,\sigma^2_u)$ is globally identifiable and $\hat{\theta}$ is a consistent estimator of $\theta$.
\end{l5}
\begin{remark6}
For assumption (iv), with row-normalized $\Phi$, $S^{-1}(\rho)$ is uniformly bounded in row sums in any closed subset of (-1,1). In this case, $P$ can be taken as a single closed set contained in $(-1,1)$.
\end{remark6}

\begin{l6}[Consistency of $\hat{C}_{it}=\hat{\lambda}_i^T\hat{F}_t$]
\label{L:FM}
By Results A and C in \cite{bai2008large}, if the following conditions are satisfied:
\begin{itemize}
    \item $E||F_t^0||^4\leq M$ and $\frac{1}{T}\sum\limits_{t=1}^TF_t^0{F_t^0}^T\rightarrow_p\Sigma_F>0$ for an $r\times r$ non-random matrix $\Sigma_F$.
    \item $\lambda_i^0$ is either deterministic such that $||\lambda_i^0||\leq M$, or it is stochastic such that $E||\lambda_i^0||^4\leq M$. In either case, $N^{-1}{\Lambda^0}^T\Lambda^0\rightarrow_{p}\Sigma_{\Lambda}>0$ for an $r\times r$ non-random matrix $\Sigma_\Lambda$, as $N\rightarrow \infty$.
    \item
(a) $E(u_{it})=0$, $E|u_{it}|^8 \leq M$.

(b) $E(u_{it}u_{js})=\sigma_{ij,ts}$, $|\sigma_{ij,ts}|\leq \bar{\sigma}_{ij}$ for all $(t,s)$ and $|\sigma_{ij,ts}|\leq \tau_{ts}$ for all $(i,j)$ such that $\frac{1}{N}\sum\limits_{i,j=1}^N\bar{\sigma}_{ij}\leq M$, $\frac{1}{T}\sum\limits_{t,s=1}^T\tau_{ts}\leq M$ and $\frac{1}{NT}\sum\limits_{i,j,t,s=1}|\sigma_{ij,ts}|\leq M$.

(c) For every $(t,s)$, $E|N^{-1/2}\sum\limits_{i=1}^N[u_{it}u_{is}-E(u_{it}u_{is})]|^4\leq M$.

(d) For each t, $\frac{1}{\sqrt{N}}\sum\limits_{i=1}^N\lambda_iu_{it}\rightarrow_d N(0, \Gamma_t)$, as $N\rightarrow \infty$ where $\Gamma_t=\mathop{lim}\limits_{N\rightarrow \infty}\frac{1}{N}\sum\limits_{i=1}^N\sum\limits_{j=1}^NE(\lambda_i\lambda_j^Tu_{it}u_{jt})$.

(e) For each i, $\frac{1}{\sqrt{T}}\sum\limits_{t=1}^TF_tu_{it}\rightarrow_d N(0, \Phi_i)$, as $T\rightarrow \infty$ where $\Phi_i=\mathop{lim}\limits_{T\rightarrow \infty}\frac{1}{T}\sum\limits_{s=1}^T\sum\limits_{t=1}^TE(F_t^0{F_s^0}^Tu_{is}u_{it})$.
    \item $\{\lambda_i\}$, $\{F_t\}$ and  $\{u_{it}\}$ are three mutually independent groups. Dependence within each group is allowed.
    \item For all $t\leq T$, $i\leq N$, $\sum\limits_{s=1}^T|\tau_{s,t}|\leq M$, and $\sum\limits_{i=1}^N|\bar{\sigma}_{ij}|\leq M$. 
\end{itemize}
Then,
\begin{itemize}
    \item[(i)] 
If (i) $g(N,T)\rightarrow0$ and (ii) $\min\{N,T\}g(N,T)\rightarrow 0$ as $N,T\rightarrow\infty$, then\[\hat{k}_{IC}\rightarrow_pr.\]
    \item[(ii)]
\[(N^{-1}A_{it}+T^{-1}B_{it})^{-1/2}(\Tilde{C}_{it}-C_{it}^0)\rightarrow_d N(0,1)\] where $A_{it}={\lambda_i^0}^T\Sigma_{\Lambda}^{-1}\Gamma_t\Sigma_{\Lambda}^{-1}\lambda_i^0$, $B_{it}={F_i^0}^T\Sigma_{F}^{-1}\Phi_i\Sigma_{F}^{-1}F_t^0$ and $\Phi_i$ is the variance of $T^{-1/2}\sum\limits_{t=1}^TF_t^0u_{it}$.
\end{itemize}		
\end{l6}

\section{Proof of Proposition \ref{P:MWM}}\label{pos4}
\begin{proof}
We consider $||\pi_T||_{\infty}=o(1)$ case and $||\pi_T||_{\infty}=o(\frac{1}{\sqrt{T}})$ case:
\begin{itemize}
    \item[(a)]$||\pi_T||_{\infty}=o(1)$ case:
    
By WLLN, $\mathbb{X}_t\rightarrow_p E(\mathbb{X}_t)$ for stationary $\mathbb{X}_t$.
If $||\pi_T||_{\infty}=o(1)$, $||\pi_{T}\mathbb{X}_{t-l}||_{max}=o_p(1)$, which means $||P_{t-1}||_{max}=o_p(1)$.

\begin{itemize}
\item[(1)] 
For fixed $N$, 

\begin{equation}\nonumber
\begin{split}
\hat{\beta}_{OLS}^M-\beta&=((\sum\limits_{t=1}^{T}(\mathbb{Z}_{t-1}^M)^T\mathbb{Z}_{t-1}^M)^{-1}\sum\limits_{t=1}^{T}(\mathbb{Z}_{t-1}^M)^T\mathbb{X}_t-\beta)\\
&=(\frac{1}{T}\sum\limits_{t=1}^{T}(\mathbb{Z}_{t-1}+P_{t-1})^T(\mathbb{Z}_{t-1}+P_{t-1}))^{-1}\frac{1}{T}\sum\limits_{t=1}^{T}(\mathbb{Z}_{t-1}+P_{t-1})^T(\mathbb{Z}_{t-1}\beta+\epsilon_t)-\beta.\\
&=(\frac{1}{T}\sum\limits_{t=1}^{T}(\mathbb{Z}_{t-1}+P_{t-1})^T(\mathbb{Z}_{t-1}+P_{t-1}))^{-1}\frac{1}{T}\sum\limits_{t=1}^{T}(\mathbb{Z}_{t-1}+P_{t-1})^T(\mathbb{Z}_{t-1}\beta)-\beta\\
&+(\frac{1}{T}\sum\limits_{t=1}^{T}(\mathbb{Z}_{t-1}+P_{t-1})^T(\mathbb{Z}_{t-1}+P_{t-1}))^{-1}\frac{1}{T}\sum\limits_{t=1}^{T}(\mathbb{Z}_{t-1}+P_{t-1})^T\epsilon_t\\
&=-(\frac{1}{T}\sum\limits_{t=1}^{T}(\mathbb{Z}_{t-1}+P_{t-1})^T(\mathbb{Z}_{t-1}+P_{t-1}))^{-1}\frac{1}{T}\sum\limits_{t=1}^{T}(\mathbb{Z}_{t-1}+P_{t-1})^T(P_{t-1}\beta)\\
&+(\frac{1}{T}\sum\limits_{t=1}^{T}(\mathbb{Z}_{t-1}+P_{t-1})^T(\mathbb{Z}_{t-1}+P_{t-1}))^{-1}\frac{1}{T}\sum\limits_{t=1}^{T}(\mathbb{Z}_{t-1}+P_{t-1})^T\epsilon_t.
\end{split}
\end{equation}

First, we prove \[\frac{1}{T}\sum\limits_{t=1}^{T}(\mathbb{Z}_{t-1}+P_{t-1})^T(\mathbb{Z}_{t-1}+P_{t-1})\rightarrow_p E(\mathbb{Z}_{t}^T\mathbb{Z}_{t}),\]
and then we show  \[\frac{1}{T}\sum\limits_{t=1}^{T}(\mathbb{Z}_{t-1}+P_{t-1})^T(P_{t-1}\beta) \rightarrow_p 0\]
and
\[\frac{1}{T}\sum\limits_{t=1}^{T}(\mathbb{Z}_{t-1}+P_{t-1})^T\epsilon_t\rightarrow_p 0.\]
\begin{itemize}
    \item [(i)]

$\frac{1}{T}\sum\limits_{t=1}^{T}(\mathbb{Z}_{t-1}+P_{t-1})^T(\mathbb{Z}_{t-1}+P_{t-1})=\frac{1}{T}\sum\limits_{t=1}^{T}(\mathbb{Z}_{t-1}^T\mathbb{Z}_{t-1}+\mathbb{Z}_{t-1}^TP_{t-1}+P_{t-1}^T\mathbb{Z}_{t-1}+P_{t-1}^TP_{t-1}).$ 

Since $P_{t-1}:=\begin{bmatrix}\boldsymbol{0}_{N\times N} &\mathop{diag}\{\pi_{T}\mathbb{X}_{t-1}\}&\cdots&\boldsymbol{0}_{N\times N} &\mathop{diag}\{\pi_{T}\mathbb{X}_{t-q}\}&\boldsymbol{0}_{N\times Np}\end{bmatrix}$, 
\[\mathbb{Z}_{t-1}^TP_{t-1},\ P_{t-1}^T\mathbb{Z}_{t-1},P_{t-1}^TP_{t-1}=o_p(1).\]
Therefore,
\[\frac{1}{T}\sum\limits_{t=1}^{T}(\mathbb{Z}_{t-1}+P_{t-1})^T(\mathbb{Z}_{t-1}+P_{t-1})\rightarrow_p E(\mathbb{Z}_{t}^T\mathbb{Z}_{t}).\]
\item [(ii)]
Since $P_{t-1}^T\epsilon_t\rightarrow_p 0$ and $P_{t-1}\beta\rightarrow 0$ as $t\rightarrow\infty$,
\begin{equation}\nonumber
\begin{split}
\frac{1}{T}\sum\limits_{t=1}^{T}(\mathbb{Z}_{t-1}+P_{t-1})^T(P_{t-1}\beta)\rightarrow_p 0
\end{split}
\end{equation}
and 
\[\frac{1}{T}\sum\limits_{t=1}^{T}P_{t-1}^T\epsilon_t\rightarrow_p 0,\]
by Lemma \ref{L:CLTOLS},
\[\frac{1}{\sqrt{T}}\sum\limits_{t=1}^{T}\mathbb{Z}_{t-1}^T\epsilon_{t}\rightarrow_{d}N(0, E(\mathbb{Z}_{t}^T\Sigma_{\epsilon}\mathbb{Z}_{t}) ).\]
Therefore,
 \[\frac{1}{T}(\sum_{t=1}^{T}\mathbb{Z}_{t-1}^T\epsilon_{t}) \rightarrow_p 0.\]
Finally,
\begin{equation}\nonumber
\begin{split}
\hat{\beta}_{OLS}^M-\beta&=((\sum\limits_{t=1}^{T}(\mathbb{Z}_{t-1}^M)^T\mathbb{Z}_{t-1}^M)^{-1}\sum\limits_{t=1}^{T}(\mathbb{Z}_{t-1}^M)^T\mathbb{X}_t-\beta)\\
&=-(\frac{1}{T}\sum\limits_{t=1}^{T}(\mathbb{Z}_{t-1}+P_{t-1})^T(\mathbb{Z}_{t-1}+P_{t-1}))^{-1}\frac{1}{T}\sum\limits_{t=1}^{T}(\mathbb{Z}_{t-1}+P_{t-1})^T(P_{t-1}\beta)\\
&+(\frac{1}{T}\sum\limits_{t=1}^{T}(\mathbb{Z}_{t-1}+P_{t-1})^T(\mathbb{Z}_{t-1}+P_{t-1}))^{-1}\frac{1}{T}\sum\limits_{t=1}^{T}(\mathbb{Z}_{t-1}+P_{t-1})^T\epsilon_t\\
&\rightarrow_p 0.
\end{split}
\end{equation}
\end{itemize}

\item[(2)] 
For diverging $N$,

\begin{equation*}
\begin{split}
 &\ \ \ \ \frac{1}{T}D(\sum_{t=1}^{T}(\mathbb{Z}_{t-1}^M)^T\mathbb{Z}_{t-1}^M)(\hat{\beta}_{OLS}^M-\beta)\\
 &=\frac{1}{T}D(\sum_{t=1}^{T}(\mathbb{Z}_{t-1}^M)^T\mathbb{Z}_{t-1}^M)((\sum\limits_{t=1}^{T}(\mathbb{Z}_{t-1}^M)^T\mathbb{Z}_{t-1}^M)^{-1}\sum\limits_{t=1}^{T}(\mathbb{Z}_{t-1}^M)^T\mathbb{X}_t-\beta)\\
 &=-\frac{1}{T}D(\sum_{t=1}^{T}(\mathbb{Z}_{t-1}+P_{t-1})^TP_{t-1}\beta)+\frac{1}{T}D(\sum_{t=1}^{T}(\mathbb{Z}_{t-1}+P_{t-1})^T\epsilon_{t}).
 \end{split}
 \end{equation*}
Since $||P_{t-1}||_{max}=o_p(1)$, 
\[\mathbb{Z}_{t-1}+P_{t-1}\rightarrow_p E(\mathbb{Z}_{t}),\]
\[-\frac{1}{T}D(\sum_{t=1}^{T}(\mathbb{Z}_{t-1}+P_{t-1})^TP_{t-1}\beta)\rightarrow_p 0,\]
and 
\[\frac{1}{T}D\sum_{t=1}^{T}P_{t-1}^T\epsilon_{t} \rightarrow_p 0,\]
by Lemma \ref{L:CLTDOLS},
\[\frac{1}{\sqrt{T}}D\sum\limits_{t=1}^{T}\mathbb{Z}_{t-1}^T\epsilon_{t}\rightarrow_{d}N(0, DE(\mathbb{Z}_{t}^T\Sigma_{\epsilon}\mathbb{Z}_{t})D^T ).\]
Thus,
\[\frac{1}{T}D\sum\limits_{t=1}^{T}\mathbb{Z}_{t-1}^T\epsilon_{t}\rightarrow_{p}0.\]
Finally,
\begin{equation*}
\begin{split}
 &\ \ \frac{1}{T}D(\sum_{t=1}^{T}(\mathbb{Z}_{t-1}^M)^T\mathbb{Z}_{t-1}^M)(\hat{\beta}_{OLS}^M-\beta)\\
 &=-\frac{1}{T}D(\sum_{t=1}^{T}(\mathbb{Z}_{t-1}+P_{t-1})^TP_{t-1}\beta)+\frac{1}{T}D(\sum_{t=1}^{T}(\mathbb{Z}_{t-1}+P_{t-1})^T\epsilon_{t})\\
 &\rightarrow_p 0.
 \end{split}
 \end{equation*}
\end{itemize}

\item[(b)] $||\pi_T||_{\infty}=o(\frac{1}{\sqrt{T}})$ case:

By the Weak Law of Large Numbers $\mathbb{X}_t\rightarrow_p E(\mathbb{X}_t)$ for stationary $\mathbb{X}_t$.
If $||\pi_T||_{\infty}=o(\frac{1}{\sqrt{T}})$, $||\pi_{T}\mathbb{X}_{t-l}||_{max}=o_p(\frac{1}{\sqrt{T}})$, which implies $||P_{t-1}||_{max}=o_p(\frac{1}{\sqrt{T}})$.

\begin{itemize}
\item[(1)] 
For fixed $N$, 

\begin{equation}\nonumber
\begin{split}
\sqrt{T}(\hat{\beta}_{OLS}^M-\beta)&=\sqrt{T}((\sum\limits_{t=1}^{T}(\mathbb{Z}_{t-1}^M)^T\mathbb{Z}_{t-1}^M)^{-1}\sum\limits_{t=1}^{T}(\mathbb{Z}_{t-1}^M)^T\mathbb{X}_t-\beta)\\
&=(\frac{1}{T}\sum\limits_{t=1}^{T}(\mathbb{Z}_{t-1}+P_{t-1})^T(\mathbb{Z}_{t-1}+P_{t-1}))^{-1}\frac{1}{\sqrt{T}}\sum\limits_{t=1}^{T}(\mathbb{Z}_{t-1}+P_{t-1})^T(\mathbb{Z}_{t-1}\beta+\epsilon_t)-\sqrt{T}\beta.\\
&=(\frac{1}{T}\sum\limits_{t=1}^{T}(\mathbb{Z}_{t-1}+P_{t-1})^T(\mathbb{Z}_{t-1}+P_{t-1}))^{-1}\frac{1}{\sqrt{T}}\sum\limits_{t=1}^{T}(\mathbb{Z}_{t-1}+P_{t-1})^T(\mathbb{Z}_{t-1}\beta)-\sqrt{T}\beta\\
&+(\frac{1}{T}\sum\limits_{t=1}^{T}(\mathbb{Z}_{t-1}+P_{t-1})^T(\mathbb{Z}_{t-1}+P_{t-1}))^{-1}\frac{1}{\sqrt{T}}\sum\limits_{t=1}^{T}(\mathbb{Z}_{t-1}+P_{t-1})^T\epsilon_t\\
&=-(\frac{1}{T}\sum\limits_{t=1}^{T}(\mathbb{Z}_{t-1}+P_{t-1})^T(\mathbb{Z}_{t-1}+P_{t-1}))^{-1}\frac{1}{\sqrt{T}}\sum\limits_{t=1}^{T}(\mathbb{Z}_{t-1}+P_{t-1})^T(P_{t-1}\beta)\\
&+(\frac{1}{T}\sum\limits_{t=1}^{T}(\mathbb{Z}_{t-1}+P_{t-1})^T(\mathbb{Z}_{t-1}+P_{t-1}))^{-1}\frac{1}{\sqrt{T}}\sum\limits_{t=1}^{T}(\mathbb{Z}_{t-1}+P_{t-1})^T\epsilon_t
\end{split}
\end{equation}

By (a)(1), \[\frac{1}{T}\sum\limits_{t=1}^{T}(\mathbb{Z}_{t-1}+P_{t-1})^T(\mathbb{Z}_{t-1}+P_{t-1})\rightarrow_p E(\mathbb{Z}_{t}^T\mathbb{Z}_{t}),\]
it suffices to show  \[\frac{1}{\sqrt{T}}\sum\limits_{t=1}^{T}(\mathbb{Z}_{t-1}+P_{t-1})^T(P_{t-1}\beta) \rightarrow_p 0\]
and
\[\frac{1}{\sqrt{T}}\sum\limits_{t=1}^{T}P_{t-1}^T\epsilon_t\rightarrow_p 0.\]

Since $P_{t-1}^T\epsilon_t=o_p(\frac{1}{\sqrt{T}})$ and $P_{t-1}\beta=o_p(\frac{1}{\sqrt{T}})$,
\[
\frac{1}{\sqrt{T}}\sum\limits_{t=1}^{T}(\mathbb{Z}_{t-1}+P_{t-1})^T(P_{t-1}\beta)\rightarrow_p 0
\]
and 
\[\frac{1}{\sqrt{T}}\sum\limits_{t=1}^{T}P_{t-1}^T\epsilon_t\rightarrow_p 0,\]
by Lemma \ref{L:CLTOLS},
\[\frac{1}{\sqrt{T}}\sum\limits_{t=1}^{T}\mathbb{Z}_{t-1}^T\epsilon_{t}\rightarrow_{d}N(0, E(\mathbb{Z}_{t}^T\Sigma_{\epsilon}\mathbb{Z}_{t}) ).\]
Hence
\begin{equation}\nonumber
\begin{split}
\sqrt{T}(\hat{\beta}_{OLS}^M-\beta)&=\sqrt{T}((\sum\limits_{t=1}^{T}(\mathbb{Z}_{t-1}^M)^T\mathbb{Z}_{t-1}^M)^{-1}\sum\limits_{t=1}^{T}(\mathbb{Z}_{t-1}^M)^T\mathbb{X}_t-\beta)\\
&=-(\frac{1}{T}\sum\limits_{t=1}^{T}(\mathbb{Z}_{t-1}+P_{t-1})^T(\mathbb{Z}_{t-1}+P_{t-1}))^{-1}\frac{1}{\sqrt{T}}\sum\limits_{t=1}^{T}(\mathbb{Z}_{t-1}+P_{t-1})^T(P_{t-1}\beta)\\
&+(\frac{1}{T}\sum\limits_{t=1}^{T}(\mathbb{Z}_{t-1}+P_{t-1})^T(\mathbb{Z}_{t-1}+P_{t-1}))^{-1}\frac{1}{\sqrt{T}}\sum\limits_{t=1}^{T}(\mathbb{Z}_{t-1}+P_{t-1})^T\epsilon_t\\
&\rightarrow_d N(0, P^{-1}QP^{-1} ).
\end{split}
\end{equation}

\item[(2)] 
For diverging $N$,

\begin{equation*}
\begin{split}
 &\ \ \ \ \frac{1}{\sqrt{T}}D(\sum_{t=1}^{T}(\mathbb{Z}_{t-1}^M)^T\mathbb{Z}_{t-1}^M)(\hat{\beta}_{OLS}^M-\beta)\\
 &=\frac{1}{\sqrt{T}}D(\sum_{t=1}^{T}(\mathbb{Z}_{t-1}^M)^T\mathbb{Z}_{t-1}^M)((\sum\limits_{t=1}^{T}(\mathbb{Z}_{t-1}^M)^T\mathbb{Z}_{t-1}^M)^{-1}\sum\limits_{t=1}^{T}(\mathbb{Z}_{t-1}^M)^T\mathbb{X}_t-\beta)\\
 &=-\frac{1}{\sqrt{T}}D(\sum_{t=1}^{T}(\mathbb{Z}_{t-1}+P_{t-1})^TP_{t-1}\beta)+\frac{1}{\sqrt{T}}D(\sum_{t=1}^{T}(\mathbb{Z}_{t-1}+P_{t-1})^T\epsilon_{t}).
 \end{split}
 \end{equation*}
Since $||P_{t-1}||_{max}=o_p(\frac{1}{\sqrt{T}})$, 
\[\mathbb{Z}_{t-1}+P_{t-1}\rightarrow_p E(\mathbb{Z}_{t}),\]
\[-\frac{1}{\sqrt{T}}D(\sum_{t=1}^{T}(\mathbb{Z}_{t-1}+P_{t-1})^TP_{t-1}\beta)\rightarrow_p 0,\]
and 
\[\frac{1}{\sqrt{T}}D\sum_{t=1}^{T}P_{t-1}^T\epsilon_{t} \rightarrow_p 0,\]
by Lemma \ref{L:CLTDOLS},
\[\frac{1}{\sqrt{T}}D\sum\limits_{t=1}^{T}\mathbb{Z}_{t-1}^T\epsilon_{t}\rightarrow_{d}N(0, DE(\mathbb{Z}_{t}^T\Sigma_{\epsilon}\mathbb{Z}_{t})D^T ).\]
Thus,
\begin{equation*}
\begin{split}
 &\ \ \ \ \frac{1}{\sqrt{T}}D(\sum_{t=1}^{T}(\mathbb{Z}_{t-1}^M)^T\mathbb{Z}_{t-1}^M)(\hat{\beta}_{OLS}^M-\beta)\\
 &=-\frac{1}{\sqrt{T}}D(\sum_{t=1}^{T}(\mathbb{Z}_{t-1}+P_{t-1})^TP_{t-1}\beta)+\frac{1}{\sqrt{T}}D(\sum_{t=1}^{T}(\mathbb{Z}_{t-1}+P_{t-1})^T\epsilon_{t})\\
 &\rightarrow_d N(0, DQD^T).
 \end{split}
 \end{equation*}
\end{itemize}

\end{itemize}
The proof for the GLS estimator follows along similar lines to that of the OLS estimator.
\end{proof}

\section{Tables and Figures} \label{tf}

\begin{table}[H]
\centering
\caption{$N=100$, $q_1=q_2=1$, $\rho=0.5$, $\Phi$ banded matrix of width 5 and $T=400$. Performance evaluation for band width of $W$ equal to 1, 5, 25 and 50.}
\resizebox{\columnwidth}{!}{
\begin{tabular}{c|c|c|c|c|c|c|c|c|c}

\hline
&Estimator&\multicolumn{2}{|c|}{BW=1}&\multicolumn{2}{|c|}{BW=5}&\multicolumn{2}{|c|}{BW=25}&\multicolumn{2}{|c}{BW=50}\\ \hline
&True value&Estimate&RMSE&Estimate&RMSE&Estimate&RMSE&Estimate&RMSE\\ \hline
$a_1\sim a_{25}$&0.1&0.097&0.400 &0.097&0.373 &0.098&0.373 &0.098&0.385 \\ 
$a_{26}\sim a_{50}$&0.2&0.197&0.192 &0.196&0.181 &0.196&0.178 &0.195&0.183 \\
$a_{51}\sim a_{75}$&0.3&0.296&0.122 &0.295&0.114 &0.295&0.113 &0.297&0.112 \\ 
$a_{76}\sim a_{100}$&0.4&0.396&0.086 &0.396&0.080 &0.397&0.079 &0.395&0.079 \\ \hline
$b_1\sim b_{25}$&0.4&0.399&0.126 &0.399&0.185 &0.397&0.313 &0.374&0.462 \\ 
$b_{26}\sim b_{50}$&0.3&0.300&0.167 &0.301&0.255 &0.294&0.493 &0.278&0.618 \\ 
$b_{51}\sim b_{75}$&0.2&0.200&0.237 &0.198&0.376 &0.191&0.740 &0.172&0.946 \\ 
$b_{76}\sim b_{100}$&0.1&0.101&0.450 &0.100&0.713 &0.095&1.274 &0.075&1.879 \\ \hline
$\gamma_1\sim \gamma_{3}$&-0.8&-0.803&0.025 &-0.804&0.024 &-0.804&0.025 &-0.803&0.027 \\ 
$\gamma_{4}\sim \gamma_{5}$&-0.4&-0.402&0.031 &-0.401&0.031 &-0.403&0.031 &-0.401&0.034 \\ 
$\gamma_{6}\sim \gamma_{7}$&0.4&0.402&0.048 &0.402&0.047 &0.402&0.050 &0.401&0.054 \\ 
$\gamma_{8}\sim \gamma_{10}$&0.8&0.803&0.032 &0.804&0.032 &0.803&0.033 &0.804&0.036 \\ \hline

\end{tabular}
}
\label{t6}
\end{table}

\begin{table}[H] 
\centering
\caption{$N=100$, $q_1=q_2=1$, $T=400$ and$ W$ and $\Phi$ banded matrices of width 5. Performance evaluation for $\rho=0.2,0.4,0.6,0.8$.}
\resizebox{\columnwidth}{!}{
\begin{tabular}{c|c|c|c|c|c|c|c|c|c}
 
\hline
&Estimator&\multicolumn{2}{|c|}{$\rho=0.2$}&\multicolumn{2}{|c|}{$\rho=0.4$}&\multicolumn{2}{|c|}{$\rho=0.6$}&\multicolumn{2}{|c}{$\rho=0.8$}\\ \hline
&True value&Estimate&RMSE&Estimate&RMSE&Estimate&RMSE&Estimate&RMSE\\ \hline
$a_1\sim a_{25}$&0.1&0.097&0.389 &0.097&0.380 &0.097&0.361 &0.097&0.340 \\ 
$a_{26}\sim a_{50}$&0.2&0.197&0.189 &0.197&0.182 &0.196&0.175 &0.197&0.162 \\ 
$a_{51}\sim a_{75}$&0.3&0.296&0.121 &0.296&0.118 &0.296&0.111 &0.296&0.101 \\ 
$a_{76}\sim a_{100}$&0.4&0.396&0.085 &0.396&0.082 &0.396&0.076 &0.396&0.071 \\ \hline
$b_1\sim b_{25}$&0.4&0.398&0.227 &0.398&0.203 &0.399&0.168 &0.398&0.131 \\ 
$b_{26}\sim b_{50}$&0.3&0.298&0.319 &0.300&0.273 &0.300&0.231 &0.298&0.176 \\ 
$b_{51}\sim b_{75}$&0.2&0.201&0.468 &0.199&0.415 &0.199&0.339 &0.201&0.253 \\ 
$b_{76}\sim b_{100}$&0.1&0.099&0.910 &0.101&0.775 &0.100&0.644 &0.101&0.478 \\ \hline
$\gamma_1\sim \gamma_{3}$&-0.8&-0.803&0.027 &-0.803&0.025 &-0.803&0.023 &-0.803&0.020 \\ 
$\gamma_{4}\sim \gamma_{5}$&-0.4&-0.403&0.036 &-0.401&0.033 &-0.402&0.029 &-0.401&0.026 \\ 
$\gamma_{6}\sim \gamma_{7}$&0.4&0.402&0.055 &0.401&0.052 &0.402&0.046 &0.401&0.040 \\ 
$\gamma_{8}\sim \gamma_{10}$&0.8&0.803&0.035 &0.804&0.033 &0.804&0.030 &0.804&0.027 \\ \hline
 
\end{tabular}
}
\label{t7}
\end{table}

\begin{table}[H] 
\centering
\caption{$N=100$, $q_1=q_2=1$, $T=400$ and $W$ banded matrix of width 5. Performance evaluation for factor model with $k=1,3,5,7$.}
\resizebox{\columnwidth}{!}{
\begin{tabular}{c|c|c|c|c|c|c|c|c|c}
 
\hline
&Estimator&\multicolumn{2}{|c|}{k=1}&\multicolumn{2}{|c|}{k=3}&\multicolumn{2}{|c|}{k=5}&\multicolumn{2}{|c}{k=7}\\ \hline
&True value&Estimate&RMSE&Estimate&RMSE&Estimate&RMSE&Estimate&RMSE\\ \hline
$a_1\sim a_{25}$&0.1&0.100&0.208 &0.100&0.203 &0.100&0.203 &0.100&0.200 \\ 
$a_{26}\sim a_{50}$&0.2&0.200&0.102 &0.200&0.100 &0.200&0.100 &0.200&0.098 \\ 
$a_{51}\sim a_{75}$&0.3&0.300&0.067 &0.299&0.066 &0.300&0.065 &0.299&0.064 \\ 
$a_{76}\sim a_{100}$&0.4&0.400&0.048 &0.399&0.048 &0.399&0.047 &0.399&0.047 \\ \hline
$b_1\sim b_{25}$&0.4&0.400&0.125 &0.400&0.104 &0.400&0.094 &0.400&0.087 \\ 
$b_{26}\sim b_{50}$&0.3&0.300&0.171 &0.300&0.139 &0.300&0.124 &0.300&0.118 \\ 
$b_{51}\sim b_{75}$&0.2&0.200&0.252 &0.200&0.207 &0.200&0.187 &0.200&0.173 \\ 
$b_{76}\sim b_{100}$&0.1&0.101&0.486 &0.101&0.400 &0.101&0.363 &0.100&0.341 \\ \hline
$\gamma_1\sim \gamma_{3}$&-0.8&-0.800&0.006 &-0.800&0.005 &-0.800&0.005 &-0.800&0.006 \\ 
$\gamma_{4}\sim \gamma_{5}$&-0.4&-0.400&0.007 &-0.400&0.007 &-0.400&0.007 &-0.400&0.007 \\ 
$\gamma_{6}\sim \gamma_{7}$&0.4&0.400&0.012 &0.400&0.011 &0.400&0.011 &0.400&0.011 \\ 
$\gamma_{8}\sim \gamma_{10}$&0.8&0.800&0.007 &0.800&0.007 &0.800&0.007 &0.800&0.007 \\ \hline
 
\end{tabular}
}
\label{t12}
\end{table}

\begin{table}[H] 
\centering
\caption{$N=100$, $q_1=q_2=1$, $T=400$, $\rho=0.5$ and $W$ being a banded matrix of width 5. Performance evaluation with bandwidth of $\Phi$ being 1, 5, 25 and 50.}
\resizebox{\columnwidth}{!}{
\begin{tabular}{c|c|c|c|c|c|c|c|c|c}
 
\hline
&Estimator&\multicolumn{2}{|c|}{BW($\Phi$)=1}&\multicolumn{2}{|c|}{BW($\Phi$)=5}&\multicolumn{2}{|c|}{BW($\Phi$)=25}&\multicolumn{2}{|c}{BW($\Phi$)=50}\\ \hline
&True value&Estimate&RMSE&Estimate&RMSE&Estimate&RMSE&Estimate&RMSE\\ \hline
$a_1\sim a_{25}$&0.1&0.097&0.322 &0.097&0.373 &0.098&0.386 &0.098&0.407 \\ 
$a_{26}\sim a_{50}$&0.2&0.197&0.154 &0.196&0.181 &0.196&0.188 &0.195&0.196 \\ 
$a_{51}\sim a_{75}$&0.3&0.297&0.099 &0.295&0.114 &0.296&0.120 &0.297&0.123 \\ 
$a_{76}\sim a_{100}$&0.4&0.396&0.070 &0.396&0.080 &0.397&0.084 &0.395&0.087 \\ \hline
$b_1\sim b_{25}$&0.4&0.396&0.175 &0.399&0.185 &0.396&0.212 &0.399&0.230 \\ 
$b_{26}\sim b_{50}$&0.3&0.302&0.239 &0.301&0.255 &0.297&0.302 &0.299&0.317 \\ 
$b_{51}\sim b_{75}$&0.2&0.200&0.359 &0.198&0.376 &0.196&0.448 &0.197&0.479 \\ 
$b_{76}\sim b_{100}$&0.1&0.097&0.684 &0.100&0.713 &0.095&0.839 &0.100&0.907 \\ \hline
$\gamma_1\sim \gamma_{3}$&-0.8&-0.803&0.023 &-0.804&0.024 &-0.804&0.026 &-0.803&0.028 \\ 
$\gamma_{4}\sim \gamma_{5}$&-0.4&-0.402&0.029 &-0.401&0.031 &-0.402&0.035 &-0.403&0.033 \\ 
$\gamma_{6}\sim \gamma_{7}$&0.4&0.401&0.044 &0.402&0.047 &0.402&0.054 &0.402&0.055 \\ 
$\gamma_{8}\sim \gamma_{10}$&0.8&0.803&0.029 &0.804&0.032 &0.804&0.036 &0.803&0.036 \\ \hline
 
\end{tabular}
}
\label{t13}
\end{table}

\begin{table}[H] 
\centering
\caption{$N=100$, $\rho=0.5$, $q_1=q_2=2$ and $W$ and $\Phi$ being banded matrices of width 5. Performance evaluation for $T=150,300,450,600$.}
\resizebox{\columnwidth}{!}{
\begin{tabular}{c|c|c|c|c|c|c|c|c|c}
 
\hline
&Estimator&\multicolumn{2}{|c|}{T=150}&\multicolumn{2}{|c|}{T=300}&\multicolumn{2}{|c|}{T=450}&\multicolumn{2}{|c}{T=600}\\ \hline
&True value&Estimate&RMSE&Estimate&RMSE&Estimate&RMSE&Estimate&RMSE\\ \hline
$A_1$ &0.3& 0.298 & 0.143 & 0.299 & 0.100 & 0.299 & 0.082 & 0.300 & 0.071 \\ 
$A_2$ &0.3& 0.300 & 0.372 & 0.300 & 0.261 & 0.300 & 0.212 & 0.300 & 0.184 \\ 
$B_1$ &0.15& 0.146 & 1.068 & 0.148 & 1.034 & 0.149 & 1.022 & 0.149 & 1.016 \\ 
$B_2$ &0.15& 0.148 & 1.251 & 0.149 & 1.133 & 0.150 & 1.088 & 0.149 & 1.068 \\ 
$\gamma$ &0.5& 0.500 & 0.016 & 0.500 & 0.011 & 0.500 & 0.009 & 0.500 & 0.008 \\ \hline
 
\end{tabular}
}
\label{t30}
\end{table}

\begin{table}[H] 
\centering
\caption{$N=100$, $q_1=q_2=1$, $\rho=0.5$ and $W$ and $\Phi$ being banded matrices of width 5. Length of confidence interval and coverage probability for $T=150,300,450$.}
\begin{tabular}{c|c|c|c|c|c|c|c}
 
\hline
&Estimator&\multicolumn{2}{|c|}{T=150}&\multicolumn{2}{|c|}{T=300}&\multicolumn{2}{|c}{T=450}\\ \hline
&&CI&CP&CI&CP&CI&CP\\ \hline
 & OLS&  0.197 & 0.951 & 0.138 & 0.952 & 0.113 & 0.949 \\ 
  $a_i$ &GLS&  0.182 & 0.950 & 0.128 & 0.951 & 0.104 & 0.951 \\ 
  &EGLS& 0.182 & 0.950 & 0.128 & 0.951 & 0.104 & 0.951 \\ \hline
   & OLS& 0.387 & 0.948 & 0.271 & 0.951 & 0.220 & 0.950 \\ 
  $b_i$ &GLS& 0.370 & 0.950 & 0.259 & 0.951 & 0.210 & 0.950 \\ 
   &EGLS&0.370 & 0.950 & 0.259 & 0.951 & 0.210 & 0.950 \\ \hline
   & OLS& 0.035 & 0.954 & 0.024 & 0.948 & 0.020 & 0.948 \\ 
 $\gamma_i$ &GLS & 0.032 & 0.952 & 0.022 & 0.948 & 0.018 & 0.947 \\ 
   &EGLS& 0.032 & 0.952 & 0.022 & 0.948 & 0.018 & 0.948 \\ \hline
 
\end{tabular}
\label{t1}
\end{table}

\begin{table}[H] 
\centering
\caption{$N=100$, $q_1=q_2=1$, $\rho=0.5$, $\Phi$ band matrix of width 5 and $T=400$. Length of confidence interval and coverage probability for band width of $W$ equal to 1, 5, 25 and 50.}
\begin{tabular}{c|c|c|c|c|c|c|c|c|c}
 
\hline
&Estimator&\multicolumn{2}{|c|}{BW=1}&\multicolumn{2}{|c|}{BW=5}&\multicolumn{2}{|c|}{BW=25}&\multicolumn{2}{|c}{BW=50} \\ \hline
&&CI&CP&CI&CP&CI&CP&CI&CP\\ \hline
 & OLS& 0.113 & 0.952 & 0.119 & 0.952 & 0.118 & 0.952 & 0.117 & 0.951 \\ 
  $a_i$ &GLS& 0.105 & 0.952 & 0.110 & 0.953 & 0.108 & 0.952 & 0.108 & 0.952 \\ 
  &EGLS&  0.105 & 0.952 & 0.110 & 0.953 & 0.108 & 0.952 & 0.108 & 0.952 \\ \hline
   & OLS& 0.143 & 0.950 & 0.234 & 0.951 & 0.407 & 0.951 & 0.546 & 0.951 \\ 
  $b_i$ &GLS&  0.133 & 0.950 & 0.224 & 0.950 & 0.402 & 0.952 & 0.545 & 0.950 \\ 
   &EGLS& 0.133 & 0.950 & 0.224 & 0.950 & 0.402 & 0.952 & 0.544 & 0.950 \\ \hline
   & OLS& 0.021 & 0.951 & 0.021 & 0.949 & 0.021 & 0.949 & 0.021 & 0.950 \\ 
 $\gamma_i$ &GLS &0.019 & 0.952 & 0.019 & 0.952 & 0.019 & 0.951 & 0.019 & 0.952 \\ 
   &EGLS& 0.019 & 0.953 & 0.019 & 0.952 & 0.019 & 0.951 & 0.019 & 0.952 \\ \hline
 
\end{tabular}
\label{t2}
\end{table}

\begin{table}[H] 
\centering
\caption{$N=100$, $q_1=q_2=1$, $T=400$ and W band matrix of width 5. Length of confidence interval and coverage probability for $\rho=0.2,0.4,0.6,0.8.$.}
\begin{tabular}{c|c|c|c|c|c|c|c|c|c}
 
\hline
&Estimator&\multicolumn{2}{|c|}{$\rho=0.2$}&\multicolumn{2}{|c|}{$\rho=0.4$}&\multicolumn{2}{|c|}{$\rho=0.6$}&\multicolumn{2}{|c}{$\rho=0.8$} \\ \hline
&&CI&CP&CI&CP&CI&CP&CI&CP\\ \hline
 & OLS&0.112 & 0.951 & 0.116 & 0.948 & 0.126 & 0.949 & 0.163 & 0.949 \\ 
  $a_i$ &GLS&0.111 & 0.950 & 0.111 & 0.949 & 0.110 & 0.950 & 0.108 & 0.950 \\ 
  &EGLS&  0.111 & 0.950 & 0.111 & 0.949 & 0.110 & 0.950 & 0.108 & 0.950 \\ \hline
   & OLS& 0.258 & 0.949 & 0.242 & 0.949 & 0.226 & 0.949 & 0.226 & 0.948 \\ 
  $b_i$ &GLS&  0.257 & 0.949 & 0.237 & 0.950 & 0.208 & 0.950 & 0.169 & 0.951 \\ 
   &EGLS& 0.257 & 0.949 & 0.237 & 0.950 & 0.208 & 0.950 & 0.169 & 0.951 \\ \hline
   & OLS&0.020 & 0.945 & 0.020 & 0.947 & 0.022 & 0.951 & 0.029 & 0.949 \\ 
 $\gamma_i$ &GLS &0.020 & 0.946 & 0.019 & 0.948 & 0.019 & 0.947 & 0.019 & 0.945 \\ 
   &EGLS& 0.020 & 0.945 & 0.019 & 0.948 & 0.019 & 0.947 & 0.019 & 0.945 \\ \hline
 
\end{tabular}
\label{t3}
\end{table}

\begin{table}[H] 
\centering
\caption{$N=100$, $q_1=q_2=1$, $T=400$ and $W$ being a banded matrix of width 5. Length of confidence interval and coverage probability for factor model with $k=1,3,5,7$, $F_t\sim N(0,1)$ and $\lambda_i \sim U(0,1)$.}
\begin{tabular}{c|c|c|c|c|c|c|c|c|c}
 
\hline
&Estimator&\multicolumn{2}{|c|}{k=1}&\multicolumn{2}{|c|}{k=3}&\multicolumn{2}{|c|}{k=5}&\multicolumn{2}{|c}{k=7} \\ \hline
&&CI&CP&CI&CP&CI&CP&CI&CP\\ \hline
 & OLS&0.125 & 0.949 & 0.148 & 0.952 & 0.166 & 0.950 & 0.181 & 0.953 \\ 
  $a_i$ &GLS& 0.109 & 0.949 & 0.107 & 0.949 & 0.105 & 0.950 & 0.103 & 0.951 \\ 
  &EGLS& 0.109 & 0.949 & 0.107 & 0.951 & 0.105 & 0.950 & 0.103 & 0.951 \\ \hline
   & OLS& 0.228 & 0.953 & 0.218 & 0.952 & 0.224 & 0.950 & 0.232 & 0.952 \\ 
  $b_i$ &GLS& 0.199 & 0.951 & 0.158 & 0.951 & 0.144 & 0.950 & 0.136 & 0.949 \\ 
   &EGLS& 0.199 & 0.951 & 0.158 & 0.951 & 0.145 & 0.950 & 0.137 & 0.950 \\ \hline
   & OLS& 0.023 & 0.945 & 0.028 & 0.955 & 0.032 & 0.953 & 0.036 & 0.949 \\ 
 $\gamma_i$ &GLS &0.020 & 0.945 & 0.020 & 0.951 & 0.020 & 0.955 & 0.020 & 0.954 \\ 
   &EGLS& 0.020 & 0.944 & 0.020 & 0.951 & 0.020 & 0.957 & 0.020 & 0.958 \\ \hline
 
\end{tabular}
\label{t4}
\end{table}

\begin{table}[H] 
\centering
\caption{$N=100$, $q_1=q_2=1$, $T=400$, $\rho=0.5$ and $W$ being a banded matrix of width 5. Length of confidence interval and coverage probability with $\Phi$ a banded matrix with bandwidth 1, 5, 25 and 50.}
\begin{tabular}{c|c|c|c|c|c|c|c|c|c}
 
\hline
&Estimator&\multicolumn{2}{|c|}{BW($\Phi$)=1}&\multicolumn{2}{|c|}{BW($\Phi$)=5}&\multicolumn{2}{|c|}{BW($\Phi$)=25}&\multicolumn{2}{|c}{BW($\Phi$)=50} \\ \hline
&&CI&CP&CI&CP&CI&CP&CI&CP\\ \hline
 & OLS&0.129 & 0.950 & 0.119 & 0.952 & 0.113 & 0.951 & 0.112 & 0.950 \\ 
  $a_i$ &GLS& 0.099 & 0.949 & 0.110 & 0.953 & 0.110 & 0.951 & 0.111 & 0.950 \\ 
   &EGLS& 0.099 & 0.949 & 0.110 & 0.953 & 0.110 & 0.951 & 0.111 & 0.950 \\ \hline
   & OLS&0.255 & 0.947 & 0.234 & 0.951 & 0.249 & 0.950 & 0.259 & 0.950 \\ 
  $b_i$ &GLS& 0.233 & 0.949 & 0.224 & 0.950 & 0.244 & 0.951 & 0.256 & 0.950 \\ 
   &EGLS& 0.232 & 0.948 & 0.224 & 0.950 & 0.244 & 0.951 & 0.256 & 0.950 \\ \hline
   & OLS&0.024 & 0.956 & 0.021 & 0.949 & 0.020 & 0.945 & 0.020 & 0.947 \\ 
 $\gamma_i$ &GLS & 0.019 & 0.954 & 0.019 & 0.952 & 0.020 & 0.945 & 0.020 & 0.945 \\ 
   &EGLS& 0.019 & 0.954 & 0.019 & 0.952 & 0.020 & 0.944 & 0.020 & 0.945 \\ \hline
 
\end{tabular}
\label{t10}
\end{table}

\begin{table}[H]
\centering
\caption{Confidence intervals of exogenous variables for Spring\label{Tab:lEGLSCI1}}
\begin{tabular}{rrrr}
  \hline
  &lower bound & estimate & upper bound\\
  \hline
air temperature & 0.000 & 0.001 & 0.002 \\ 
wind speed rate & -0.050 & -0.047 & -0.043 \\ 
sky condition total coverage & -0.014 & -0.012 & -0.009 \\ 
relative humidity & -0.003 & -0.003 & -0.002 \\ 
   \hline
\end{tabular}
\end{table}

\begin{table}[H]
\centering
\caption{Confidence intervals of exogenous variables for Summer\label{Tab:lEGLSCI2}}
\begin{tabular}{rrrr}
  \hline
  &lower bound & estimate & upper bound\\
    \hline
air temperature & -0.008 & -0.006 & -0.005 \\ 
wind speed rate & -0.047 & -0.043 & -0.040 \\ 
sky condition total coverage & -0.011 & -0.009 & -0.007 \\ 
relative humidity & -0.004 & -0.003 & -0.003 \\ 
   \hline
\end{tabular}
\end{table}

\begin{table}[H]
\centering
\caption{Confidence intervals of exogenous variables for Fall\label{Tab:lEGLSCI3}}

\begin{tabular}{rrrr}
  \hline
  &lower bound & estimate & upper bound\\
    \hline
air temperature & -0.007 & -0.006 & -0.005 \\ 
wind speed rate & -0.069 & -0.066 & -0.062 \\ 
sky condition total coverage & -0.008 & -0.006 & -0.004 \\ 
relative humidity & -0.004 & -0.004 & -0.004 \\ 
   \hline
\end{tabular}
\end{table}

\begin{table}[H]
\centering
\caption{Confidence intervals of exogenous variables for Winter\label{Tab:lEGLSCI4}}

\begin{tabular}{rrrr}
  \hline
  &lower bound & estimate & upper bound\\
    \hline
air temperature & -0.013 & -0.011 & -0.010 \\ 
wind speed rate & -0.062 & -0.058 & -0.053 \\ 
sky condition total coverage & -0.010 & -0.008 & -0.006 \\ 
relative humidity & -0.003 & -0.003 & -0.002 \\ 
   \hline
\end{tabular}
\end{table}

\section{The Residual Bootstrap for the NAR model}
\label{Appendix-bootstrap}

Consider the NAR$(q_1,q_2)$ model:
\[ \mathbb{X}_{t}=A \mathbb{X}_{t-1} + B W \mathbb{X}_{t-1} +\mathbb{Y}_{t-1}\gamma+\epsilon_{t}=\mathbb{Y}_{t-1}\gamma+G\mathbb{X}_{t-1}+\epsilon_{t}=\mathbb{Z}_{t-1}\beta+ \epsilon_{t}\]

The main steps for applying the residual bootstrap to obtain the asymptotic distribution of the OLS estimator are:
\begin{itemize}
    \item[Step 1:] $\hat{\beta}_{OLS}=(\sum_{t=1}^{T}\mathbb{Z}_{t-1}^T\mathbb{Z}_{t-1})^{-1}\sum_{t=1}^{T}\mathbb{Z}_{t-1}^T\mathbb{X}_t$.
    \item[Step 2:] $\hat{\epsilon}_{t}=\mathbb{X}_t-\mathbb{Z}_{t-1}\hat{\beta}_{OLS}$, $t=1,2,\cdots,T$. Compute centered residuals $\hat{\epsilon}_t-\Bar{\epsilon}$ where $\Bar{\epsilon}=\frac{1}{T}\sum\limits_{t=1}^{T}\hat{\epsilon}_t$.
     \item[Step 3:] Draw $\epsilon^*_t$ ($t=0,1,\cdots,T$) with replacement from $\hat{\epsilon}_t-\Bar{\epsilon}$.
     \item[Step 4:] Define 
     \[\mathbb{X}^*_0=\epsilon^*_1\]\[\mathbb{X}^*_t=\mathbb{Z}^*_{t-1}\hat{\beta}_{OLS}+ \epsilon^*_{t},\ t=1,\cdots,N.\]
     Further, $\hat{\beta}^*_{OLS}=(\sum_{t=1}^{T}\mathbb{Z}_{t-1}^{*T}\mathbb{Z}^*_{t-1})^{-1}\sum_{t=1}^{T}\mathbb{Z}_{t-1}^{*T}\mathbb{X}^*_t$.
     \item[Step 5:] Repeat Step 3 and 4 a large number of times and get the empirical distribution of $\hat{\beta}^*_{OLS}$.
\end{itemize}

The main steps for applying the residual bootstrap to obtain the asymptotic distribution of the EGLS estimator are:
\begin{itemize}
    \item[Step 1:] $\hat{\beta}_{OLS}=(\sum_{t=1}^{T}\mathbb{Z}_{t-1}^T\mathbb{Z}_{t-1})^{-1}\sum_{t=1}^{T}\mathbb{Z}_{t-1}^T\mathbb{X}_t$.
    \item[Step 2:] $\hat{\epsilon}_{t,OLS}=\mathbb{X}_t-\mathbb{Z}_{t-1}\hat{\beta}_{OLS}$, $t=1,2,\cdots,T$ and use $\hat{\epsilon}_{t,OLS}$ to estimate $\Sigma_\epsilon(\hat\rho)$.
    \item[Step 3:] $\hat{\beta}_{EGLS}=(\sum_{t=1}^{T}\mathbb{Z}_{t-1}^T\Sigma_\epsilon(\hat\rho)^{-1}\mathbb{Z}_{t-1})^{-1}\sum_{t=1}^{T}\mathbb{Z}_{t-1}^T\Sigma_\epsilon(\hat\rho)^{-1}\mathbb{X}_t$.
     \item[Step 4:] $\hat{\epsilon}_{t,EGLS}=\mathbb{X}_t-\mathbb{Z}_{t-1}\hat{\beta}_{EGLS}$, $t=1,2,\cdots,T$. Compute centered residuals $\hat{\epsilon}_{t,EGLS}-\Bar{\epsilon}$ where $\Bar{\epsilon}=\frac{1}{T}\sum\limits_{t=1}^{T}\hat{\epsilon}_{t,EGLS}$.
     \item[Step 5:] Draw $\epsilon^*_t$ ($t=0,1,\cdots,T$) with replacement from $\hat{\epsilon}_{t,EGLS}-\Bar{\epsilon}$. 
     \item[Step 6:] Define 
     \[\mathbb{X}^*_0=\epsilon^*_1\]\[\mathbb{X}^*_t=\mathbb{Z}^*_{t-1}\hat{\beta}_{EGLS}+ \epsilon^*_{t},\ t=1,\cdots,N.\]
     Further, $\hat{\beta}^*_{EGLS}=(\sum_{t=1}^{T}\mathbb{Z}_{t-1}^{*T}\Sigma_\epsilon(\hat\rho)^{-1}\mathbb{Z}^*_{t-1})^{-1}\sum_{t=1}^{T}\mathbb{Z}_{t-1}^{*T}\Sigma_\epsilon(\hat\rho)^{-1}\mathbb{X}^*_t$.
     \item[Step 7:] Repeat Steps 5 and 6 a large number of times and get the empirical distribution of $\hat{\beta}^*_{EGLS}$.
\end{itemize}

The residual bootstrap is a popular resampling method for constructing confidence intervals for the model parameters in both parametric linear and non-linear auto-regressive time series models \citep{kreiss2012bootstrap}. The proposed NAR model fits well into this framework, since it is a linear and parametric auto-regressive model, while the existence of the network effect parameters reduces the parameter space dimension. Hence, utilizing the residual bootstrap for NAR model is appropriate, and also supported by the numerical results presented in Section \ref{sec:influence-error-term}.

\end{appendix}

\end{document}